\documentclass[11pt, oneside]{book}
\usepackage[a4paper, top=3cm, bottom=3cm]{geometry}
\usepackage{fancyhdr}
\usepackage{hyperref}
\usepackage{cite}
\usepackage{xcolor}
\usepackage{varioref}
\usepackage{mathrsfs}
\usepackage{amsmath}
\usepackage{amssymb}
\usepackage{setspace}
\makeatletter
\def\cleardoublepage{\clearpage\if@twoside \ifodd\c@page\else%
    \hbox{}%
    \thispagestyle{empty}
    \newpage%
    \if@twocolumn\hbox{}\newpage\fi\fi\fi}
\makeatother
\usepackage{fancyhdr}
\usepackage{tocloft}
\pdfoutput =1
\usepackage{graphics}
\usepackage{graphicx}
\DeclareGraphicsExtensions{.pdf}
\usepackage{float} 
\usepackage{amsmath}
\usepackage{amsfonts}   
\usepackage{amssymb}
\labelformat{chapter}{Chapter #1}
\labelformat{section}{Section #1}
\labelformat{subsection}{Section #1}
\labelformat{subsubsection}{Section #1}
\labelformat{equation}{Eq.~(#1)}

\begin{document}
\date{}

\begin{titlepage}
\begin{center}

\begin{huge}
	\textbf{Some Aspects of Quantum Mechanics and
		Quantum Field Theory on Quantum Space-Time \\\vspace {0.75cm}}
\end{huge}
\vfill

\normalsize
{\Large Thesis Submitted For The Degree Of}\\[2.2ex]
\textbf{\Large Doctor of Philosophy (Science)}\\[2ex]
In\\[2ex] 
{\Large \textbf{Physics (Theoretical)} }\\[2ex]
By\\[2ex]
\textbf{\Large Partha Nandi}

\vfill

\vfill

\textbf{{\Large \textbf{\large University of Calcutta, India} }}\\[2ex]
{\large 2021}

\end{center}
\end{titlepage}


\newpage

\begin{flushright}
\vspace*{\fill}
{\large{\it{To my beloved Mother ....}}}
\vspace*{\fill}
\end{flushright}
\thispagestyle{empty}
\newpage

\chapter*{Acknowledgments}
It is a pleasant duty to acknowledge my
indebtedness to all those who have contributed in different ways to the completion of this thesis. I thank them all for their co-operation, my heartfelt gratitude is to my supervisor Professor Biswajit Chakraborty (S. N. Bose National Centre for Basic Sciences) whom I owe my understanding of Physics. He is the person who suggested to me the ideas that have been developed in the following pages and actively guided me all along till the completion of this task.\\

My heartfelt thanks go out to my great teacher Prof. Debashis Chatterjee (Vijaygarh Joytish Roy College). I sincerely thank him for all those intense academic discussions.  I would like to thank Prof. Frederik G. Scholtz (Universiteit Stellenbosch) for sharing with me his insightful and innovative ideas which led to some very fruitful discussions. I thank Prof A. P. Balachandran and  Prof. Michel Berry, Prof. V.P Nair for always being there to help me in my work. I take this opportunity to convey my sincere gratitude to Prof. Sibasish Ghosh (IMSc, India ) for organising a short visit and holding valuable discussions. I am also grateful to Prof. Rabin Banerjee for his enthusiastic discussions on physics to clarify very elementary but at same time subtle concepts in physics. \vspace{0.5cm}
  
  I would also like to express my gratitude Prof. Amitava Lahiri, Prof. Archan S. Majumdar and Prof. Manu Mathur and Prof. Sunandan Gangopadhyay for helping me in my academics at this Centre. I am indebted to  Prof. Anirban Kundu, Prof. Indrajit Mitra and Prof. Debnarayana Jana, University of Calcutta, for giving me crucial advices and guidance, whenever I required.\\
  
  I am indebted to Prof. S.K Roy, ex. Director of SNBNCBS, for many facilities I enjoyed in this centre. I thank all the academic and administrative staff of S.N Bose centre for helping me in many ways. I am thankful to them for their services.
  This works been made possible through a fellowship of
  the S. N. Bose National Centre for Basic Sciences.
  I have been fortunate enough to count some very bright and brilliant persons namely Sayan Pal, Saptarshi Biswas, Anwesha Chakraborty, Sankarshan Sahu. It has been a rewarding experience working with them.\\
  
  My thesis will be incomplete if I don't acknowledge the role my mother has played. Without her unflinching support, this journey would have been impossible.  I dedicate this thesis to her.

\chapter*{List of publications}
\thispagestyle{empty}



\begin{enumerate}

\item \textit{``A note on broken dilatation symmetry in planar noncommutative theory",}\\
{\textbf{Partha Nandi}, Sankarshan Sahu, Sayan Kumar Pal}\\
e-Print:  arXiv:2101.07076v3 [hep-th] (To appear in \textbf{Nucl.Phys.B}).\\

\item \textit{``Emergence of a geometric phase shift in planar noncommutative quantum mechanics''}\\
{Saptarshi Biswas, \textbf{Partha Nandi}, Biswajit Chakraborty}\\
Published in: \textbf{ Phys.Rev.A 102 (2020) 2, 022231 }\\
e-Print: arXiv: 1911.03196 [hep-th].\\

\item \textit{``Revisiting quantum mechanics on non-commutative space-time''}\\
{\textbf{Partha Nandi}, , Sayan Kumar Pal,   Aritra N. Bose,  Biswajit Chakraborty }\\
Published in: \textbf{  Annals Phys. 386 (2017) 305-326.}\\
e-Print: arXiv: arXiv:1708.04769v1 .\\

\item \textit{``Particle dynamics and Lie-algebraic type of non-commutativity of space–time''}\\{\textbf{Partha Nandi},  Sayan Kumar Pal,   Ravikant Verma}\\
 Published in: \textbf{  Nucl.Phys.B 935 (2018) 183-197.}\\
 e-print: 1807.05062 [physics.gen-ph].\\

\item \textit{``Connecting dissipation and noncommutativity: A Bateman system case study''}\\ Sayan Kumar Pal,{\textbf{Partha Nandi}, Biswajit Chakraborty}\\
Published in: \textbf{ Phys.Rev.A 97 (2018) 6, 062110 }\\
e-print: arXiv: 1803.03334 [quant-ph].\\

\item \textit{``Effect of Non-commutativity of space-time on Thermodynamics of Photon gas''}\\ { Ravikant Verma, \textbf{Partha Nandi}}\\
Published in: \textbf{ Gen.Rel.Grav. 51 (2019) 11, 143  }\\
e-print: arXiv: 1810.07005 [physics.gen-ph].\\\\\\\\

\item \textit{``Effect of dynamical noncommutativity on the limiting mass of white dwarfs''}\\ { Sayan Kumar Pal, \textbf{Partha Nandi}}\\
Published in: \textbf{ Phys.Lett.B 797 (2019) 134859   }\\
e-print: arXiv: 1908.11206 [gr-qc].\\
\item \textit{``Fuzzy Classical Dynamics as a Paradigm for Emerging Lorentz Geometries''}\\ { F.G. Scholtz, \textbf{Partha Nandi}, S.K. Pal, B. Chakraborty }\\
e-print: arXiv: 1905.05018 [hep-th] (Submitted).\\
\item \textit{``Emergent entropy of exotic oscillators and squeezing in three-wave mixing process''}\\ { Sayan Kumar Pal, \textbf{Partha Nandi}, Sibasish Ghosh, Frederik G. Scholtz, Biswajit Chakraborty }\\
Published in: \textbf{Phys.Lett.A 403 (2021) 127397   }\\
e-print: arXiv: 2012.07166 [hep-th].\\

\item \textit{``A Conformally Invariant Unified Theory of Maxwell Fields and Linearized Gravity as Emergent Fields''}\\ {  \textbf{Partha Nandi}, Partha Ghose }\\
e-print: arXiv: 2106.07028 [gr-qc] (Submitted).\\

 $^{*}$This thesis is based on the publications numbered by [1,2,3].

\end{enumerate}
\newpage
\thispagestyle{empty}

~~~~~~~~~~~~~~~~~ ~~~~~~~~~~~~~~~~~~~~ ~~~~~~~~~~~~~

~~~~~~~~~~~~~~~~~~~~~~~~ ~~~~~~~~~~~~~~~~~~~~~~~~~~~~~~

~~~~~~~~~~~~~~~~~~~~~~~~~~~~~~~ ~~~~~~~~~~~~~~~~~~~~~~~~

~~~~~~~~~~~~~~~~~~~~~~~~ ~~~~~~~~~~~~~~~~~~~~~~~~~~~~~~~~~

~~~~~~~~~~~~~~~~~~~~~~~~~~~~~~~~~~~~~~~~~~~~ ~~~~~~~~~~~~~~~~~~

~~~~~~~~~~~~~~~~~~~~~~~~~~~~~~~~~~~~~~~~~~~~~~~~~~~~~~~~~~~~~~

~~~~~~~~~~~~~~~~~~~~~~~~~~~~~~~~~~~~~~~~~~~ ~~~~~~~~~~~~~~~~~~~

~~~~~~~~~~~~~~~~~~~~~~~~~~~~~~~~~~~~~ ~~~~~~~~~~~~~~~~~~~~~~~~




\newpage
\tableofcontents


\chapter{Introduction}

A reconciliation of a plausible structure of space-time at a short length scale i.e. in the vicinity of Planck
scale and the eventual emergence of space-time as we perceive it at a longer length scale as a pseudo-Riemannian differentiable manifold,
is arguably one of the most challenging problems in modern physics: the so-called `` holy grail" of contemporary theoretical physics.
 In order to formulate a full quantum theory of gravity, one needs a more systematic and sophisticated approach to confront the question regarding the nature of space-time itself at very short length scales or equivalently very high energy scales, like in the vicinity of Planck scale. As far as we are aware, it was Bronstein \cite{mpb} who was first person to advance the argument suggesting that any attempt to localize an event down to such a length scale will inevitably result in gravitational collapse, forming a black hole with an attendant event horizon. This, in turn, will hide all the sought-after information, thereby rendering these information inaccessible to any outside observer. This was carried forward further by Sergio Doplicher, Klaus Fredenhagen, and John Roberts \cite{dfr,db} in a more rigorous approach, where they could demonstrate the inevitability of such a scenario, if the basic tenets of both general relativity and quantum theory are extrapolated to that scale. They pointed out the need for a quantum notion of space -time at such short length scales where the quantum nature of space-time manifested through the non-commuting nature of the coordinates, which are now promoted to the level of operators. \\

On the other hand,  one can understand, some what heuristically, the main discrepancy between quantum mechanics and general relativity from Einstein’s equations: 
\begin{equation}
G_{\mu\nu}+\lambda g_{\mu\nu}=8\pi G T_{\mu\nu},
\label{gh}
\end{equation}
where the gravitational field, i.e. that
of the geometry of space-time, is captured by the curvature tensor (and hence by the metric tensor) which occurs in the left hand side of the above equation through a certain contracted form of the curvature tensor and gets related to the matter content of the system and is given by the energy-momentum tensor ($T_{\mu\nu}$) that occurs in the right hand side. But quantum theory implies that the energy-momentum tensor is necessarily operator valued and will have a significant role to play at a very high energy scale. This then readily implies that
the left-hand side of the equation (\ref{gh}) also must be quantized at this scale. Hence, a quantum structure of space-time seems quite plausible \cite{dbw}. The geometry of such quantum spaces possibly can only be described by the non-commutative geometry, as formulated by Alain Connes \cite{acon}.\\\\

Although envisaged initially to be an alternative and novel approach to quantum gravity, Connes and his collaborators, however, applied this framework of Non-commutative geometry to develop a completely unified description of the Standard model of particle physics \cite{ah, ahc,WDV,Lizzi,Ts,Ts1}, albeit at the classical level. Among its many successes includes a unified description of Higgs field and other
Yang-Mills gauge fields present in the model, which eventually facilitated the exact computation of Higgs mass \cite{A.HC,A.HC1}. Despite all these successes, there are certain shortcomings, which are
still present in the construction. Primary among them is that the model takes the space-time to be a $4D$ compact Riemannian differentiable manifold $M$ of Euclidean signature-rather than Lorentzian. However, one should note that modelling
the space-time, thus by a Riemannian manifold $M$, of course serves its purpose adequately at the scale of
Standard model, as one can expect to see any granular structure, associated with the above mentioned
quantum space-time only in the vicinity of Planck scale. One can recall in this context that the Standard
model scale is many orders below the GUT scale and which, in turn, is a couple of orders below the Planck
scale. And for them the total space is taken to be the so-called ``Almost commutative space" (AC): $M\times F$,
where $F$ is the space which is described by suitable matrix algebra, which is responsible for the
non-commutative structure in the whole construction. Gauge symmetries appear here through the group
of inner automorphism , which is a normal sub-group of the total automorphism group of $M \times F$,
while the outer automorphism group, obtained by quotienting,  gets identified with the group of
diffeomorphism. Despite its beautiful mathematical and conceptual edifice, however, it is clear that the
quantum structure of space-time should be taken seriously and be incorporated suitably, so that one
can go beyond this  framework of AC space, even if one tries to build a tentative and toy model of quantum gravity.\\

Before the success of the renormalization procedure had gained acceptance, the idea of quantized space-time was first proposed by Heisenberg himself in the late 1930s with the hope of taming the short-distance i.e. ultraviolet divergences of quantum
fields. He shared his idea with Rudolf Peierls, who eventually made use of it in electronic systems. On the other hand, Rudolf Peierls informed  Pauli about this idea, who
in turn told J. Robert Oppenheimer about it and finally, in 1947, his student Hartland Snyder, using this idea, published his articles on the subject\cite{hsn,hsn2}.
In the same
year C. N. Yang \cite{cny}  generalized it to curved de-Sitter space-time. However, due to the
success of the renormalization scheme, Snyder's work(s) did not receive much attention among his contemporaries. But eventually when activities towards the formulation of quantum gravity
was initiated again, it started to become quite transparent gradually that the usual models of space-time as a differentiable manifold has to given away and possibly replaced by some form of quantized space-time for which an intrinsic non-commutative space-time can be a good example. Furthermore, it was anticipated that the  quantized version of gravity may act as a regulator of quantum field theories. In fact, it has been pointed out in \cite{aak} that the  classical self-energy of a charged particle is automatically regularized in the presence of classical gravity itself. But if one goes further to incorporate the quantum effects to the self-energy of the charged
particle, this will then introduce additional divergences and to regulate these divergences, one will most probably require a theory of quantum gravity and a non-commutative geometry stemming from a non-commutative structure of space-time may become one of the strong
contenders in this case.\\

Apart from the quantum gravity point of view, the concept of space-time noncommutativity (with
time being an ordinary c-numbered parameter) has found application in various condensed matter phenomenon \cite{ysm,hpa,chernj,ipa,FDM} as well. The physics of the anomalous quantum Hall effect \cite{aeqh}, spin Hall effect \cite{spino,spio}
are other areas in which the non-commutativity among the coordinates show up. Besides, a Berry curvature in momentum space for a semi-classical Bloch electron has been used to explain the anomalous/spin/optical Hall effects. In fact, constant Berry curvature effects in this system arising due to the breaking of time-reversal symmetry induce a canonical type of non-commutativity among the spatial coordinates \cite{xio}. On the other hand, a rigorous derivation of the quantization of the Hall conductivity in quantum Hall effect used methods of non-commutative geometry\cite{JbE}. In the $2+1$ dimension, the non-commutative geometry of the lowest Landau level also plays a vital role in the recent work of Haldane \cite{Ghm} on the geometrical description of the fractional quantum Hall effect. Also, the effects of coordinate noncommutativity in $2+1$ quantum gravity can be realized by the introduction of anyons with fractional spin wherein non-commutative parameter is related to the fractional spin parameter, as has been discussed in \cite{PG1}.  In fact, a close correspondence between space-time quantization and $2+1$ dimensional gravity was shown by G.'t Hooft in \cite{PG2}.\\

Furthermore, in quantum space-time, spatial noncommutativity can be correlated to a possible curvature of momentum space \cite{Smj}, just as we know the
curvature in configuration space enforces momentum-space operators to satisfy non-trivial
commutation algebra. Indeed the non-commutative coordinates are realized as translation generators on momentum space.  A few years back, Townsend \cite{OPK} had shown that Planck length also appears in gravity due to the noncommutativity between the generators of space-time translations. It has been naturally suggested that, along with the spatial components of space-time coordinates, the canonically conjugate momentum components too should satisfy a non-commutative algebraic structure \cite{Tpsing} at the quantum level because momentum noncommutativity arise naturally as a consequence of coordinate noncommutativity. The reciprocity theorem proposed by Max Born in 1938 \cite{mborn} had earlier hinted at this. In a similar vein, \cite{gac} discovered that the noncommutative structure of the spatial and momentum components can be connected to the respective curvatures in momentum and coordinate spaces.\\

String theory also exhibits space-time uncertainty, as first pointed out in \cite{mli}, which is another approach to describe the quantum theory of gravity. Not only that, the method of non-commutative geometry has been used rigorously to study the duality symmetries of string theory \cite{2l,3l}.  It has also been shown that non-commutative geometry appears due to the toroidal compactification of the matrix model\cite{A1}. In fact, these matrix models lead to a non-commutative version of Yang-Mills theory as an effective quantum field theory. Seiberg and Witten in
their seminal paper \cite{A2} showed how this space-time noncommutativity can also appear as a low energy effect in string theory as a kind of effective theory by studying the physics of $D$ branes in the background of nonzero Neveu-Schwarz $B_{\mu\nu}$ field.
They further demonstrated, by explicit construction, the existence of a nontrivial transformation connecting ordinary gauge fields and non-commutative gauge fields, which is usually called the Seiberg-Witten map and using which one can construct an equivalent commutative description of non-commutative gauge theory. Another important feature of this map is concerned with the notion of the emergent gravity phenomenon.  But, there are some novel ambiguities \cite{T1,T2} in the Seiberg-Witten map which are particularly more  significant in the presence of matter fields. In fact, a few years back, it has been shown by Scholtz et al. in \cite{FG2} that this map does not always generate a spectrum preserving transformation for interacting quantum Hall system. All these studies, nevertheless,  indicates
that space and time will become non-commutative operators in a more fundamental formulation of string/M theory. This triggered renewed interest in quantum (non-commutative) space-time and formulation on non-commutative quantum mechanics, and quantum field theories, on such spaces.\\

To describe field theory on a  quantum (non-commutative) space-time, there are two approaches.
In one approach, we need to define fields whose base manifold is non-commutative and treated as operators in some Hilbert space. In the other method, fields are considered as some functions of usual commutative space-time variables, and noncommutativity
among these variables is incorporated through an appropriate ``star" ($\star$) product \cite{czacos,A3}, which corresponds to a deformed multiplication map and is obtained by deforming point wise multiplication of functions. Perhaps the most commonly studied case is that of the Moyal space-time: $[\hat{x}^{\mu},\hat{x}^{\nu}]=i\theta^{\mu\nu}$; i. e. when the non-commutative parameter ($\theta$) is taken as a constant matrix. Field theories defined on such a  quantum (non-commutative) space-time are based on the Weyl-Wigner correspondence, in which all products are simply replaced by the Moyal ``star" products ($\star_{M}$) \cite{A4}, defined by
\begin{equation}
(O_{1}\star_{M}O_{2})(x)=e^{\frac{i}{2}\theta^{\mu\nu}\partial_{\mu}\partial^{'}_{\nu}}O_{1}(x)O_{2}(x^{'})|_{x^{'}\rightarrow x}
\end{equation}

 As a result the mass terms in the action are not modified by the $\star_{M}$-product because it contributes up to an additional surface term in the action; terms beyond quadratic interaction terms are only modified. Hence one may arrive at an undeformed free propagator. Historically the first unexpected behavior of non-commutative quantum field theories was pointed out by Minwalla, van Raamsdonk and Seiberg \cite{smr}. Using Moyal star product formulation, these authors could show that the perturbative analysis of quantum field theories on non-commutative space-time are afflicted with some novel kind of divergences, where the  ultraviolet and infrared divergences (UV/IR mixing) gets mixed up. In other words, the high and low energy scales of energy get entangled due to presence of noncommutativity among space-time coordinates which may affect the renormalization procedure.
In fact this UV/IR mixing problem is not only the 
hallmark of relativistic fields,  non-relativistic scalar field theory over $2+1$ dimensional non-commutative manifolds also shows this behavior \cite{jmgk}. On the other hand, in a more recent paper \cite{basup}, it has been pointed out that the Voros star product is more appropriate in dealing with quantum mechanics on Moyal plane, as it lends itself to a probabilistic interpretation, unlike Moyal star product. The analysis of \cite{basup} were, of course, restricted to the level of non-commutative quantum mechanics, where the position variables where the hermitian observable. On the other hand, at the level of QFT on non-commutative spaces also similar results were obtained by \cite{F4}, although the position variables are no longer observables there and are merely the labels of continuous degrees of freedom. This should be pointed out in this context that the Voros star product naturally arises whenever coherent states are used as basis in the quantum Hilbert space. However, a slight variant of this coherent state-based formulation of quantum field theory on non-commutative space-time has been developed recently in \cite{jac}. In this approach, instead of using star product,  any element of algebra generated by the non-commutative space-time operators are replaced by the coherent state expectation values of the non-commutative space-time and introduce a modified coherent state representation of plane wave solution in an ad-hoc way which may not satisfy the orthonormality condition without Voros start product which leads to a nontrivial modification at the level of free propagators. The basic advantage of this new star-product independent formulation is that the whole formalism of the perturbative field theory is UV finite which indicates that there is no UV/IR mixing problem \cite{jac1}. But, a major disadvantage of this formulation is that it does not provide a straightforward reduction to single-particle non-commutative quantum mechanics. \\

It is worthwhile to mention at this stage that, like in the commutative case \cite{TPd}, non-commutative quantum mechanics can also be recovered from relativistic non-commutative quantum field theory through some suitable non-relativistic limit, as has been extensively studied in \cite{pmh}. Except for the new feature that particles of opposite charges should be associated with opposite non-commutative parameter ($\theta$). On the other hand, a rigorous formulation of non-commutative quantum mechanics with spatial noncommutativity, making use of Hilbert-Schmidt (HS) operators has been studied in great detail \cite{F1,F2,F3}. The advantage of this HS operatorial formulation is that it bypasses the use of any star product and enables one to confront head-on the operatorial nature of spatial coordinates and evade any kind of ambiguities that may arise from the use of different kinds of star product, like that Moyal or Voros \cite{F5,F6,F7}. However, there was no follow-up work to extend this HS operatorial formulation beyond the so-called first quantization to second quantization or for that matter the full-fledged relativistic quantum field theory (QFT). Despite this, one can expect to capture some of the surviving features, if not all, of this completely non-commutative
relativistic QFT even at the level of non-commutative quantum mechanics. This is simply because of the fact that, like in the commutative case \cite{A D Boozer}, here too one can definitely expect to interpret non commutative quantum mechanics as $0+1$ dimensional QFT. In other words, non-commutative quantum mechanics may be
considered as a laboratory where  some of the properties of non-commutative quantum field theories can still be studied in a simpler setting. In fact, one can go other way around and try to formulate QFT, as a sort of a unique and inevitable formulation, starting from that of relativistic quantum mechanics, in a bottom up approach , as was shown by Steven Weinberg in his famous book on QFT \cite{swg}. In other words, the way we know about QFT today, is the only possibility, where a merger of both the principles of special relativity and quantum mechanics is envisaged.\\

\section{ Organization of the thesis}
In the present thesis, however, which is mainly based on the works \cite{pn1,pn2,pn3}, no attempts will be made to address any of the issues related to quantum gravity directly. Rather, this thesis is
devoted to studying various aspects of quantum mechanics on non-commutative space-time and to capture some of the surviving aspects of symmetries of quantum field theory on such space-time, illustrated through toy models in $(0 + 1)$ dimension. This allows one to gain some insights into this and other related issues in a more transparent manner, as we have explained above. The outline of this thesis is as follows:\\

In chapter 2, we discuss the emergence of quantum (non-commutative) space by considering a quantum Hall system involving a pair of interacting oppositely charged
particles in a region of high magnetic field (uniform and static) in the low mass limit. We then show how the Lagrangian of the system can be directly mapped to a harmonic oscillator in Moyal plane, referred to here as exotic oscillators. Therefore, noncommutativity is an emergent feature of this physical model here, unlike in many other works where it is conferred with a fundamental status. Subsequently, a quantum mechanical analysis of this model is carried out using
the Hilbert-Schmidt operator formulation of non-commutative quantum mechanics. Then a functional integral approach is adopted to show that Ward-Takahashi (W-T) identities corresponding to the dilation transformations are anomalous in the sense that additional contributions in the form of quantum corrections from the Jacobian due to time dilatation transformation is obtained. This is over and above the classical dilatation symmetry breaking term. To the best of our knowledge, this has remained unexplored in the literature from the point of view of $0 + 1$ dimension nonlocal field theory. The anomalous term is then computed and regularized following Fujikawa's method.\\

The appearance of the adiabatic geometrical phase in the system of a planar isotropic time-dependent harmonic oscillator on a non-commutative phase space has been discussed in chapter 3. Using a non-canonical phase space transformation we study the effect of noncommutativity analytically. The emergent nature of geometric phase-related issues is addressed in Heisenberg picture. It is shown that the additional geometrical phase over and above the dynamical phase is then found to be  given in terms of a product of the non-commutative parameters. Then we also discuss the correspondence between Berry's phase and classical Hannay angle shifts.\\

In the last chapter (Chapter 4) we develop the concept of quantum space-time in $1 + 1$ dimension, taken to be of Moyal type. Then we discuss non-relativistic quantum mechanics on such a space-time. It is demonstrated that utilising a coherent state basis, an effective commutative theory can be written down starting from an abstract form of the Schr\``odinger equation using Hilbert-Schmidt operators. The coherent state basis offers a very natural description of canonical type non-commutative space-time wherein one can obtain the commutative limit in an explicit manner, which also provides some calculation advantages. These are, apart from the fact that in this coherent state based formulation, the probabilistic interpretation comes out in a very transparent manner.  We, next, concern ourselves with several problems that are often associated with quantum mechanics. Finally, we end up with some concluding remarks in chapter 5.\\

	\chapter{Planar noncommutativity and broken dilatation symmetry}
	\label{quan-model}
By now there is already a huge body of literature dedicated to the study of quantum theories
on the background of quantum spacetime. Postulating a simple structure for the non-vanishing
commutator algebra between operator-valued space-time coordinates is expected to shed
some insights into the future theory of quantum gravity \cite{jGi}, through these quantum theories. Normally people introduce noncommutativity between the spatial coordinates by hand. However, in certain non-relativistic planar system, non-commutative coordinates arise naturally. For example, the Landau problem in a strong magnetic field with the lowest Landau-level projection \cite{PG,for}. Also, the quantum space (non-commutative space) indicates the existence of a minimal length scale in the theory. On the other hand, scale or dilatation symmetry (exact or broken) in the corresponding commutative theory has a long history \cite{fS. Coleman,R. Jackiw}. In this chapter, we would like to study the fate of dilatation symmetry in the planar non-commutative theory both in classical and quantum level.
Note that the breaking of dilatation symmetry is somewhat expected at the classical level itself, because of the occurrence of a natural scale coming from the non-commutative parameter. What we are interested in to is study here is to ascertain whether there is any additional contribution to the breaking term, arising from the process of quantization i.e over and above the classical effect, which we can refer to as the anomaly term. And for that we consider here a non-relativistic interacting quantum Hall system \cite{SZABO,RJ} with the usual constant magnetic field. Here noncommutativity between spatial coordinates emerges naturally as a consequence of the deformed symplectic structure among the spatial coordinates at a large magnetic field limit.\\

The chapter is organized as follows. In sec. 2.1, following the ``Peierls substitution" scheme \cite{WE}, we demonstrate how deformed symplectic structure arises naturally at the
classical level itself, among the spatial coordinates of a two-dimensional harmonic oscillator in a planar
system of electric dipole under a very strong magnetic field, normal to the plane, where an additional harmonic interaction between the charges with an impurity is also considered.
As we shall see that a first-order formulation is quite adequate in this case i.e. in the limit of large magnetic field. In sec.2.2  we will move on to discuss the quantum picture of our considered model where we  will give a short review on the formulation of non-commutative quantum mechanics for a two-dimensional harmonic oscillator in the framework of Hilbert-Schmidt operators. This is completely an operatorial formulation of non-commutative quantum mechanics, initiated mainly in \cite{fg}. Next, in sec 2.3, by using the coherent state approach, we construct a path-integral action and functional integral for the non-commutative
dynamical system. This also makes a connection with Chern-Simons quantum mechanics. Symplectic brackets can then be computed, rather trivially from this
path-integral action and the corresponding non-commutative structure between the position coordinates is seen to to emerge naturally here and therefore also serves as a consistency check. In section 2.4,  we treat the non-commutative quantum mechanics as a (0+1) dimensional non-canonical (non-commutative) field theory. Then by exploiting Darboux's theorem at the level of functional integral we explicitly compute Ward-Takahashi identities due to the time dilatation symmetry of an equivalent commutative effective theory.
Thereafter we shall demonstrate that the Ward-Takahashi identities associated with dilatation symmetry (broken) are anomalous. Also from these identities we infer that there are three fold breaking of dilatation symmetry in non-commutative theory: (i) explicit
breaking due to presence of spring constant at the classical level itself, (ii) deformation in symplectic structure
also provides another classical breaking, and finally (iii) the third is due to the anomaly (a
purely quantum effects) which is induced from noncommutativity. The anomaly term
needs to be regularized and is done in section 2.5 using Fujikawa's regulator, which is then
found to be proportional to the product of the spring constant of the harmonic oscillator and the non-commutative parameter. Sec.2.6 is reserved for the concluding remarks.\\

\section{The classical model of non-commutative space}

We consider a planar system consisting of a pair of interacting oppositely charged particles having the same mass $m$ and subjected to a constant magnetic field $B$ (ignoring the other's effect like Coulomb and radiation interaction) pointed along the normal to the plane. The system is described by the following standard Lagrangian in C.G.S units :
\begin{equation}
L= \frac{1}{2}m(\dot{x}^{2}_{i}+\dot{y}^{2}_{i})+ \frac{eB}{2c}\epsilon_{ij}(x_{j}\dot{x}_{i}-y_{j}\dot{y}_{i})-\frac{k_{0}}{2}(x_{i}-y_{i})^{2}-\frac{k_{1}}{2}x^{2}_{i}
\label{hall} 
\end{equation}
where  $c$  denotes the speed of light in vacuum, and $x_{i}$ and $y_{i}$ are coordinates of the positive and negative charge respectively. Apart from the kinetic term and magnetic interaction the third term of the above Lagrangian (\ref{hall}) represent the harmonic interaction between the two charges and the fourth term describes additional interactions of the negative charge with an impurity.\\

Here, we will be interested only in the limit of the strong magnetic field $B$ and small mass $m$, in which the kinetic term is quite redundant and potential energy playing an important role \cite{Peirels}. Hence the kinetic terms from (\ref{hall}) can be effectively ignored \cite{for,S. Hellerman} in the limit $\frac{m}{eB}\rightarrow 0$.  Thus the effective Lagrangian describing the high magnetic field limit essentially reduces to the following form:-
\begin{equation}
L_{0}=\frac{eB}{2c}(\epsilon_{ij}x_{j}\dot{x}_{i}-\epsilon_{ij}y_{j}\dot{y}_{i})-V(x_{i},y_{i})~,
\label{sp1}
\end{equation}
with $V(x_{i},y_{i})=\frac{k_{0}}{2}(x_{i}-y_{i})^{2}+\frac{k_{1}}{2}x^{2}_{i}$.
Thus the classical equations of motion associated with the above first-order Lagrangian (\ref{sp1}) is given by -
\begin{equation}
\dot{x_{i}}=\frac{c}{eB} \epsilon_{ij}\frac{\partial V}{\partial x_{j}}~~~; \dot{y_{i}}=-\frac{c}{eB} \epsilon_{ij}\frac{\partial V}{\partial y_{j}}.
\label{la}
\end{equation}
The  canonical Hamiltonian corresponding to (\ref{sp1}) is constructed by the usual Legendre transformation :
\begin{equation}
H_{0}=\frac{\partial L_{0}}{\partial \dot{x}_{i}}\dot{x}_{i}+\frac{\partial L_{0}}{\partial \dot{y}_{i}}\dot{y}_{i}-L_{0}=V(x_{i},y_{i}).
\label{S}		
\end{equation}

Now, in order to establish the equivalence between the Hamiltonian
and Lagrangian formalisms \cite{Ro,RB}, we consider the Hamilton's equations of motion:

\begin{equation}
\begin{alignedat}{1}\dot{x_{i}} & =\{x_{i},H_{0} \}=\{x_{i},V \}\\
\dot{y_{i}} & =\{y_{i},H_0 \}=\{y_{i},V \} ,
\end{alignedat}
\label{bps2l}
\end{equation}

with the potential $V(x_{i},y_{i})$ playing the role of the Hamiltonian \cite{Y}.
The symplectic structure can readily be obtained now by comparing the Lagrangian equations of motion (\ref{la}) with the form of Hamilton's equations of motion  (\ref{bps2l})  to yield the following brackets :

\begin{equation}\label{sy1}
\{x_{i},x_{j}\}=\frac{c}{eB} \epsilon_{ij};~~\{y_{i},y_{j}\}=-\frac{c}{eB} \epsilon_{ij};~\{x_{i},y_{j}\}=0.
\end{equation} 
Here we can eliminate the degrees of freedom of the negative charge ($y_{i}$) by defining a new pair of variables  as,
\begin{equation}
p_{i}=\frac{eB}{c}\epsilon_{ij}(x_{j}-y_{j})~~and ~~x_{i},
\label{deg}
\end{equation}
satisfying the following symplectic structure :
\begin{equation}
\{x_{i},x_{j}\}=\frac{c}{eB} \epsilon_{ij};~~\{x_{i},p_{j}\}= \delta_{ij};~\{p_{i},p_{j}\}=0,
\label{sm}
\end{equation}
where $p_{i}$ play the role of abelian translation generators of the cordinates $x_{i}$. Thus at very high magnetic field and low mass limit, the canonical Hamiltonian (\ref{S}) can be rewritten as -
\begin{equation}
H_{0}=\frac{p_{i}^2}{2m_{B}}+\frac{1}{2}k_{1}x_{i}^2,
\label{hamil}
\end{equation}
where $ m_{B}=\frac{e^2B^2}{k_{0}}$. Therefore at large magnetic field limit, the dynamical system (\ref{hall}) is governed by a  bi-dimensional harmonic oscillator with the non-canonical symplectic structure (\ref{sm}). Notice that noncommutativity of the coordinates has been already established at the classical level as the symplectic bracket between
coordinates $x_{i}$ is nonzero. In the next section we will discuss the quantum picture of this theory.

\section{Noncommutativity in quantum picture}

We now proceed ahead to describe the quantum theory of the above model at sufficiently strong magnetic field ($B$)limit in a systematic
manner with the Hamiltonian operator (\ref{hamil}):
\begin{equation}
\hat{H}=\frac{\hat{p}_{i}^{2}}{2m_{B}}+ \frac{1}{2} k_{1} \hat{x}_{i}^{2},
\label{Ham}
\end{equation}
where  phase-space variables (operators) ($ \hat{x}_{i}, \hat{p}_{i}$) satisfy the following planar non-commutative Heisenberg algebra (NCHA) :
\begin{equation}
[\hat{x}_{i},\hat{x}_{j}]=i\theta \epsilon_{ij},~ [\hat{x}_{i},\hat{p}_{j}]=i\hbar\delta_{ij}, ~[\hat{p}_{i},\hat{p}_{j}]=0;~~ for~ i,j=1,2
\label{hup} 
\end{equation}
with $\theta=\frac{\hbar c}{eB}>0$. So, our system (\ref{Ham}) of interest is  nothing but a two-dimensional isotropic harmonic oscillator living in effective non-commutative plane \cite{BM}.\\

\subsection{Operator-based formulation of planar non-commutative quantum mechanics}

It has been pointed out recently in \cite{F1,fg,KK,RR} that non-commutative quantum
mechanics may be formulated in a completely abstract operatorial level, where quantum systems can be identified with the space of rays on the Hilbert space of Hilbert-Schmidt operators, which act on an auxiliary Hilbert space. This latter auxiliary Hilbert space will be referred to as the configuration space, as its sole purpose is to serve as a space, which furnishes a representation of just the coordinate sub-algebra. Here, we present a very short review of the formulation of the Hilbert-Schmidt operator approach in an appropriate physical setting which also will then pave the way for a path-integral
formulation.\\

In two dimension, the NC coordinate sub-algebra is given by :
\begin{equation}\label{coordinatealg}
[\hat{x}_{i},\hat{x}_{j}]=i\theta \epsilon_{ij};  ~for~i=j=1,2
\end{equation}
 A Hilbert space $\mathcal{H}_{c}$, furnishing a representation of this coordinate sub-algebra, i.e. the configuration space in our case, can be easily identified to be the one, which is isomorphic to that of one-dimensional harmonic oscillator in the usual quantum mechanics. Thus the construction of this Hilbert space proceeds along the same line. We thus begin by first introducing the annihilation and creation operators as

\begin{equation}
\hat{b}=\frac{1}{\sqrt{2\theta}}(\hat{x}_{1}+i\hat{x}_{2}),~~\hat{b}^{
	\dagger}=\frac{1}{\sqrt{2\theta}}(\hat{x}_{1}-i\hat{x}_{2}),
\end{equation}
satisfying the commutation relation $[\hat{b},\hat{b}^{\dagger}]=\mathbb{I}_{c}$. One then defines non-commutative configuration space which is isomorphic to boson Fock space as :

\begin{equation}
\mathcal{H}_{c}=~span\{\left|n\right\rangle= \frac{1}{\sqrt{n!}}(\hat{b}^{\dagger})^{n}\left|0\right\rangle\}
\label{hc}
\end{equation}
where the span is understood to be over the complex numbers. Here the ground state $\left|0\right\rangle$ such that $\hat{b}\left|0\right\rangle=0$ so that
\begin{equation}
\hat{b}\left|n\right\rangle=\sqrt{n}\left|n-1\right\rangle; ~~\hat{b}^{\dagger}\left|n\right\rangle=\sqrt{n+1}\left|n+1\right\rangle
\end{equation}
We would like to emphasize here that this $\mathcal{H}_{c}$ furnishes only the representation of the coordinate algebra (\ref{coordinatealg})
and not of the entire non-commutative Heisenberg algebra (\ref{hup}) and the reasons for that will become clear very soon. Now, in order to construct a corresponding Hilbert space, which furnishes a representation of this entire non-commutative Heisenberg algebra, we need to consider the Hilbert space formed by the set of all Hilbert-Schmidt (HS) operators acting on this configuration space $\mathcal{H}_{c}$. The abstract quantum states of our system can be identified with these HS operators. And a generic HS operator $\psi(\hat{x}_{1},\hat{x}_{2})=\psi(\hat{b},\hat{b}^{\dagger})$ is an algebra element, generated by $(\hat{x}_{1},\hat{x}_{2})$  or equivalently by $(\hat{b},\hat{b}^{\dagger})$, having a finite HS norm:
\begin{equation}
\Vert\psi\Vert_{HS}:= \sqrt{tr_{c}(\psi^{\dagger}\psi)} <\infty,
\end{equation}
which in turn is a dense subspace of space bounded operators ($\mathcal{B}(\mathcal{H}_{c})$) acting on $\mathcal{H}_{c}.$ \\

This space of HS operators forms a Hilbert space of its own and is our ultimate quantum Hilbert space $\mathcal{H}_{q}$ and is defined as:
\begin{equation}
\mathcal{H}_{q}=span\{\ \psi(\hat{x}_{1},\hat{x}_{2})=|\psi):\psi(\hat{x}_{1},\hat{x}_{2})\in \mathcal{B}(\mathcal{H}_{c}), \Vert\psi\Vert_{HS} <\infty\}\,
\label{hq}
\end{equation}\normalsize
This space is equipped with the inner product:
\begin{equation}
(\phi|\psi)=tr_{c}(\phi^{\dagger}(\hat{x}_{1},\hat{x}_{2})\psi(\hat{x}_{1},\hat{x}_{2})),
\label{inn}
\end{equation}
where the subscript $c$ refers to tracing over $\mathcal{H}_{c}$.\\

Note also these HS operators $\Psi(\hat{x}_{1},\hat{x}_{2})$ can also be expanded as,
\begin{equation}
\psi(\hat{x}_{1},\hat{x}_{2})=\psi(\hat{b},\hat{b}^{\dagger})=\sum_{n,m}c_{n,m}|n><m|,
\end{equation}
where $c_{n,m}=<n|\psi(\hat{b},\hat{b}^{\dagger})|m>\in \mathbb{C} $ with $\sum_{n,m}\mid c_{n,m}\mid^{2}<\infty.$
This shows that $\mathcal{H}_{q}$ can also be identified as $\mathcal{H}_{c} \otimes \mathcal{H}^{*}_{c}$, with $\mathcal{H}^{*}_{c}$ being the dual of $\mathcal{H}_{c}$ (\ref{hc}) and this is the space which can furnish the representation of the entire NCHA, as can be seen easily by defining appropriate actions of phase space operators on it. But before we discuss about this point, we pause for a while to introduce the notations we are going to employ to designate the states in $\mathcal{H}_{c}$ and $\mathcal{H}_{q}$:
(i) states in $\mathcal{H}_c$
are denoted by angular kets $|.>$, and (ii) states in the $\mathcal{H}_{q}$
are denoted by a ``round ket"  $\psi(\hat{x}_{1},\hat{x}_{2})=|\psi)$.\\

Now like in the commutative case, here too we look for a unitary representation of the non-commutative  Heisenberg algebra (\ref{hup}) on quantum Hilbert space in terms of operators $\hat{X}_{i}$, $\hat{P}_{i}$, whose actions on $\mathcal{H}_{q}$ are given as follows in the following way -
 \begin{equation}\label{qh1}
 \hat{X}_{i} |\psi)=|\hat{x}_{i}\psi),~~ \hat{P}_{i}=\frac{\hbar}{\theta}\epsilon_{ij}|[\hat{x}_{j},\psi])=\frac{\hbar}{\theta}\epsilon_{ij} [\hat{X}_{j}-\hat{X}^{R}_{j}]||\psi).
 \end{equation}
 Here we have used the capital letters $\hat{X}_{i}$ and $\hat{P}_{i}$  to denote the  representations of the operators $\hat{x}_{i}$ and $\hat{p}_{i}$ acting on $\mathcal{H}_{q}$. Note that here we have taken the action of $\hat{X}_{i}$ by
  direct left action on an abstract state vector $|\psi)$, by default, and the momentum operator is taken to act
  adjointly, which is clearly seen to involve the right action on the quantum Hilbert space and is defined in the
  following way :
 \begin{equation}\label{qh2}
 \hat{X}^{R}_{i}|\psi)=|\psi \hat{x}_{i});~ [\hat{X}^{R}_{i},\hat{X}^{R}_{j}]|\psi)=|\psi[\hat{x}_{j},\hat{x}_{i}])=-\theta \epsilon_{ij}|\psi)~~\forall \psi \in \mathcal{H}_{q}
 \end{equation}
 It can be checked easily that the left and right action commutes:
 \begin{equation}
 [\hat{X}_{i},\hat{X}^{R}_{j}]=0
 \end{equation}
 We can recognize at this stage this $\hat{X}^{R}_{i}$ can be interpreted as an additional degree of freedom and can be identified with the quantum counterpart of $y_{i}$ in (\ref{deg}) and (\ref{qh1}, \ref{qh2}) are analogous to (\ref{sy1}, \ref{deg}) in the quantum setting. Therefore, the classical picture discussed in the previous section provides a clear physical realization of the abstract formulation of non-commutative quantum mechanics in terms of Hilbert-Schmidt operators. We introduce one further notational convention for any  operator $ \hat{O}$  acting on the quantum Hilbert space (\ref{hq}), one may
define left and right action (denoted by superscripted L and R) as follows:
\begin{equation}
\hat{O}^{L}|\psi)=\hat{O}|\psi),~\hat{O}^{R}|\psi)=|\psi)\hat{O};~~\forall |\psi)=\psi(\hat{x}_{1},\hat{x}_{2})\in \mathcal{H}_{q}
\end{equation}
Finally, it is a matter of straightforward verification to see that the phase space operators acting on the quantum Hilbert space indeed obey the following the commutation relations, which are precisely the NCHA: 
\begin{equation}
[\hat{X}_{i},\hat{X}_{j}]=i\theta \epsilon_{ij},~[\hat{X}_{i},\hat{P}_{j}]=i\hbar\delta_{ij},~[\hat{P}_{i},\hat{P}_{j}]=0~~ for~ i,j=1,2;
\label{bd1}
\end{equation} 
It is now useful to define the following non-hermitian operators on the quantum Hilbert space which we shall require later:
\begin{equation}
\hat{B}=\frac{\hat{X}_{1}+i\hat{X}_{2}}{\sqrt{2\theta}},~ \hat{B}^{\ddagger}=\frac{\hat{X}_{1}-i\hat{X}_{2}}{\sqrt{2\theta}},~\hat{P}=\hat{P}_{1}+i\hat{P}_{2},~ \hat{P}^{\ddagger}=\hat{P}_{1}-i\hat{P}_{2};~~ [\hat{B},\hat{B}^{\ddagger}]=\mathbb{I}_{q}
\label{cx}
\end{equation} 
with corresponding actions
\begin{equation}
\hat{B}\psi(\hat{x}_{1},\hat{x}_{2})=\hat{b}\psi(\hat{x}_{1},\hat{x}_{2}),~~\hat{B}^{\ddagger}\psi(\hat{x}_{1},\hat{x}_{2})=\hat{b}^{\dagger}\psi(\hat{x}_{1},\hat{x}_{2}),
\end{equation}
where we have used the symbol $\ddagger$ specifically for the operator adjoint on $\mathcal{H}_{q}$.\\


Therefore, the representation of our system Hamiltonian (\ref{Ham}) on $\mathcal{H}_{q}$ is given by
\begin{equation}
\hat{H}=\frac{1}{2m_{B}}(\hat{P}_{1}^{2}+\hat{P}_{2}^{2})+ \frac{1}{2} k_{1} (\hat{X}_{1}^{2}+\hat{X}_{2}^{2}),
\label{ham}
\end{equation}

	where phase-space operators $\hat{X}_{i} $ and $\hat{P}_{i}$ satisfy the same i.e. isomorphic algebra (\ref{bd1})
	while acting on the quantum Hilbert space $\mathcal{H}_{q}.$ \\
	
	Further it is noted that - 
	\begin{equation}
	\hat{X}_{1}=\sqrt{\frac{\theta}{2}}(\hat{B}+\hat{B}^{\ddagger}),~\hat{X}_{2}=-i\sqrt{\frac{\theta}{2}}(\hat{B}-\hat{B}^{\ddagger}),~\hat{P}_{1}=\frac{\hbar}{\theta}(\hat{X}_{2}-\hat{X}^{R}_{2}),~ \hat{P}_{2}=-\frac{\hbar}{\theta}(\hat{X}_{1}-\hat{X}^{R}_{1}) 
	\end{equation}
	On the other hand, for instance, the complex momenta $\hat{P}$ in (\ref{cx}) can be written as -
	
	\begin{equation}
	\hat{P}=i\hbar\sqrt{\frac{2}{\theta}}[\hat{B}^{R}-\hat{B}],~~\hat{P}^{\ddagger}=i\hbar\sqrt{\frac{2}{\theta}}[\hat{B}^{\ddagger}-\hat{B}^{R\ddagger}],
	\end{equation}
	where $\hat{B}^{R}=\frac{\hat{X}^{R}_{1}+i\hat{X}^{R}_{2}}{\sqrt{2\theta}}$ and  it can be easily checked that -
	\begin{equation}
	[\hat{B}^{R},\hat{B}^{R \ddagger}]=-1;~~ [\hat{B},\hat{B}^{R}]=0
	\end{equation}
Accordingly the system Hamiltonian (\ref{ham}) can be re-casted in a much simpler form
	
	\begin{normalsize}
		\begin{equation}
		\begin{alignedat}{1}\hat{H} & =\frac{1}{2m_{B}}\hat{P}^{\ddagger}\hat{P}+ k_{1}\theta (\hat{B}^{\ddagger}\hat{B}+\frac{1}{2})\\
		& =\alpha_{1}\hat{B}^{\ddagger}\hat{B}+\alpha_{2} \hat{B}^{R\dagger}\hat{B}^{R}-\alpha_{3}(\hat{B}^{\dagger}\hat{B}^{R}+\hat{B}^{R \ddagger}\hat{B})+\alpha_{4},
		\end{alignedat}
		\end{equation}
	\end{normalsize}
	where
	\begin{equation}
	\alpha_{1}=k_{1}\theta+\frac{\hbar^{2}}{m_{B}\theta},~\alpha_{2}=\frac{\hbar^{2}}{m_{B}\theta},~\alpha_{3}=\frac{\hbar^{2}}{2m_{B}},~~\alpha_{4}=\frac{k_{1}\theta}{2}.
	\end{equation}
	On recognizing that the system Hamiltonian is obtainable from that for the standard quadratic Hermitian form by  a canonical
	transformation, namely,
	
	\begin{equation}
	\begin{alignedat}{1}\hat{B}^{'} & =\sqrt{\frac{m_{B}\omega \theta}{2\hbar\alpha}}\left[ (\frac{1}{2}+\frac{\hbar\alpha}{m_{B}\omega\theta})\hat{B}+(\frac{1}{2}-\frac{\alpha \hbar}{m_{B}\omega \theta}) \hat{B}^{R} \right]\\
	\hat{B}^{'R} & = \sqrt{\frac{m_{B}\omega \theta}{2\hbar\alpha}}\left[ (\frac{1}{2}-\frac{\hbar\alpha}{m_{B}\omega\theta})\hat{B}+(\frac{1}{2}+\frac{\alpha \hbar}{m_{B}\omega \theta}) \hat{B}^{R} \right],
	\end{alignedat}
	\label{bps1}
	\end{equation}

	with $\omega=\sqrt{\frac{k_{1}}{m_{B}}} $ and  $\alpha=\sqrt{1+\frac{m^{2}_{B}\omega^{2}\theta^{2}}{4\hbar^{2}}}, $ satisfying the commutation relation $[\hat{B}^{'},\hat{B}^{'\ddagger}]= \mathbb{I}_{q}$  and  $[\hat{B}^{'R},\hat{B}^{'R\ddagger}]=- \mathbb{I}_{q}$ like their un-primed counterparts, renders the system Hamiltonian to decompose into two decoupled harmonic oscillators of frequencies $\omega_{+}$ and $\omega_{-}$:
	
	\begin{equation}
	\hat{H}=\hbar\omega_{+} \hat{B}^{'\ddagger}\hat{B}^{'} +\hbar\omega_{-}\hat{B}^{'R}\hat{B}^{'R\ddagger}+\frac{\hbar}{2}( \omega_{+}+ \omega_{-})\mathbb{I}_{q}
	\label{new}
	\end{equation}
	
	where the characteristic frequencies $\omega_{\pm}$ are given by, $ \omega_{\pm}=\alpha \omega\pm \frac{m_{B}\omega^{2}\theta}{2\hbar}$. Here the quadratic Hermitian form of the Hamiltonian (\ref{new}) act on the quantum Hilbert space $\mathcal{H}_{q}$ (\ref{hq}), and this vector space can be represented by another complete orthonormal set of basis vector $|n_{1},n_{2})^{'}$: 
	\begin{equation}
	\mathcal{H}_{q}=span\{|n_{1},n_{2})^{'}= \left|n_{1} \right\rangle^{'} \langle n_{2}|^{'}= \frac{(\hat{B}^{'\ddagger})^{n_{1}} (\hat{B}^{'R})^{n_{2}}}{\sqrt{n_{1}~!n_{2}!}} \left|0 \right\rangle^{'}\langle 0|^{'} \}^{\infty}_{n_{1}=n_{2}=0}~~,
	\label{adj}
	\end{equation}
	with the inner product 
	\begin{equation}
	^{'}(\tilde{n}_{1},\tilde{n}_{2}|n_{1},n_{2})^{'}=tr_{c}[(\left|\tilde{n}_{1} \right\rangle^{'} \langle \tilde{n}_{2}|^{'})^{\ddagger}\left|n_{1} \right\rangle^{'} \langle n_{2}|^{'}]=\delta_{\tilde{n}_{1},n_{1}}\delta_{\tilde{n}_{2},n_{2}}.
	\end{equation}
	The basis vectors of $\mathcal{H}_{q}$ constitute a complete set and accordingly the completeness relation reads:
	\begin{equation}
	\sum_{n_{1},n_{2}}|n_{1},n_{2})^{'}(n_{1},n_{2}|^{'}=\mathbb{I}_{q}
	\end{equation}

	Here the operators  $\hat{B}^{'\ddagger}$ and $\hat{B}^{'R}$ appearing in (\ref{adj}) and their respective Hermitian conjugates  act on the left and right sectors of the states of quantum  Hilbert space (\ref{adj}) respectively, and $ \left|0 \right\rangle^{'} \langle 0|$ standing for the vacuum state of $\mathcal{H}_{q}$:

	
	\begin{equation}
	\begin{alignedat}{1}
	\hat{B}^{'}\left|0 \right\rangle^{'}\langle 0|^{'}& =0\\
	\hat{B}^{'R\ddagger}\left|0 \right\rangle^{'} \langle 0|^{'}&=\left|0 \right\rangle^{'} \langle 0|^{'}\hat{B}^{\dagger}=0
	\end{alignedat}
	\end{equation} 
	
	The eigenvalue equation of this Hamiltonian is,
	
	\begin{equation}
	\hat{H}\left|n_{1},n_{2}\right)^{'} =E_{n_{1}n_{2}}\left|n_{1},n_{2}\right)^{'} ,
	\label{drf}
	\end{equation}
	

	whose  solution spectrum  can virtually be read-off from as,
	\begin{equation}
	E_{n_{1},n_{2}}= (n_{1}\hbar\omega_{+}+n_{2}\hbar\omega_{-})+E_{0,0},  
	\end{equation}
	
	where the additional constant $E_{0,0}$ is
	\begin{equation}
	E_{0,0}=\frac{1}{2}\hbar (\omega_{+}+\omega_{-}),
	\end{equation}
which is nothing but the finite zero point energy of the system Hamiltonian. This, however, being a constant $c$-number and not an operator will give the same value for all states and can easily be disposed off by giving a shift to the zero of energy as,
	\begin{equation}
	\hat{\widetilde{H}}:=\hat{H}-\mathbb{I}_{q} E_{0,0},
	\label{nh}
	\end{equation}
	Therefore 
	\begin{equation}
	\hat{\tilde{H}}\left|n_{1},n_{2}\right)^{'} =\tilde{E}_{n_{1}n_{2}}\left|n_{1},n_{2}\right)^{'} ,
	\label{nope}
	\end{equation}
	with $\hat{\widetilde{H}}|0,0)^{'}=0$ and $\tilde{E}_{n_{1}n_{2}}=n_{1}\hbar\omega_{+}+n_{2}\hbar\omega_{-}$. From now on, we only work with $\hat{\widetilde{H}}.$ We now, as discussed at the outset, proceed to carry out the path-integral of the system in order to study its scaling properties.

 \section{Coherent state based path-integral approach}
 
Like in commutative case, here too we would like to introduce the notion of position states, however, in view of the absence of common eigenstates of $\hat{x}_{1} $ and $ \hat{x}_{2}$ operators, the best we can do is to introduce the minimal uncertainty states i.e. maximally
 localized states (coherent states) on $\mathcal{H}_{c}$ \cite{m} as,
 \begin{equation}
 \left|z \right\rangle=e^{-\bar{z}\hat{b}+z \hat{b}^{\dagger}} \left|0 \right\rangle= e^{-\frac{\mid z\mid^{2}}{2}} e^{z\hat{b}^{\dagger}}\left|0 \right\rangle ~\in \mathcal{H}_{c},
 \label{coh}
 \end{equation}
 where $z=\frac{x_{1}+i x_{2}}{\sqrt{2\theta}}$ is a dimensionless complex number. As is well known that these coherent state $ \left|z \right\rangle$ is an eigenstate of the annihilation operator,
 \begin{equation}
 \hat{b} \left|z \right\rangle=z \left|z \right\rangle
 \end{equation}

  These states provide a complete (albeit over-complete) set on $\mathcal{H}_{c}$ as,
 \begin{equation}
 \int \frac{dz d\bar{z}}{\pi}  \left|z \right\rangle \langle z|=\mathbb{I}_{c},
 \end{equation}
 where the integration is over the entire complex $z$ plane. And, these states constitutes a minimum uncertainty states, namely
 
 \begin{equation}
 \Delta x_{1}\Delta x_{2}=\frac{\theta}{2}
 \end{equation}
 As these are minimum uncertainty states, the real and imaginary parts of $z$ are the closest we can get to points in space obtained from measurement. 
 
\subsection{A position and momentum basis for the quantum Hilbert space } 
 
The quantum Hilbert space was defined in sec 2.1 as the space of HS  operators acting on the configuration space (Fock space) $\mathcal{H}_{c}$. Using the coherent state states (\ref{coh}), we can construct a coherent states (operator) in quantum Hilbert space $\mathcal{H}_{q}$ as follows :
 \begin{equation}
 |z,\bar{z})=|x_{1},x_{2})=\frac{1}{\sqrt{2\pi\theta}}\left|z \right\rangle \langle z|~\in \mathcal{H}_{q};~~~(w.\bar{w}|z,\bar{z})=\frac{1}{2\pi\theta}e^{-\mid w-z\mid^{2}},
 \label{position}
 \end{equation}
 satisfying  
 \begin{equation}
 \hat{B}|z,\bar{z})=\hat{b}\left|z \right\rangle \langle z|=z|z,\bar{z}).
 \end{equation}
 
 Furthermore, the coherent state ("position") representation of a general state $|\psi)=\psi(\hat{x}_{1},\hat{x}_{2})$ can be expressed as,
 \begin{equation}
 \psi(x_{1},x_{2})=(z,\bar{z}|\psi)=\frac{1}{\sqrt{2\pi\theta}}tr_{c}( \left|z \right\rangle \langle z|\psi(\hat{x}_{1},\hat{x}_{2})) = \frac{1}{\sqrt{2\pi\theta}}\langle z|\psi(\hat{x}_{1},\hat{x}_{2})\left|z \right\rangle.
 \end{equation}
 
 This looks exactly like the coordinate representation of the state vectors in commutative quantum mechanics and the states $|z,\bar{z})$ playing a role analogous to position basis in commutative quantum mechanics. From now on, we will speak loosely of the states $|z,\bar{z})=|x_{1},x_{2})$ as position states, keeping in mind that this is merely an analogy.\\\\\\\\

\textbf{Claim:}
 
 The resolution of identity of the quantum Hilbert space $\mathcal{H}_{q}$ can be given in terms of the coherent
 states (\ref{position}) through the following equation
 \begin{equation}
  \int 2\theta dzd\bar{z} ~|z,\bar{z})\star_{V} (z,\bar{z}|= \int dx_{1}dx_{2} ~|x_{1},x_{2})\star_{V} (x_{1},x_{2}|= \mathbb{I}_{q},
  \label{cor}
  \end{equation}
  where $\star_{V}=e^{\stackrel{\leftarrow}{\partial_{\bar{z}}}
  	\stackrel{\rightarrow}{\partial_z}}$  is called Voros \cite{voros} star product. 
  
  \textbf{Proof:}
  
  In order to derive the  resolution of the identity in (\ref{position}), let us consider the following relations
  \begin{eqnarray}
 |\psi)&=&\mathbb I_{c}\psi(\hat{x}_{1},\hat{x}_{2})\mathbb I_{c}\nonumber\\
  &=& \frac{1}{\pi^{2}}\int dzd\bar{z}dwd\bar{w}
  |z\rangle \langle w| \langle z|\psi|w \rangle \nonumber\\
  &=&\frac{1}{\pi^{2}}\int dzd\bar{z}dvd\bar{v}
  |z\rangle \langle z+v| \langle z|\psi|z+v \rangle\nonumber\\
  &=&\frac{1}{\pi}\int dzd\bar{z}\frac{1}{\pi}\int d^{2}v e^{-|v|^{2}}|z\rangle\langle z|
  e^{\bar v\overleftarrow{\partial_{\bar z}} + v\overrightarrow{\partial_{z}}}\langle z|\psi|z\rangle.\nonumber \\
  \label{+}
  \end{eqnarray}
  where we have made use of  $w=z+v$ with $d^{2}w = d^{2}v $,  and   $e^{v \partial_{z}} f(z) = f(z+v)$.
  
  Now, consider the integral over $v,\bar{v}$ first:
  \begin{equation}
  \int d^{2}v e^{-\mid v\mid^{2}}e^{\bar v\overleftarrow{\partial_{\bar z}} + v\overrightarrow{\partial_{z}}}
  \end{equation}
 
 where $d^{2}v=dv d\bar{v}$ represents an element of area in the
 complex $v$ plane and the integration is to be carried out over
 the entire plane. Resorting to a polar representation of and we
 have
 \begin{eqnarray}
 \int d^{2}v e^{-\mid v\mid^{2}}e^{\bar v\overleftarrow{\partial_{\bar z}} + v\overrightarrow{\partial_{z}}}
 &=&\int^{\infty}_{0}r dr\int^{\pi}_{-\pi}d\theta e^{-r^{2}}\sum_{n,m}\frac{1}{n!m!}(r\overleftarrow{\partial_{\bar z}})^{n}(r\overrightarrow{\partial_{z}})^{m}e^{i\theta (n-m)}\nonumber\\
 &=&2\pi \sum_{n}\frac{1}{n!n!}(\overleftarrow{\partial_{\bar z}}\overrightarrow{\partial_{z}}) ^{n} \frac{n!}{2}\nonumber\\
 &=&\pi e^{\stackrel{\leftarrow}{\partial_{\bar{z}}}
 	\stackrel{\rightarrow}{\partial_z}}.\nonumber\\
\label{in}
 \end{eqnarray}
 Therefore, using (\ref{+})
 \begin{eqnarray}
 |\psi)&=&\frac{1}{\pi}\int dzd\bar{z}\frac{1}{\pi}\int d^{2}v e^{-|v|^{2}}|z\rangle\langle z|
 e^{\bar v\overleftarrow{\partial_{\bar z}} + v\overrightarrow{\partial_{z}}}\langle z|\psi|z\rangle\nonumber \\
 &=& \frac{1}{\pi}\int 2\theta dzd\bar{z}|z,\bar{z})e^{\overleftarrow{\partial_{\bar z}}\overrightarrow{\partial_{z}}} (z,\bar{z}|\psi)=\mathbb I_{q}|\psi).
\label{inftm}
 \end{eqnarray}
 
 Since the state $|\psi)$ is arbitrary, we can conclude that the identity operator on quantum Hilbert space may be represented as
 \begin{equation}
 \int 2\theta dzd\bar{z} ~|z,\bar{z})e^{\overleftarrow{\partial_{\bar z}}\overrightarrow{\partial_{z}}} (z,\bar{z}|= \mathbb{I}_{q},
 \label{c}
 \end{equation}
 
 which completes the proof.\\
 
Here we recognize that the above exponential operator to be the Voros star-product operator between two functions $f(z,\bar{z})$ and $g(z,\bar{z})$, which is defined as-
 \begin{equation}
 f(z,\bar{z})\star_{V} g(z,\bar{z})=f(z,\bar{z}) e^{\stackrel{\leftarrow}{\partial_{\bar{z}}}
 	\stackrel{\rightarrow}{\partial_z}}  g(z,\bar{z}) 
 \end{equation}

 	We now introduce a natural definition of a momentum basis \cite{SG} in the quantum Hilbert space such that -
 	\begin{equation}
 	|p,\bar{p})=\sqrt{\frac{\theta}{2\pi\hbar^{2}}} e^{i\sqrt{\frac{\theta}{2\hbar^{2}}}(\bar{p}\hat{b}+p\hat{b}^{\dagger})};~\hat{P}_{i}|p,\bar{p})=p_{i}|p,\bar{p}),~ p=p_{1}+ip_{2}
 	\end{equation}
 	and these states satisfy the usual resolution of identity and orthonormality condition -
 		\begin{equation}
 		\int dp~d\bar{p} ~|p,\bar{p})(p,\bar{p}|= \mathbb{I}_{q} ,~~ ~(p^{'},\bar{p}^{'}|p,\bar{p})=\delta(p^{'}-p)
 		\label{l}
 		\end{equation}
 		The overlap of a state $|z,\bar{z})$ with momentum states on the non-commutative plane is
 	\begin{equation}
 	(z,\bar{z}|p,\bar{p})=\frac{1}{\sqrt{2\pi\hbar^{2}}}e^{-\frac{\theta}{4\hbar^{2}}p\bar{p}} e^{i\sqrt{\frac{\theta}{2\hbar^{2}}}(p\bar{z}+\bar{p}z)}.
 	\end{equation}
In this context, we would like to mention that the authors in  \cite{E,F} have also pointed out that the effective low energy dynamics of relativistic quantum field theories in the presence of sufficiently strong magnetic field can be described by the Voros star product.\\
 	\subsection{ The Path integral action and generating functional }
 	
Now, following \cite{SG}, and using the completeness relations for the momentum (\ref{l}) and coordinate basis (\ref{cor}) in place,  we proceed to write down the path integral representation of the overlap between a pair of coherent states on the two dimensional Moyal plane as,
 	\footnotesize
 	\begin{equation}
 	(z_{f},t_{f}|z_{0},t_{0})=\lim_{n\to\infty}\int^{\infty}_{-\infty} (2\theta)^{n}\prod_{j=1}^{n}  dz_{j}d\bar{z}_{j}(z_{f},t_{f}|z_{n},t_{n})\star_{V_n}(z_{n},t_{n}|...........|z_{1},t_{1})\star_{V_{1}}(z_{1},t_{1}|z_{0},t_{0})
 	\label{ker}
 	\end{equation}
 	\normalsize
  The propagation over an infinitesimal  step $\tau=\frac{t_{f}-t_{0}}{n+1}$  are
 	\small
 	\begin{eqnarray}
 	(z_{j+1}, t_{j+1}|z_j, t_j)&=&(z_{j+1}|e^{-\frac{i}{\hbar}\tau\hat{\widetilde{H}}}|z_j)\nonumber\\
 	&=&\int_{-\infty}^{+\infty}d^{2}p_j~e^{-\frac{\theta}{2\hbar^{2}}\bar{p}_j p_{j}}
 	e^{i\sqrt{\frac{\theta}{2\hbar^{2}}}\left[p_{j}(\bar{z}_{j+1}-\bar{z}_{j})+\bar{p}_{j}(z_{j+1}-z_{j})\right]}\nonumber\\
 	&&~~~~~~~~~~~~~~~~~~~~~~~~~~~~~~\times e^{-\frac{i}{\hbar}\epsilon[\frac{\bar{p}_j p_{j}}{2m_{B}}+\frac{k_{1} \theta}{2}(2\bar{z}_{j+1}z_{j}+c_{0})]}
 	\label{inftm}
 	\end{eqnarray}
 	\normalsize
 	where $c_{0}=1-\frac{2}{\theta k_{1}} E_{0,0}$ . Now substituting the above expression in (\ref{ker}) and computing star products explicitly, we obtain (apart from a constant factor) - 
 	\small
 	\begin{equation}
 	\begin{alignedat}{1}
 	& (z_f, t_f|z_0, t_0)\\ & =e^{-\vec{\partial}_{z_f}\vec{\partial}_{\bar{z}_0}}\lim_{n\rightarrow\infty}\int \prod_{j=1}^{n} (dz_{j}d\bar{z}_{j})
 	\prod_{j=0}^{n}d^{2}p_{j}  \exp\sum_{j=0}^{n}\tau\left[\frac{i}{\hbar}\sqrt{\frac{\theta}{2}}\left[p_{j}\left\{\frac{\bar{z}_{j+1}-\bar{z}_{j}}{\tau}\right\}+\bar{p}_{j}\left\{\frac{z_{j+1}-z_{j}}{\tau}\right\}-\sqrt{2\theta} k_{1}\bar{z}_{j+1}z_{j}\right]\right]\\
 	&\exp\sum_{j=0}^{n}\tau\left[-\frac{i}{2m\hbar} \bar{p}_{j}p_{j}+\frac{\theta}{2\hbar^2}\frac{(p_{j+1}-p_{j})}{\tau}\bar{p}_{j}\right],
 	\end{alignedat}
 	\label{ax}
 	\end{equation}
 	\normalsize
 	where, $\sigma=-i(\frac{\tau}{2m_{B}\hbar}-i\frac{\theta}{2\hbar^{2}})$ and $z_{n+1}=z_{f}$. Finally, in the $\tau\rightarrow0$ or $n\rightarrow \infty $ limit, symbolically, we write the path integral kernel in phase space representation as,
 	\small
 	\begin{equation}
 	(z_f, t_f|z_0, t_0)=e^{-\vec{\partial}_{z_f}\vec{\partial}_{\bar{z}_0}}\int^{z(t_{f})=z_{f}}_{z(t_{0})=z_{0}} \mathcal{D}\bar{z}(t)\mathcal{D} z(t)\int \mathcal{D}\bar{p}(t)\mathcal{D} p(t)e^{\frac{i}{\hbar}S_0 [z(t),\bar{z}(t),p(t),\bar{p}(t)]}
 	\end{equation}
 	\normalsize
 	where $S_0$ is the phase-space action given by -
 	\begin{equation}
 	S_0=\int^{t_{f}}_{t_{0}} dt \left[ \sqrt{\frac{\theta}{2}} (p\dot{\bar{z}}+\bar{p}\dot{z})- \frac{i\theta}{2\hbar}\bar{p}\dot{p}-\frac{1}{2m_{B}}\bar{p}p-\theta k_{1}\bar{z}z \right].
 	\label{action}
 	\end{equation}

It may be noted that the action (\ref{action}) contains an exotic $\bar{p}\dot{p}$ term which is a Chern-Simons like term in momentum space and is responsible for rendering the configuration space action to be highly non-local \cite{sf}. On recognizing that the first-order action (\ref{action}) can be taken to be the one written in an   extended phase space, where both $z(t)$ and $p(t)$ are considered as configuration space variables of the system, enables us to recast the action in the so-called symplectic form \cite{fr} as -
 	\begin{equation}
 	S_0=\int^{t_{f}}_{t_{i}}dt \left[ \frac{1}{2}\xi^{\alpha} f_{\alpha \beta}\dot{\xi^{\beta}}-V(\xi)\right] ,~~~~\xi^{\alpha}:=\{z, \bar{z}, p,\bar{p}\};~~~\alpha=1,2..,4
 	\end{equation}   
 	with  
 	\begin{equation}
 	f = \begin{pmatrix}
 	\huge0 & -\sqrt{\frac{\theta}{2}}\sigma_{1}   \\
 	\sqrt{\frac{\theta}{2}}\sigma_{1} & -\frac{\theta}{2}\sigma_{2}
 	\end{pmatrix}.
 	\end{equation} 
 	where $\sigma_{1}$, $\sigma_{2}$ are the usual Pauli matrices, and 
 	\begin{equation}
 	V(\xi)=\frac{1}{2m_{B}}\bar{p}p+\theta k_{1}\bar{z}z 
 	\end{equation}
 Therefore one can easily read off the symplectic brackets by using the Faddeev-Jackiw formalism \cite{fr} as:-
 	\begin{equation}
 	\{z,\bar{z}\}=-\frac{i}{\hbar},~~~~~\{z,\bar{p}\}=\{\bar{z},p\}=\sqrt{\frac{2}{\theta}},~~~~ \{p,\bar{p}\}=0
 	\label{symplectic}
 	\end{equation}
 	These symplectic brackets \ref{symplectic} are consistent with the classical version
 	of the non-commutative Heisenberg algebra  given in (\ref{sm}).\\
 Looking at the phase space representation of the path integral kernel, one can now re-interpret non-commutative quantum mechanics as a  $0+1$ dimensions complex scalar field theory where the coordinates are complex fields and which depends on time as $z(t)$. Now after providing this brief review of the path-integral formalism in Moyal plane, our next task will be to write down the corresponding matrix elements of the
 	time-ordered product of a source dependent term for all $t$ with $t_{i}<t<t_{f}$ as a path integral -
 	\small
 	\begin{equation}
 	\begin{alignedat}{1} (z_{f},t_{f}|z_{0},t_{0})_{J,\bar{J}} & =(z_{f},t_{f}|\large{T}exp\left[\frac{i}{\hbar}\int^{t_{f}}_{t_{i}} \sqrt{\frac{\theta}{2}}(J(t)\hat{B}^{\dagger}+\bar{J}(t)\hat{B}) dt  \right]|z_{0},t_{0})\\
 	& =e^{-\vec{\partial}_{z_f}\vec{\partial}_{\bar{z}_0}}\int^{z(t_{f})=z_{f}}_{z(t_{0})=z_{0}}\mathcal{D}\bar{z}\mathcal{D} z \mathcal{D}\bar{p}\mathcal{D} p~~ exp\left[\frac{i}{\hbar}(S_0+\int^{t_{f}}_{t_{i}}\sqrt{\frac{\theta}{2}}[\bar{z}(t)J(t)+z(t)\bar{J}(t)]\right]\\
 	&
 	=e^{-\vec{\partial}_{z_f}\vec{\partial}_{\bar{z}_0}}\int^{z(t_{f})=z_{f}}_{z(t_{0})=z_{0}}\mathcal{D}\bar{z}\mathcal{D} z \mathcal{D}\bar{p}\mathcal{D} p~~ exp\left(\frac{i}{\hbar}S_{J,\bar{J}}\right),
 	\end{alignedat}
 	\label{oci}
 	\end{equation}
 	\normalsize
 	with 
 	\begin{equation}
 	S_{J,\bar{J}}= \int ^{t_{f}}_{t_{0}} dt \left[ \sqrt{\frac{\theta}{2}} (p\dot{\bar{z}}+\bar{p}\dot{z})- \frac{i\theta}{2\hbar}\bar{p}\dot{p}-\frac{1}{2m_{B}}\bar{p}p-\theta k_{1}\bar{z}z  +\sqrt{\frac{\theta}{2}}[\bar{z}(t)J(t)+z(t)\bar{J}(t)\right],
 	\end{equation}
 	where $J(t)=J_{1}(t)+iJ_{2}(t)$  is a time dependent source vanishing  at $ t \rightarrow \pm\infty$.
 	Now in quantum theories, the prime interest is the vacuum to 
 	vacuum persistence amplitude in the presence of an adiabatically varying external source. A standard way to obtain this is to go back to the propagation kernel (\ref{oci}) and introduce complete sets of energy states as :-
 	\small
 	\begin{equation}
 	\begin{alignedat}{1}
 	& (z_{f},t_{f}|z_{0},t_{0})_{J,\bar{J}} \\
 	& =\sum_{n_{1},n_{2}}\sum_{m_{1},m_{2}}(z_{f},t_{f}|n_{1},n_{2}) (n_{1},n_{2}|\large{T}\exp\left[\frac{i}{\hbar}\int^{t_{f}}_{t_{0}} \sqrt{\frac{\theta}{2}}(J(t)\hat{B}^{\dagger}+\bar{J}(t)\hat{B}) dt  \right]|m_{1},m_{2})\times\\
 	& (m_{1},m_{2}|z_{0},t_{0})\\
 	& =\sum_{n_{1},n_{2}}\sum_{m_{1},m_{2}}(n_{1},n_{2}|\large{T}\exp\left[\frac{i}{\hbar}\int^{t_{f}}_{t_{0}} \sqrt{\frac{\theta}{2}}(J(t)\hat{B}^{\dagger}+\bar{J}(t)\hat{B}) dt  \right]|m_{1},m_{2})\\
 	&~~~~~~~~~~~~~~~\times(m_{1},m_{2}|e^{\frac{i}{\hbar}t_{0}\hat{\tilde{H}}}|z_{0}) (z_{f}|e^{-\frac{i}{\hbar}t_{f}\hat{\tilde{H}}}|n_{1},n_{2})\\
 	&
 	=(0,0|\large{T}\exp\left[\frac{i}{\hbar}\int^{t_{f}}_{t_{0}} \sqrt{\frac{\theta}{2}}(J(t)\hat{B}^{\dagger}+\bar{J}(t)\hat{B}) dt  \right]|0,0) \times(0,0|z_{0}) (z_{f}|0,0)\\
 	&~~~~~~~+\sum_{n_{1},n_{2}\neq 0}\sum_{m_{1},m_{2}\neq 0}(n_{1},n_{2}|\large{T}\exp\left[\frac{i}{\hbar}\int^{t_{f}}_{t_{i}} \sqrt{\frac{\theta}{2}}(J(t)\hat{B}^{\dagger}+\bar{J}(t)\hat{B}) dt  \right]|m_{1},m_{2})\\
 	&~~~~~~~~~~~~~~~\times(m_{1},m_{2}|z_{0}) (z_{f}|n_{1},n_{2}) e^{-\frac{i}{\hbar}t_{f}\tilde{E}_{n_{1},n_{2}}+\frac{i}{\hbar}t_{0}\tilde{E}_{m_{1},m_{2}}}~~~,  
 	\end{alignedat}
 	\label{oct}
 	\end{equation}
 	\normalsize
 	where we have used the fact of (\ref{nope}). To accomplish the projection onto the ground state (vacuum) of the system Hamiltonian \ref{Ham}, we now introduce a parameter $T$ with units of time and replace -
 	\begin{equation}
 t_{f}\rightarrow	t_{f}=T(1-i\epsilon), ~~~~~~~t_{0}\rightarrow t_{0}=-T(1-i\epsilon),
 	\end{equation}
 	and take $T\rightarrow \infty$, with $\epsilon$ being a very small positive constant. In the end,  we will take the limit $\epsilon \rightarrow 0$, but only after the infinite limit of $T$ is carried out. In the limit $T\rightarrow \infty$, the exponentials in (\ref{oct}) oscillate out to zero except for the ground
 	state. Hence in this asymptotic limit, we obtain
 	\small
 	\begin{equation}
 	\begin{alignedat}{1}
 	\lim_{\epsilon\to 0}\lim_{\substack{t_{f}\rightarrow\infty(1-i\epsilon)\\t_{0}\rightarrow-\infty(1-i\epsilon)}}(z_{f},t_{f}|z_{0},t_{0})_{J,\bar{J}}&
 	=(0,0|\large{T}exp\left[\frac{i}{\hbar}\int^{\infty}_{-\infty} \sqrt{\frac{\theta}{2}}(J(t)\hat{B}^{\dagger}+\bar{J}(t)\hat{B}) dt  \right]|0,0) \\
 	&~~~~~~~~~~~~~~~~~~\times(0,0|z_{0}) (z_{f}|0,0).
 	\end{alignedat}
 	\end{equation}
 	\normalsize
 	However, one can also observe that -
 	\begin{equation}
 	\lim_{\epsilon\to 0}\lim_{\substack{t_{f}\rightarrow\infty(1-i\epsilon)\\t_{0}\rightarrow-\infty(1-i\epsilon)}}(z_{f},t_{f}|z_{0},t_{0})_{J=\bar{J}=0}=(0,0|z_{0}) (z_{f}|0,0)
 	\end{equation}
 	Consequently, we can write -
 	\small
 	\begin{equation}
 	(0,0|T\exp\left[\frac{i}{\hbar}\int^{\infty}_{-\infty} \sqrt{\frac{\theta}{2}}(J(t)\hat{B}^{\dagger}+\bar{J}(t)\hat{B})dt  \right]|0,0) =\lim_{\epsilon\to 0}\lim_{\substack{t_{f}\rightarrow\infty(1-i\epsilon)\\t_{0}\rightarrow-\infty(1-i\epsilon)}}\frac{(z_{f},t_{f}|z_{0},t_{0})_{J,\bar{J}}}{(z_{f},t_{f}|z_{0},t_{0})_{J=\bar{J}=0}}
 	\label{9}
 	\end{equation} 
 	\normalsize
 	Since  the left hand side of (\ref{9}) is not sensitive to the boundary condition $z_{0} $ and $z_{f}$ imposed at $t\rightarrow\pm\infty$, the right-hand side should also be  independent of the boundary conditions imposed at $t\rightarrow\pm\infty$, if one chooses the same boundary conditions in the numerator and the
 	denominator. Moreover, the right hand side has the structure of a functional integral
 	and we can write (\ref{9}) as,
 	\small
 	\begin{equation}
 	\begin{alignedat}{1}(0,0|\large{T}exp\left[\frac{i}{\hbar}\int^{\infty}_{-\infty} \sqrt{\frac{\theta}{2}}(J(t)\hat{B}^{\dagger}+\bar{J}(t)\hat{B}) dt  \right]|0,0)&
 	=(0,0;\infty|0,0;-\infty)_{J,\bar{J}}\\
 	=N^{-1} \int \mathcal{D}\bar{z}\mathcal{D} z \mathcal{D}\bar{p}\mathcal{D} p~~ e^{\frac{i}{\hbar}\left[ S_{0}+\int^{t_{f}=\infty}_{t_{0}=-\infty}\sqrt{\frac{\theta}{2}}[\bar{z}(t)J(t)+z(t)\bar{J}(t)dt ]\right]}, 
 	\label{a}
 	\end{alignedat}
 	\end{equation}
 	\normalsize
 	where the normalization factor $N$ in (\ref{a}) is fixed so as to ensure the normalized ground state ($(0,0|0,0)=1$).\\
 	
 	Therefore, the generating functional or the vacuum persistence amplitude in
 	presence of  external sources $J(t)$ and $\bar{J}(t)$ is defined as,
 	\begin{equation}
 	Z[J,\bar{J}]\equiv Z[J_{1},J_{2}]= N^{-1} \int \mathcal{D}\bar{z}\mathcal{D} z \mathcal{D}\bar{p}\mathcal{D} p~~ e^{\frac{i}{\hbar}\left[ S_{0}+\int^{t_{f}=\infty}_{t_{0}=-\infty}\sqrt{\frac{\theta}{2}}[\bar{z}(t)J(t)+z(t)\bar{J}(t)]dt \right]}.
 	\label{me} 
 	\end{equation}
 	\subsection{Equivalent commutative description}
 	Here we notice that the non-canonical  brackets (\ref{symplectic}) can be realized  in terms of canonical variables $\alpha, p$ as,   
 	\begin{equation}
 	\begin{alignedat}{1}z(t) & =\alpha(t)+\frac{i}{2\hbar}\sqrt{\frac{\theta}{2}}p(t)\\
 	\bar{z}(t) & =\bar{\alpha}(t)-\frac{i}{2\hbar}\sqrt{\frac{\theta}{2}}\bar{p}(t),
 	\end{alignedat}
 	\label{bps2}
 	\end{equation}
 	satisfying 
 	\begin{equation}
 	\{\alpha,\bar{\alpha}\}=0,~~~~~\{\alpha,\bar{p}\}=\{\bar{\alpha},p\}=\sqrt{\frac{2}{\theta}},~~~~ \{p,\bar{p}\}=0
 	\end{equation}
However, in order to avoid the non-locality \cite{deg} in the first-order action (\ref{action}), we may apply the change of variables (\ref{bps2}) (change of variables, which are dummy in nature is quite harmless in the path-integral framework and this leaves the integration measure invariant) in the generating functional (\ref{me}) to re-write it as
 	\small
 	\begin{equation}
 	Z[J,\bar{J}]= N^{-1} \int \mathcal{D}\bar{\alpha}\mathcal{D} \alpha \mathcal{D}\bar{p}\mathcal{D} p~ \exp(\frac{i}{\hbar}S_{J,\bar{J}} [\alpha,\bar{\alpha},p,\bar{p}]),
 	\label{le} 
 	\end{equation}
 	where
 	\begin{eqnarray}
 	\label{nctrans}
 	S_{J,\bar{J}} [\alpha,\bar{\alpha},p,\bar{p}]=\int_{-\infty}^{\infty}dt &&\!\!\!\!\!\!\!\!\!\  \left[\sqrt{\frac{\theta}{2}} (\dot{\bar{\alpha}}-\frac{i\theta}{2\hbar}k_{1}\bar{\alpha}+\frac{i}{\hbar}\sqrt{\frac{\theta}{2}}\bar{J})p+ \sqrt{\frac{\theta}{2}} (\dot{\alpha}+\frac{i\theta}{2\hbar}k_{1}\alpha-\frac{i}{\hbar}\sqrt{\frac{\theta}{2}}J)\bar{p}\right. \nonumber\\
 	&& \left. -\frac{1}{2m_{B}}(1+\frac{k_{1}\theta^{2}m_{B}}{4\hbar^{2}})\bar{p}p-\theta k_{1}\bar{\alpha}\alpha +\sqrt{\frac{\theta}{2}}(\alpha \bar{J}+\bar{\alpha}J)\right]
 	\end{eqnarray}
 	\normalsize
In this context, it is important to mention that the canonical phase space variables ($\alpha,\bar{\alpha}$ and $p$, $\bar{p}$) can be identified with the center of mass coordinates and their corresponding canonical conjugate momenta of our two particles system (\ref{sp}). This has been pointed out in \cite{yin,Ree,hom}. However, the equivalence between two actions (\ref{nctrans} and \ref{action}) in absence of source has already been established \cite{subir} from Batalin-Tyutin point of view. But of course this change of variables follows from the Faddeev-Jackiw reduction scheme \cite{Ro,EE} to carry out the standard canonical quantization scheme of a theory initially described in terms of non-canonical (constrained) variables. This method relies on Darboux's theorem \cite{J} which asserts that any non-standard symplectic two-form can always be brought back to the standard canonical form through a redefinition of the phase-space variables, see \cite{WP, kn} for related applications in this direction. This procedure yields a canonical (equivalent commutative) description of the given constrained (noncommutative) theory. Here, one can shift the burden of deformation from the symplectic structure to the Hamiltonian which then acquires a new interaction term induced from the deformation parameter ($\theta$) which will be made explicit as we move ahead. Now for simplicity,  under
 rescaling the dynamical variables as -
 	\begin{equation}
 	\begin{alignedat}{1}\alpha(t)\rightarrow Z(t)  & = \frac{1}{\sqrt{1+\frac{k_{1}\theta^{2}m_{B}}{4\hbar^{2}}}}\alpha(t)\\
 	p(t)\rightarrow P(t) & =\sqrt{1+\frac{k_{1}\theta^{2}m_{B}}{4\hbar^{2}}}p(t)~,
 	\end{alignedat}
 	\label{bps1}
 	\end{equation}
 	the functional integral (\ref{le}) yields :-
 	\begin{equation}
 	Z[J,\bar{J}]= N^{-1} \int \mathcal{D}\bar{Z}\mathcal{D} Z \mathcal{D}\bar{P}\mathcal{D} P~e^{\frac{i}{\hbar}S_{J,\bar{J}} [Z,\bar{Z},P,\bar{P}]}
 	\label{dge} 
 	\end{equation}
 	with
 	\small
 	\begin{eqnarray}
 	\label{nct}
 	S_{J,\bar{J}} [Z,\bar{Z},P,\bar{P}]&&=\int_{-\infty}^{\infty}dt\left[\sqrt{\frac{\theta}{2}} (\dot{\bar{Z}}-\frac{i\theta}{2\hbar}k_{1}\bar{Z}+\frac{i\lambda}{2\hbar}\sqrt{\frac{\theta}{2}}\bar{J})P+ \sqrt{\frac{\theta}{2}} (\dot{Z}+\frac{i\theta}{2\hbar}k_{1}Z-\frac{i\lambda}{2\hbar}\sqrt{\frac{\theta}{2}}J)\bar{P}\right. \nonumber\\
 	&& \left. -\frac{1}{2m_{B}}\bar{P}P-\theta k_{1}(1+\frac{k_{1}\theta^{2}m_{B}}{4\hbar^{2}})\bar{Z}Z+\lambda(1+\frac{k_{1}\theta^{2}m_{B}}{4\hbar^{2}})\sqrt{\frac{\theta}{2}}(Z \bar{J}+\bar{Z}J)\right]
 	\label{dbi}
 	\end{eqnarray}
 	\normalsize
 	where $\lambda=(1+\frac{k_{1}\theta^{2}m_{B}}{4\hbar^{2}})^{-\frac{1}{2}}.$  The generating functional in configuration space is now easily derived by integrating over the momenta variables. Indeed, the dependence on momenta in the exponent of (\ref{dge}) is at most quadratic in nature and one may perform the gaussian integrations over the momenta to obtain -
 	\begin{equation}
 	Z[J,\bar{J}]= \tilde{N}^{-1} \int \mathcal{D}\bar{Z}\mathcal{D} Z ~ e^{\frac{i}{\hbar}S_{eff} [Z,\bar{Z}]} e^{\frac{i}{\hbar}S_I[Z,\bar{Z},J,\bar{J}]},
 	\label{magt}
 	\end{equation}
 	where $\tilde{N}$ is a normalization constant which arises from the integration of momenta, and 
 	\begin{eqnarray}
 	\label{nct0}
 	S_{eff} [Z,\bar{Z}]=\int_{-\infty}^{\infty}dt &&\!\!\!\!\!\!\!\!\!\  \left[m_{B}\theta \dot{\bar{Z}}\dot{Z} -\frac{im_{B}\theta^{2}k_{1}}{2\hbar}(\bar{Z}\dot{Z}-\dot{\bar{Z}}Z)-k_{1}\theta \bar{Z}Z\right];
 	\end{eqnarray}

 	\begin{eqnarray}
 	\label{yu}
 	S_I[Z,\bar{Z},J,\bar{J}]=\int_{-\infty}^{\infty}dt \lambda\sqrt{\frac{\theta}{2}}    \left[\bar{Z}(J+\frac{i m_{B}\theta}{2\hbar}\dot{J})+Z(\bar{J}-\frac{i m_{B}\theta}{2\hbar}\dot{\bar{J}})+\frac{m_{B}\lambda\theta}{2\hbar^{2}}\sqrt{\frac{\theta}{2}}\bar{J}J\right]~~~
   \end{eqnarray}

 	Now, writing $J(t)=J_1(t)+iJ_2(t)$,~$ Z(t)=\frac{q_{1}(t)+iq_{2}(t)}{\sqrt{2\theta}}$, the above interacting part of the action (\ref{yu}) can be rewritten as -
 	\begin{eqnarray}
 	\label{octe}
 	S_I =\int_{-\infty}^{\infty}dt{\lambda} (J_iq_i-\epsilon_{ij}\frac{m_{B}\theta}{2\hbar}\dot{J}_jq_i)+\int_{-\infty}^{\infty}dt\left(\frac{m_{B}\lambda^2{\theta}^2}{4\hbar^{2}}J_i^2\right)
 	\end{eqnarray} 
 	Here we note that the interaction part of the action (\ref{octe}) contains three terms: (i) standard source terms ($J_iq_i$), (ii) a new quadratic source term and, (iii) a time derivative of source term coupled with coordinates ($\epsilon_{ij}\dot{J}_jq_i$).\\
 	
 	Thus the functional integral can be written in terms of an effective theory comprising the usual variables as,
 	\begin{equation}\label{SOB}
 	Z[J_1,J_2]= \tilde{N}^{-1}e^{\frac{im_{B}}{\hbar}(\frac{\lambda \theta}{2\hbar})^{2}\int_{-\infty}^{\infty}dtJ_i^2}\int\mathcal{D}q_{1}(t)\mathcal{D}q_{2}(t) ~~~e^{\frac{i}{\hbar}\int^{\infty}_{-\infty} dt\left[ L_{eff}+\lambda(J_iq_i-\epsilon_{ij}\frac{m_{B}\theta}{2\hbar}\dot{J}_jq_i)\right]}
 	\end{equation}
 	where,
 	\begin{equation}
 	L_{eff}=\frac{1}{2}m_{B}\dot{q}^{2}_{i}-\frac{m_{B}\theta k_{1}}{2\hbar}\epsilon_{ij}q_{j}\dot{q}_{i}-\frac{1}{2}k_{1} q^{2}_{i}
 	\label{effctive}
 	\end{equation}
 	Defining the new column vectors:-
 	\begin{equation}
 	J=\begin{pmatrix}
 	J_1&&J_2
 	\end{pmatrix}^T ~~;~~ 	X=\begin{pmatrix}
 	q_{1}&&q_{2}
 	\end{pmatrix}^T
 	\end{equation}
     we can	recast the form of the action (\ref{SOB}) in the following way:

 	\begin{eqnarray}\label{connec}
 	Z[J_1,J_2]&=&\tilde{N}^{-1}e^{\frac{im_{B}}{\hbar}(\frac{\lambda \theta}{2\hbar})^{2}\int_{-\infty}^{\infty}dt{J}^{T}{J}}\int\mathcal{D}X(t) ~e^{\frac{i}{\hbar}\int^{\infty}_{-\infty} dt\left[ X^{\dagger}\mathcal{R}X+\lambda J^{T}X-\lambda\frac{m_{B}\theta}{2\hbar}\tilde{J}^{T}X\right]}\nonumber\\
 	&=&\tilde{N}^{-1}e^{\frac{i}{\hbar}\int_{-\infty}^{\infty}dt\bigg(\frac{m_{B}{\lambda}^2{\theta}^2}{4\hbar^{2}}{J}^{T}{J}-\frac{{\lambda'}^2}{{4}} \tilde{J}^{T}\mathcal{R}^{-1}\tilde{J}+\frac{\lambda\lambda '}{2}J^{T}\mathcal{R}^{-1}\tilde{J}\bigg)}\nonumber \\ 
 	&& \times \int\mathcal{D}X(t)  e^{\frac{i}{\hbar}\int^{\infty}_{-\infty} dt\left[ (X-\frac{\lambda '}{2}\mathcal{R}^{-1}\tilde{J})^{\dagger}\mathcal{R}(X-\frac{\lambda '}{2}\mathcal{R}^{-1}\tilde{J})+\lambda J^{T}(X-\frac{\lambda '}{2}\mathcal{R}^{-1}\tilde{J})\right]}\nonumber\\
 	&=& \tilde{N}^{-1}e^{\frac{im_{B}}{\hbar}(\frac{\lambda \theta}{2\hbar})^{2}\int dt {J}^{T}{J}} ~e^{\frac{i}{\hbar}\int dt (\frac{\lambda\lambda'}{2}J^T\mathcal{R}^{-1} \tilde{J}-\frac{\lambda'^2}{4}\tilde{J}^T\mathcal{R}^{-1} \tilde{J})} \int \mathcal{D}X(t)e^{\frac{i}{\hbar}\int_{-\infty}^{\infty}dt [X^{\dagger}\mathcal{R}X +\lambda J^{T}X]}~~~~~~~
 	\end{eqnarray}
 	Here we have redefined the variable (X) in the last line by giving it a suitable shift as
 	\begin{equation}
 	X\rightarrow X-\frac{\lambda '}{2}\mathcal{R}^{-1}\tilde{J}
 	\end{equation}
 	with $\tilde{J}=\begin{pmatrix}
 	\dot{J}_{2} && -\dot{J}_{1}
 	\end{pmatrix}^T$;
 	$\mathcal{R} = -(m_{B}\frac{d^2}{dt^2}+k_{1})I+im_{B}\sigma_{y}k_{1}\frac{\theta}{\hbar}\frac{d}{dt} $; and  $\lambda '=\frac{m_{B}\theta}{2\hbar}\lambda$.\\
 	
 	 Finally, we can identify the effective commutative generating functional in terms of $q_{i}$'s as,
 	\begin{equation}\label{parteff}
 	Z[J_1,J_2]=\tilde{N}^{-1}\mathcal{K}[J_1,J_2]Z_{eff}[J_1,J_2;\lambda]~~~;~~~~~~~~Z_{eff}[J_1,J_2;\lambda]=\int \mathcal{D}q_{1}(t)\mathcal{D}q_{2}(t)~e^{\frac{i}{\hbar}\int_{-\infty}^{\infty}dt[L_{eff}+\lambda J_iq_{i}]}
 	\end{equation}
 	where the quadratic source term and its higher order time derivatives have been denoted by $\mathcal{K}[J_1,J_2]$ and $L_{eff}$ is the same as in (\ref{effctive}).
 	\\
 	
 	Thus at strong magnetic field limit, our non-commutative system (\ref{Ham}) can be described through an effective  Lagrangian (\ref{effctive}), which can be interpreted as the Lagrangian of a charged particle moving in the commutative $q_{1}$-$ q_{2}$ plane in presence of a constant effective magnetic field with an additional quadratic potential where the second term in the right hand side of (\ref{effctive}) describes the interaction of a
 	charged particle (of charge e) with a constant magnetic field $(\vec{B}_{eff})$  pointing along the normal to the plane. The components of the
 	corresponding vector potential $\vec{A}_{eff}$ in symmetric gauge can be read off from (\ref{effctive}), 
 	\begin{equation}
 	(A_{eff})_{i}=-\frac{m_{B}\theta k_{1}c}{2 \hbar e} \epsilon_{ij}q_{j},
 	\end{equation}
 	with $\vec{B}_{eff} = \vec{\nabla}\times \vec{A}_{eff}$. Thus, we have an exact mapping between the planar non-commutative system and ``generalized" Landau problem \cite{Sm}, where the magnetic field is a manifestation of the presence of non-commutativity. Therefore, the classical dynamics of the planar non-commutative system may be  described by the following action :
 	\begin{equation}
 	S_{eff}=\int dt \left[\frac{1}{2}m_{B}\dot{q}^{2}_{i}+\frac{e}{c}(A_{eff})_{i}\dot{q}_{i}-\frac{1}{2}k_{1} q^{2}_{i}\right].
 	\label{nwe ac}
 	\end{equation}\\
 	This form of the effective (commutative equivalent) action reminds us of that of a charged particle interacting with planar magnetic point vertex, written in terms of some appropriate variables for facilitating an alternate consistent quantization procedure \cite{vor}. Also, the corresponding action was shown there to be scale or dilatation invariant. We now move on to study the scale symmetry  of the action (\ref{nwe ac}).

\section{ Broken dilatation (scale) symmetry and Ward-Takahasi identities }

In relativistic case, space and time both transform uniformly under scale transformation, while in non-relativistic situation scale transformation can act ``anisotropically" \cite{cr}, in the sense that there is no compulsion for the space and time coordinates to have the same scaling factor and the case in which these scaling factors are really different, is known as ‘Lifshitz scaling’. This type of scaling plays a vital role in condensed matter systems\cite{fed}. In this section, we  study the time dilatation symmetry (broken) on noncommutative space by using the effective commutative description. Further we will be deriving the anomalous Ward-Takahashi (W-T) identities.

\subsection{Non-conserved dilatation charge from Noether’s theorem  }
The well-known paradigms of planar quantum mechanics \cite{Vort} of charged particles in a background magnetic field are the magnetic vortex and Landau  problems. Few years back, in a series of papers, Jackiw \cite{vor,mvb} has pointed out that the existence of  dilatation symmetry for a charged  particle interacting with magnetic point vortex and magnetic monopole at the classical level. On the other hand, recently in \cite{kd} the time dilatation symmetry of the non-relativistic Landau problem has been investigated. However, it has been found that there is no exact dilatation symmetry associated with the non-relativistic Landau problem, thus giving rise to a non-conserved dilatation charge at the classical level and further a dilatation anomaly due to quantization. It may be noted that all these problems has a similar form for the Lagrangian but their scale transformation properties are different. These observations naturally encourage us to investigate whether the generalized Landau problem respects this symmetry and the dilatation anomaly, if any, pops up.\\

 Now,
by inspection we see that only the kinetic term of the effective action (\ref{nwe ac}) is invariant under global time dilatation transformation:
\begin{equation}\label{scaletrans}
t\rightarrow t^{'}= e^{-\gamma}t,
\end{equation}
when the coordinates must concomitantly undergo a change:
\begin{equation}
q_{i}(t)\rightarrow q^{'}_{i}(t^{'})=e^{-\frac{\gamma}{2}} q_{i}(t),
\end{equation}
where $\gamma$ is a real parameter.  Following Jackiw \cite{R. Jackiw,mvb}, the Lie differential of the field $q_{i}(t)$ due to infinitesimal dilatation transformation is given by -     
\begin{equation}
\delta_{D} {q}_{i}=\gamma(t\dot{{q}}_{i}-\frac{1}{2}{q}_{i}),
\label{doc}
\end{equation}
where $\delta_{D}q_{i}(t)=q^{'}_{i}(t)-q_{i}(t)$. It is now possible to write the non-conserved dilatation charge \cite{fS. Coleman} from  Noether's analysis of (\ref{nwe ac}). To compute the generator of the  dilatation transformation, it will be convenient to allow the parameter $\gamma$ in (\ref{doc}) to depend on
time: $\gamma=\gamma(t)$ i.e., infinitesimal local  variation viz. 
\begin{equation}
\delta^{L}_{D} {q}_{i}(t)=\gamma(t)\mathcal{Q}q_{i}(t)),
\label{dla}
\end{equation}
where $\mathcal{Q}q_{i}(t)=(t\frac{\partial}{\partial t}-\frac{1}{2})q_{i}(t).$ The change in the action due to these transformations (\ref{doc}) is
\begin{align}
\delta^{L}_{D}{S}_{eff} & = \int dt\bigg[\frac{\partial L_{eff}}{\partial \dot{q}_{i}}\dot{\gamma}\mathcal{Q}q_{i}(t)+\gamma(t)\bigg(\frac{\partial L_{eff}}{\partial \dot{q}_{i}}\mathcal{Q}\dot{q}_{i}(t)+\frac{\partial L_{eff}}{\partial q_{i}}\mathcal{Q}q_{i}(t)\bigg)\bigg]\\
& =\int dt\bigg[p_{i}\dot{\gamma}\mathcal{Q}q_{i}+\gamma(t)\bigg(\frac{d}{dt}(tL_{eff})+k_{1}{q}_{i}^2+\frac{m_{B}\theta k_{1}}{2\hbar}\epsilon_{ij}q_{j}\dot{q}_{i}\bigg)\bigg],
\end{align}
where $p_{i}=\frac{\partial L_{eff}}{\partial \dot{q}_{i}}$ is the canonical momentum. Here, we choose $\gamma(t)$ to vanish asymptotically so that one can safely discard the boundary contribution and the action $S_{eff}$ then changes as :
\begin{align}
\delta^{L}_{D} {S}_{eff} & =\int dt \gamma(t)\bigg( k_{1}{q}_{i}^2+m_{B}k_{1}{\frac{\theta}{2\hbar}}\epsilon_{ij}q_{j}\dot{{q}}_{i}-\frac{d D(t)}{dt}\bigg),
\label{nd}
\end{align}
where $D(t)=tH_{eff}-\frac{1}{2}(q_{i}p_{i})$, and $H_{eff}=\dot{q}_ip_{i}-L_{eff}.$ \\
The above expression (\ref{nd}) holds for any arbitrary field configuration $q_{i}(t)$ with the specific change $\delta ^{L}_{D}q_{i}(t)$. However, when $q_{i}(t)$  obeys the classical equations of motion then $\delta S_{eff}=0$ for any $\delta q_{i}$
including the symmetry transformation (\ref{dla}) with $\gamma(t)$ a function of time. This
indicate that at the on-shell level, we have the  non-conserved dilatation charge ($D$) corresponding to the (broken) dilatation symmetry as -  
\begin{equation}
\frac{d D(t)}{dt}= \Delta_{0}(t)
\label{ncd}
\end{equation}
where $\Delta_{0}(t):=\Delta_{0}(q_{i},\dot{q}_{i};t)=k_{1}({q}_{i}^{2}+m_{B}\frac{\theta}{2\hbar}\epsilon_{ij}q_{j}\dot{q}_{i}).$\\

We can really identify the non conserved dilatation charge $D$ as the generator of the infinitesimal global scale transformations (\ref{doc}). The generator of the transformation, consistent with $ \delta_{D}q_{i}(t)=\gamma\{D(t),q_{i}(t)\} $. At this stage,
 we observe that the scale invariance is broken explicitly by the presence of parameter $k_{1}$ in the action (\ref{nwe ac}) and the
dilatation charge $ D$ acquires a non zero time derivative (\ref{ncd}). If $k_{1}$  vanishes, the scale transformation (\ref{scaletrans}) has no effect on the dynamics and therefore corresponds to a symmetry at the classical level. In presence of $k_1$, the second term proportional to $\theta$ arises primarily because of the non-zero scale dimension of $\theta$ and also for fact that the symplectic bracket between the  non-commuting coordinates (\ref{symplectic}) is not invariant under dilatations (\ref{doc}). Our observation therefore corroborates with the findings of \cite{jmp,jop}.

\subsection{Anomalous Ward-Takahashi Identities}

 We now  proceed ahead to discuss the path integral formulation of Ward-Takahashi
(W-T) identities associated with the action described in \eqref{nwe ac}. At zeroth-order, it represents the quantum-mechanical counterpart of  Noether's current conservation theorem.  In this section, we will explicitly calculate the W-T identities upto 2nd order and we will see that the W-T identities associated with broken dilatation symmetry is anomalous, indicating an existence of a non-vanishing dilation anomaly.
Here we use the word anomaly to emphasise its genuine quantum mechanical origin and its contribution is over and above the terms arising from the breaking of the symmetry at the classical level itself. During the course of the derivation of these W-T identities, we adopt
the method due to Fujikawa wherein the anomalous terms are  identified with the Jacobian factor arising
from the path-integral measure under the scale  transformation. The anomalous term arising here is eventually regularized using Fujikawa's prescription \cite{kf1,kf2}.\\

In order to present our analysis we consider the generating functional in $0+1$ dimensional QFT defined in effective commutative description  $\eqref{parteff}$ :-
\begin{normalsize}
	\begin{equation}
	\begin{alignedat}{1} & Z_{eff}\left[J_1,J_2;\lambda\right] 
	& =\int\mathcal{D}X(t) ~~~e^{\frac{i}{\hbar}\int^{\infty}_{-\infty}dt\left( X^{T}(t)\mathcal{R}X(t) + \lambda J^{T}.X \right)}
	\end{alignedat}
	\label{cik}
	\end{equation}
\end{normalsize}
 For our purpose we confine our attention to  $Z_{eff}$ instead of $Z$ as the presence/absence of solely source dependent factors ($\mathcal{K}[J_1,J_2]$) does not change the Ward identities.
Here $\mathcal{R}$ is self-adjoint and its eigenvalues are being considered as discrete real here  -
\begin{equation}
\mathcal{R} \phi_{k}(t)=\lambda_{k}\phi_{k}(t),
\end{equation}
where the eigenfunctions $\{\phi_{k}(t)\}$ are taken to satisfy the usual orthogonality and completeness conditions:
\begin{equation}
\begin{alignedat}{1}\int dt \phi_{k}^{\dagger}(t)\phi_{j}(t) & =\delta_{kj}~;\\
\sum_{k=1}^{\infty}\phi_{k}(t)\phi^{\dagger}_{k}(t^{'}) & =\delta(t-t^{'})\mathbb{I}_{2}
\end{alignedat}
\label{ps1}
\end{equation}
In order to  specify the functional measure, we expand   
\begin{equation}
X(t)=\sum_{k=1}^{\infty} a_{k}\phi_{k}(t)
\label{fey}
\end{equation}
Now, the functional-integral measure is then defined as,
\begin{equation}\label{pmeasure}
\mathcal{D}X=\prod_{k=1}^{k=\infty} da_{k}=\mathcal{D}a
\end{equation}
To obtain the W-T identities, we study the behaviour of the  action (\ref{nwe ac})
and the measure defined in (\ref{pmeasure}) under infinitesimal local dilatation transformations (\ref{dla}) of the field $q_{i}$ as 
\begin{equation}
X(t)\rightarrow X'(t)=X(t)+\delta^{L}_{D} X(t),
\label{lie}
\end{equation}
with $\delta^{L}_{D} X(t)=\gamma(t)(t\frac{\partial}{\partial t}-\frac{1}{2})X(t),$ the expansion coefficients of (\ref{fey}) change to-
\begin{align}
a_{j}\rightarrow	a_{j}' =a_{j}+\sum_{k}a_{k}\int dt\gamma(t)\phi^{\dagger}_{j}(t)(t\frac{\partial}{\partial t}-\frac{1}{2})\phi_{k}(t)).
\label{jaco}
\end{align}
Thus, the measure changes as \cite{k3} -
\begin{equation}
\mathcal{D}X\rightarrow \mathcal{D}X^{'}=\prod_{j=1}^{j=\infty} da^{'}_{j}=(\det c_{jk})\prod_{k=1}^{k=\infty} da_{k}=e^{Tr (ln \; c_{jk})}\mathcal{D}X
\label{br}
\end{equation}
where $det( c_{jk})$ can be read as the Jacobian of the transformation (\ref{jaco}), whereas, the transformation matrix elements $c_{jk}$ can be obtained by,
\begin{equation}
c_{jk}=\frac{\partial a'_{j}}{\partial a_{k}}= \delta_{jk}+\int dt \gamma(t)\phi_{j}^{\dagger}(t)(t\frac{\partial}{\partial t}-\frac{1}{2})\phi_{k}(t)
\end{equation}
Now, the above effective generating functional of connected Green's functions (\ref{cik}) can be re-written under the dilatation transformations (\ref{lie}) as,
\begin{eqnarray}\label{ci}
Z_{eff}\left[J_1,J_2;\lambda\right]
&=&\int\mathcal{D}X'(t) ~~~e^{\frac{i}{\hbar}\int^{\infty}_{-\infty} dt\left[ X^{'T}\mathcal{R}X^{'}+\lambda J^{T}.X^{'}\right]} \nonumber\\
&=&\int e^{Tr(ln \; c_{jk})}\mathcal{D}X ~~~e^{\frac{i}{\hbar}\int^{\infty}_{-\infty} dt\left[ X^{T}\mathcal{R}X+\lambda J^{T}.X\right]+ \delta^{L}_{0}S_{eff}+\left[\lambda\int_{-\infty}^{\infty}dt J^{T}(t)\delta^{L}_{0} X(t)\right]}~~~~~~~~~~~~
\end{eqnarray}
In the first equality, we have simply re-labeled $ X(t)$  by $X^{'}(t)$ as a dummy variable in
the functional integral (\ref{cik}). The second equality is nontrivial and uses the assumed coordinate transformations (\ref{lie}) and (\ref{br}). For infinitesimal local scale parameter ($\gamma(t)$) the generating functional changes to become,
\begin{align}
& Z_{eff}\left[J_1,J_2;\lambda\right] 
= \int\mathcal{D}X(t) ~e^{\frac{i}{\hbar}\int^{\infty}_{-\infty} dt\left[ X^{T}\mathcal{R}X+\lambda J^{T}.X\right]} e^{\frac{i}{\hbar}[\delta^{L}_{0}S_{eff}+\int_{-\infty}^{\infty}dt (\lambda J^{T}\delta_{0}^{L} X-i\hbar\gamma(t)A(t))]}\nonumber\\
\label{dbc}
\end{align}
where,
\begin{equation}
A(t)=\sum_{k}\phi_{k}^{\dagger}(t)(t\frac{\partial}{\partial t}-\frac{1}{2})\phi_{k}(t).
\label{Anf}
\end{equation}
This $A(t)$ is known  as anomalous term. Now reverting back to component notation and performing the Taylor series expansion of the exponential factor in (\ref{dbc}) upto terms of $\mathcal{O}(\gamma)$, we obtain :-
\begin{align}\label{R}
& Z_{eff}\left[J_1,J_2;\lambda\right]=\int \mathcal{D}q_{1}(t)\mathcal{D}q_{2}(t)e^{\frac{i}{\hbar}\left[ S_{eff}+\int^{\infty}_{-\infty} dt~ \lambda J_iq_{i}\right]}\nonumber\\
&~~~~~~~~~~~~~~  \times \left(1+\frac{i}{\hbar} \int dt \gamma(t)(\Delta_{0}-\frac{d D}{dt}
+\lambda J_i(t)\mathcal{Q}{q}_{i}(t) -i\hbar A(t))+\mathcal{O}(\gamma^{2})\right), 
\end{align}
where in the last line we have made use of eq.(\ref{nd}). The invariance of the functional integral under the infinitesimal transformations (\ref{lie}) (the change of variables in integration does not change the integral
itself) represents the W-T identities. Specifically, the variation with respect to $\gamma(t)$ must vanish since it holds for any value of $\gamma$, i.e.
\begin{equation}
\frac{\delta Z_{eff}[J_1,J_2;\lambda]}{\delta \gamma(t')}|_{\gamma=0}=0
\end{equation}
This yields
\begin{align}
& 0 =  \int \mathcal{D}q_{1}(t)\mathcal{D}q_{2}(t)e^{\frac{i}{\hbar}S_{eff}}\nonumber \\
& \times (1+\frac{i \lambda}{\hbar}\int dt_{1} J_i(t_{1})q_{i}(t_{1})-\frac{{\lambda}^2}{2{\hbar}^2}\int\int dt_{1}dt_{2}J_i(t_{1})J_k(t_{2})q_{i}(t_{1})q_{k}(t_{2})+........)\nonumber  \\
& \times \frac{i}{\hbar}\left(\Delta_{0}-\frac{d D}{dt^{'}}
+ {\lambda}J_i(t')\mathcal{Q}{q}_{i}(t') -i\hbar A(t')\right),
\end{align}\\\\
Rearrangement of the previous expression yields:			
\begin{align}\label{m}
& 0 =  \int \mathcal{D}q_{1}(t)\mathcal{D}q_{2}(t)e^{\frac{i}{\hbar}S_{eff}}(\frac{i}{\hbar}  (\Delta_{0}(t')-\frac{d D}{dt'}-i\hbar A(t'))\nonumber\\& +\frac{i\lambda}{\hbar}[\frac{i}{\hbar}\int dt_{1}J_i(t_{1})q_{i}(t_{1})(\Delta_{0}(t')-\frac{d D}{dt'}-i\hbar A(t'))\nonumber\\
& +\int dt_{1}J_i(t_{1})\mathcal{Q}q_{i}(t_{1})\delta(t_{1}-t')]\nonumber\\
& -\frac{{\lambda}^2}{2{\hbar}^2}[\int\int dt_{1}dt_{2}J_i(t_{1})J_k(t_{2})q_{i}(t_{1})\mathcal{Q}q_{k}(t_{2})\delta(t_{2}-t')\nonumber\\
&+\int\int dt_{1}dt_{2}J_i(t_{1})J_k(t_{2})\mathcal{Q}q_{i}(t_{1})q_{k}(t_{2})\delta(t_{1}-t')\nonumber\\
& +\frac{i}{\hbar}\int\int dt_{1}dt_{2}J_i(t_{1})J_k(t_{2})q_{i}(t_{1})q_{k}(t_{2})({\Delta}_{0}(t')-\frac{d D}{dt'}-i\hbar A(t'))\nonumber\\
&\times \frac{i}{\hbar}({\Delta}_{0}(t')-\frac{d D}{dt'}-i\hbar A(t'))]+...............)
\end{align}
 Since  W-T identity is actually an infinite number of identities satisfied by the Green functions, the different orders of Ward-Takahasi identities are obtained in terms of mean values of dynamical quantities \footnote{where the mean values are defined as: $\left\langle T^*[............]\right\rangle=\int\mathcal{D}q_{1}(t)\mathcal{D}q_{2}(t)[............]e^{\frac{i}{\hbar}S_eff}$ apart from the aforementioned normalization factor.} by taking the functional derivative of right hand side of (\ref{m}) with respect to the source term i.e. $\frac{\delta}{\delta J_i(t'_{1})}\frac{\delta}{\delta J_m(t'_{2})}.....$ and then setting $J_l(t)$'s to $0$.
\\
\\
Thus the zeroth-order W-T identity:-
\begin{normalsize}
	\begin{align}\label{ao}
	\frac{d}{dt'}\left\langle D(t')\right\rangle=\left\langle\Delta_{0}(t')\right\rangle-i\hbar\left\langle A(t')\right\rangle
	\end{align} 
\end{normalsize}
First-order W-T identity:-
\begin{normalsize}
	\begin{align}\label{aff}
	\frac{d}{dt'}\left\langle T^{*}[D(t')q_{i}(t'_{1})]\right\rangle  = & \left\langle T^{*}[\Delta_{0}(t')q_{i}(t'_{1})]\right\rangle-i\hbar \left\langle T^{*}[A(t')q_{i}(t'_{1})]\right\rangle\nonumber \\
	&-i\hbar\delta(t'_{1}-t')\left\langle\mathcal{Q}{q}_{i}(t_{1})\right\rangle
	\end{align}
\end{normalsize}
Second-order W-T identity:-
\begin{normalsize}
	\begin{align}\label{afff}
	\frac{d}{dt'}\left\langle T^{*}[D(t'){q}_{k}(t'_{2}){q}_{i}(t'_{1}))]\right\rangle & =\left\langle T^{*}[\Delta_{0}(t'){q}_{k}(t'_{2}){q}_{i}(t'_{1})]\right\rangle\nonumber\\
	& -i\hbar\delta(t'_{1}-t')\left\langle T^{*}[{q}_{k}(t'_{2})\mathcal{Q}{q}_{i}(t'_{1})\right\rangle\nonumber\\
	& -i\hbar\delta(t'_{2}-t')\left\langle T^{*}[\mathcal{Q}{q}_{k}(t'_{2}){q}_{i}(t'_{1})]\right\rangle\nonumber\\
	& -i\hbar \left\langle T^{*}[ A(t'){q}_{k}(t'_{2}){q}_{i}(t'_{1})]\right\rangle 
	\end{align}
\end{normalsize}
Similarly one can also derive the higher-orders Ward-Takahashi identities. Particularly, the zeroth-order W-T identity tells us that the dilatation symmetry is explicitly broken due to presence of the spring-constant $k_{1}$ in the Lagrangian (\ref{effctive}), which is consistent with that from Noether's analysis (\ref{ncd}). But at the quantum level, there is also an additional correction term on the right which is the contribution from the existence of anomaly due to quantization and it is the anomaly term $A(t)$ which is solely responsible for modifying the rate of change of dilatation charge $D(t).$
The higher-order identities also show up this anomalous behavior up to "contact terms".\\

Furthermore, it is worthwhile to observe that in this derivation of W-T identities, the  ill-defined expression of the anomaly term in (\ref{Anf}) apparently looks divergent. This is because, at each time point $t$ we are summing over an infinite number of eigenmodes $\phi_{k}(t)$. However, it is possible to give a physical meaning to this expression by the method of regularization and subsequently extract out a finite result as we will see in the next section.

\section{Computation of the dilatation anomaly: Fujikawa's method}

Now we are in good position to compute the anomalous term in W-T identities. Following Fujikawa's prescription \cite{kf2}, the dilatation anomaly can be regularized by correcting each contribution from $\phi_{k}(t)$ by a factor $e^{-\lambda_{k}^2/M^2}$ as,

\begin{equation}
A(t)\rightarrow A_{Reg.}(t):=\lim_{M\to \infty}\sum_{k} \phi_{k}^{\dagger}(t)(t\frac{\partial}{\partial t} -\frac{1}{2})\phi_{k}(t) e^{-\lambda_{k}^2/M^2}
\label{reg}
\end{equation}
Here $\phi_{k}(t)\phi^{\dagger}_{k}(t)$ is an ill-defined object and can be interpreted consistently by first separating time points  and then taking the coincident time limit at a appropriate step. A simple manipulation therefore yields:

\begin{align}\label{reganomaly1}
A_{Reg.}(t)& =  \lim_{M\to \infty}\lim_{t\to t'}~ tr \sum_{k} (t\frac{\partial}{\partial t} -\frac{1}{2})e^{-\mathcal{R}^2/M^2}\phi_{k}(t)\phi_{k}^{\dagger}(t')\nonumber \\
& =  \lim_{M\to \infty}\lim_{t\to t'}~ tr
\bigg[(t\frac{\partial}{\partial t} -\frac{1}{2})e^{-\mathcal{R}^2/M^2} \delta(t-t')\mathbb{I}_{2}\bigg]\nonumber \\
& = \lim_{M\to \infty}\lim_{t\to t'}~ tr \int_{-\infty}^{\infty} \frac{dz}{2\pi}(t\frac{\partial}{\partial t} -\frac{1}{2})e^{-\mathcal{R}^2/M^2}e^{iz(t-t')}\mathbb{I}_{2}
\end{align}
Note that in above, tr refers to only the $2\times 2$ matrix indices whereas the Tr appearing earlier in (\ref{br}) refers to both functional and matrix indices. Here we have made use of the completeness relation of $\phi_{k}(t)$ (\ref{ps1}) and in the last line, the integral representation of Dirac delta ($\delta$)-function has been used. Now we define a classical parameter ${\theta}'=\frac{\theta}{2\hbar}$ to be used from now onwards in order to remove any $\hbar$ dependency of $\theta$. The  expression of the anomalous term (\ref{reganomaly1}) after computing $\lim_{t\to t'}e^{-\frac{\mathcal{R}^2}{M^2}}e^{iz(t-t')}$ and then taking the limit $t\to t'$ becomes -

\begin{equation}
A_{Reg.}(t)=\lim_{M\to\infty}~tr~\int_{-\infty}^{\infty}\frac{dz}{2\pi}(izt-\frac{1}{2})e^{-a\bf{I}+b\bf{\sigma_{y}}}\mathbb{I}_{2},
\end{equation}
where
\begin{align*}
a & =\frac{1}{M^2}(m_{B}^2{z^4}+4m_{B}^2k_{1}^2{\theta}'^2{z^2}-2m_{B}k_{1}z^2+k_{1}^2)\\
b & = \frac{4m_{B}k_{1}{\theta}'}{M^2}(m_{B}z^3-k_{1}z)
\end{align*}
Using $e^{-a\bf{I}}=e^{-a}\bf{I}$, $e^{b\bf{\sigma_{y}}}=\bf{I}\cosh{b}+\bf{\sigma_{y}}\sinh{b}$ and completing the trace operation, we have :-
\begin{equation}
A_{Reg.}(t)=2\lim_{M\to\infty}\int_{-\infty}^{\infty}\frac{dz}{2\pi}(izt-\frac{1}{2})e^{-a}\cosh{b}
\label{frg}
\end{equation}
Since $e^{-a}\cosh{b}$ is an even function in $z$, the first term in the integral doesn't contribute. Therefore, the regularized version of the anomaly term (\ref{frg}) can be re-written in a much simpler form as,

\begin{equation}
A_{Reg.}(t)=-\lim_{M\to\infty}\frac{1}{2\pi}(I_{1}+I_{2}),
\end{equation}
where the integrals $I_1$ and $I_2$ are given by -
\begin{align*}
I_{1}=\int_{0}^{\infty}dze^{-a+b}\\
I_{2}=\int_{0}^{\infty} dz e^{-a-b}
\end{align*}
 Let $z^2-2k_{1}{\theta}'z=s$,~ $\frac{m_{B}^2}{M^2}=p$,~ $p{\omega'}^2=q$ (where ${\omega'}^2=k_{1}^2{\theta}'^2+\frac{k_{1}}{m_{B}}$) and $r=\sqrt{s+k_{1}^2{\theta}'^2}$. The integral $I_1$ becomes :
\begin{align}
I_{1} & =\int_{0}^{\infty}\frac{ds}{2\sqrt{s+k_{1}^2{\theta}'^2}} \exp{-p(s-{\omega}^2)^2}\nonumber\\
& = \int_{k_{1}^2{\theta}'^2}^{\infty}\frac{dr}{2\sqrt{r}}\exp{-p(r-\omega'^2)^2}\nonumber\\
& = e^{-p{\omega'}^4}\int_{k_{1}^2{\theta}'^2}^{\infty}\frac{dr}{2\sqrt{r}}e^{-pr^2+2qr}
\end{align}
Similarly one can show :-
\begin{equation}
I_{2} = e^{-p{\omega'}^4}\int_{k_{1}^2{\theta}'^2}^{\infty}\frac{dr}{2\sqrt{r}}e^{-pr^2+2qr}
\end{equation}
Thus the regularised anomaly factor $A_{Reg.}(t)$ simplifies to -
\begin{align}
A_{Reg.}(t) & =-\lim_{M\to\infty}\frac{1}{2\pi}e^{-p{\omega'}^4}\int_{k_{1}^2{\theta}'^2}^{\infty}dr \;{r}^{-1/2}e^{-pr^2+2qr}\nonumber\\
& =- \frac{1}{2\pi}\lim_{M\to\infty}e^{-p{\omega'}^4}\left[\int_{0}^{\infty}dr \;{r}^{-1/2}e^{-pr^2+2qr}-\int_{0}^{k_{1}^2{\theta}'^2}dr \;{r}^{-1/2}e^{-pr^2+2qr}\right] \label{finalregularised}
\end{align}
Now, taking the  limit $M\to\infty$, the 2nd integral in (\ref{finalregularised}) is convergent and gives a finite result of $2k_{1}\theta '$. The first integral can be evaluated as in \cite{ig}:
\begin{equation}
\lim_{M\to\infty}e^{-p{\omega'}^4}\int_{0}^{\infty}\frac{dr}{\sqrt{r}}e^{-pr^2+2qr}=\lim_{M\to\infty}(2p)^{-\frac{1}{4}}\Gamma\left(\frac{1}{2}\right)D_{-\frac{1}{2}}(y)e^{\frac{y^2}{2}}e^{-p{\omega '}^4}
\end{equation}
where $y=\frac{p{\omega'}^2}{\sqrt{2p}}$. We can also write $D_{-1/2}(y)$ in terms of $K_{1/4}(y)$, where $D_{-1/2}(y)$ and $K_{1/4}(y)$ represent the parabolic cylindrical functions and modified Bessel function respectively.
\begin{equation}
\lim_{M\to\infty}(2p)^{-\frac{1}{4}}\Gamma\left(\frac{1}{2}\right)D_{-\frac{1}{2}}(y)e^{\frac{y^2}{2}}=\lim_{M\to \infty}(2p)^{-\frac{1}{4}}e^{\frac{y^2}{2}}(\frac{1}{4}y^2)^{\frac{1}{4}}K_{1/4}(\frac{1}{4}y^2)
\end{equation}
Using $K_{1/4}(x)\approx\frac{1}{2}\Gamma(\frac{1}{4})(\frac{x}{2})^{-\frac{1}{4}}$ as $x\rightarrow 0$, we have :-
\begin{equation}
\lim_{M\to\infty}e^{-p{\omega'}^4}\int_{0}^{\infty}\frac{dr}{\sqrt{r}}e^{-pr^2+2qr}=\frac{1}{2}\lim_{M\to\infty} \Gamma\left(\frac{1}{4}\right)p^{-\frac{1}{4}}= \frac{1}{2}\lim_{M\to\infty} \Gamma\left(\frac{1}{4}\right)(M/m_{B})^{1/2}
\end{equation}
\begin{equation}
A_{Reg.}(t)=-\frac{1}{2\pi}\left[ \frac{1}{2}\lim_{M\to\infty} \Gamma\left(\frac{1}{4}\right)(M/m_{B})^{1/2}-2k_{1}{\theta '}\right]
\label{final}
\end{equation}

Note that the first term in the expression is independent of the coupling $k_{1}$ and is the same as in the free particle case of our effective commutative theory, which is usually considered to be non-anomalous and only the second term is independent of the cutoff parameter $M$. Therefore, following conventional wisdom \cite{mu}, we now concentrate on the nontrivial finite contribution coming from (\ref{final}). To obtain a non-divergent anomaly part, we need to renormalize the relation (\ref{final}) in free particle theory limit $k_{1}\rightarrow 0$ by simply adding a term $i\hbar \gamma(t)A_{f}$ in the Lagrangian $L_{eff}$ (\ref{dbc}), thus cancelling the divergence in (\ref{final}). Here, $A_{f}$ is equal to the Fujikawa factor for the free particle theory,

\begin{equation}
A_{f}=-\frac{1}{4\pi}\lim_{M\to\infty} \Gamma\left(\frac{1}{4}\right)(M/m_{B})^{1/2} 
\end{equation}
Therefore, after renormalization the correct anomaly is given by :
\begin{equation}
A_{renormalized}= A_{Reg.}-A_{f}=\frac{1}{\pi}k_{1}{\theta '}
\label{q}
\end{equation}        
Taking into account the above scheme of renormalization, only the finite piece of the anomaly term contributes to obtain renormalized version of the set of bare W-T identities given in (\ref{ao}-\ref{afff}). 
Furthermore,  expression of the anomaly
 term (\ref{q}) is completely independent of field configuration, so, the form of the anomalous term is the same even if we were to
 work directly with noncommutative space variables, except that the dilatation transformation would have been implemented
 in a different way. On the other hand, since the anomalous correction to the W-T identities is a first order in $\theta$, where $\theta=\frac{\hbar c}{eB}$, which suggests that for commutative limit i.e. $\theta\rightarrow 0$, one must have $\hbar\rightarrow 0$, indicating thereby  the commutative limit coincides with the classical limit. This indeed attests to the fact that in this setting, the correct quantum corrections are fully taken into account only when we switch to the non-commutative quantum framework.

\section{Discussion}

In this chapter, we've studied a system of interacting two opposite charged particles in two dimensions, subject to a strong perpendicular magnetic field and low
mass limit. We have demonstrated clearly how the first-order Lagrangian of the system gets mapped
directly to a deformed harmonic oscillator. Then we reviewed the quantum
picture for the above using Hilbert-Schmidt operators-based formulation. Here
the noncommutativity is given as an effective description, unlike in many other
works where it's postulated as a fundamental parameter. We then construct a
path-integral formulation of the functional integral using coherent states, as a theory in 0+1 dimensions.
The commutative equivalent action is then derived through path-integral which then describes a charged particle moving in a constant background magnetic field and under the influence of additional harmonic oscillator potential - the generalized Landau problem. Moreover, the non-commutative harmonic oscillator and generalized Landau problem are found to be related very closely. The important point is that the form of this action has the same form as that of a charged particle interacting with some point magnetic vortex which is scale
or dilatation invariant unlike our present case, thus suggesting the need for a radical analysis for each individual case. We compute
the W-T identities associated with the broken dilation symmetry which appear to be anomalous. To the extent, we are aware this has been remained unexplored in the literature from the point of view of non-commutative physics under scaling transformations. The anomalous term has been computed and eventually regularized following Fujikawa's method.

\chapter{ Quantum (phase)-space
	and emergent geometrical
	phase}\label{galve}
In the previous chapter, we have discussed the broken dilatation symmetry at the classical and quantum level for an interacting dipolar system  (a two-body problem) subjected to a large magnetic field at which our system effectively described as a harmonic oscillator placed in the ambient quantum (non-commutative) plane. Also, we have established a mapping between the non-commutative harmonic oscillator and a generalized Landau problem.\\

Indeed, apart from this spatial noncommutativity in the  well known Landau problem, it is well known that the non-relativistic theory in a $2+1$ dimension \cite{kb}, planar Galilean group can be endowed with two fold centrally extended Galilean algebra: (i) one involving the
commutator of the Galilean boost generators $\hat{K}_{i}$ between themselves,
which is a non-zero constant  $[\hat{K}_{i},\hat{K}_{j}]=i\hbar \kappa \epsilon_{ij}$, and (ii) the other involves the commutation relation between the Galilean boost and linear momentum, which is related to the mass ($m$) of the particle: $ [\hat{K}_{i},\hat{P}_{i}]=i\hbar m\delta_{ij}$, with other commutators taking their usual structures. Here the central charge $\kappa$ can be identified with the fractional spin of the anyons, as has been shown in \cite{jrn,er}, on using certain nonrelativistic reduction of (2+1)-dimensional
Poincare algebra $iso (2, 1)$. The price to be paid for a realization of two fold centrally extended Galilean algebra 
is that the Galilean covariant coordinates necessarily satisfy non-commutative algebra $[\hat{x}_{i},\hat{x}_{j}]=i\theta \epsilon_{ij}$, with $\hbar\kappa=m^{2}\theta.$ Now, we can further consider a modified commutator between the two components of the momentum operators $[\hat{P}_{i},\hat{P}_{j}]=iB\epsilon_{ij}$ with constant $B$. Here this $B$ can arise in presence of a uniform and constant magnetic field normal to the plane. This can be seen to correspond to the case of the coupling of non-relativistic anyons with this background magnetic field.  Various interesting points of this simultaneous noncommutativity in the coordinates and momenta has been studied in the context of fractional quantum Hall-effect(FQHE) \cite{ho}\cite{sb},\cite{pah}. Besides this, the effects of phase-space noncommutativity can have nontrivial consequences in gravitational wave detection and such a situation has been encountered in \cite{saha}. On the other-hand, non-commutative  phase-spaces also provide a physical background for exact quantization of interacting scalar field theory \cite{EL,yh}.\\

In more recent times, Xiao et.al. have shown that in the semi-classical analysis of Bloch electrons in presence of magnetic field, a nonzero Berry curvature of the band structure leads to a  noncommutativity between phase
space variables\cite{xio} and the effect of Berry Phase in deformed phase-space geometry has also been discussed in \cite{Eo}. As a result, it's important to look into whether the Berry phase exists in a quantum-mechanical system where both the position and momentum operators fulfil non-commutative algebra. We are also aware of the universal appeal of the Berry phase \cite{M.V,JM} from both theoretically such as fractional statistics  \cite{Dw,ha}, anomalies in gauge field theories \cite{LA,n,Hon} and also in several other situations  encountered in experiments \cite{Chiao,A.T}, wherein the system Hamiltonian depends on some set of parameters and the Berry phase or geometrical phase obtained through the adiabatic transport of a quantum system around a closed path in this parameter space. An important feature of this geometric phase is that it does not depend at all on the rate of traversal of the circuit provided of course the change is adiabatic. Furthermore, a classical counterpart of this quantual geometric phase shift was found by Hannay as well as Berry for a classical integrable system, which is known as Hanny's angle. A more detailed discussion is given in \cite{jh,mv}.\\

In this chapter, we study the appearance of adiabatic phase shift in the context of planar quantum mechanics for a quantum
two-dimensional simple harmonic oscillator model with constant noncommutativity in both coordinates and momenta. However, in this context, we would like to mention a similar work carried out in \cite{AP}, though
in a different system involving a gravitational quantum well. In that paper, their original model was
defined on the usual(commutative) phase-space and introduced noncommutativity just rewriting the commutative variables by their NC counterparts by exploiting inverse of the Bopp shift (or the Seiberg-Witten
map in the parlance of \cite{AP}), and found no geometric contribution in the total phase. Here we follow a different approach to obtain a non-vanishing Berry phase which does not agree with the conclusions of \cite{AP} for reasons that we explain in the sequel.\\

The different sections of the chapter are organized as follows.
In section 3.1 we introduced a parametric harmonic oscillator placed in the ambient non-commutative phase-space (both positions
and momenta are non-commutative) where mass and frequency parameter are slowly varying periodic function of time. Here the energy spectrum is computed by using a generalised non-canonical phase space transformation, which we refer to as generalised Bopp shift (see Appendix-A), which connect the non-commutative phase space variables  with their commutative
counterparts. In this context, we would like to mention that in the literature one can find claims by \cite{YZ} that this mapping from non-commutative phase space to usual (commutative) phase
space is not unique. However, as we show in the appendix- A that for a particular case, they are unitary equivalent. In the next section (3.2) we find
out, the geometric phase factor in Heisenberg picture which is  acquired by the annihilation (and corresponding creation) operators under an adiabatic transport
in parameter space of the system. In section 3.3 we discuss the usual expression of geometric phase in terms of the phase gathered by the state vector by going over from the Heisenberg picture to Sch\"odinger picture and it is our present contention that we provide a connection
between the quantal description of geometric phase shift and classical Hannay angle in a rather straightforward manner in sec 3.4. Concluding remarks  are made in sec-3.5. There are two appendices; in the Appendix-A 
we discuss different realizations
of non-commutative algebra and their equivalence while in Appendix-B, apart from reviewing some of the
necessary group theoretical aspects related to our model, we discuss
the apparent removability of geometric phase by a time-dependent unitary transformation, which nevertheless is shown to reappear in ”disguise” in the dynamical phase but retaining its geometrical nature.\\

\section{Quantum parametric oscillator in non-commutative phase-space}

In this section, we basically consider a two-dimensional non-commutative plane where a harmonic oscillator with time-dependent parameters $P(t)$, $Q(t)$ varying adiabatically with period $T$ is subjected also to momentum noncommutativity. This system is governed by the Hamiltonian

\begin{equation}
\mathcal{H}(t)=P(t)(\hat{p}_{1}^{2}+\hat{p}_{2}^{2})+Q(t)(\hat{x}_{1}^{2}+\hat{x}_{2}^{2}),\label{nch}
\end{equation}

with $P(t), Q(t)>0$ and these time-dependent slowly varying parameters are assumed to absorb
all other parameters like mass, spring constant as mentioned in the previous section. Our model is constructed in the spirit of \cite{Wp,Br,Ag,Lf,Ms} where the motivation of having time dependency of the parameters $P(t)$ and $Q(t)$ in the real physical model was demonstrated. The entire Heisenberg algebra of Non-commutative phase space is characterized by:

\begin{equation}
\begin{array}{c}
[\hat{x}_{i},\hat{x}_{j}]=i\theta\epsilon_{ij};[\hat{p}_{i},\hat{p}_{j}]=i\eta\epsilon_{ij};[\hat{x}_{i},\hat{p}_{j}]=i\hslash\delta_{ij}\end{array};\theta\eta<0\label{non-com}
\end{equation}

where $\theta$,$\eta$ are taken to be constants and we shall restrict ourself to negative value of $\theta\eta$ for consistent quantization. See for a detail discussion  \cite{Ha,Sy} and references therein. However, this deformed phase-space algebra \ref{non-com} enforces a symmetry between the coordinates $\hat{x}_{i}$ and momentum $\hat{p}_{i}$ under the following transformation
\begin{eqnarray}
\begin{array}{rcl}
&&\hat{x}_{i}\rightarrow \hat{p}_{i}\\
&&\hat{p}_{i}\rightarrow \hat{x}_{i}\\ 
&&\theta\rightarrow- \eta\\
&&\eta\rightarrow -\theta\\
&&i\rightarrow -i.
\end{array}
\label{sym1}
\end{eqnarray}

Of course, this symmetry between $\hat{x}_{i}$ and  $\hat{p}_{i}$ will be lost in absence of momentum non-commutativity.\\

In order to obtain the spectra of the model,
 we introduce a linear realization of the above non-commutative  algebra (\ref{non-com}) in terms of ordinary phase space variables ($q_{i}, p_{i}$), which we refer to as generalized Bopp's shift (See Appendix-A for other kinds of Bopp shift). These are given by,
 
 \begin{equation}
 \begin{alignedat}{1}\hat{x}_{i} & =q_{i}-\frac{\theta}{2\hslash}\epsilon_{ij}p_{j}+\frac{\sqrt{-\theta\eta}}{2\hslash}\epsilon_{ij}q_{j}\\
 \hat{p}_{i} & =p_{i}+\frac{\eta}{2\hslash}\epsilon_{ij}q_{j}+\frac{\sqrt{-\theta\eta}}{2\hslash}\epsilon_{ij}p_{j},
 \end{alignedat}
 \label{bpshift}
 \end{equation}
 
 where the canonical pair $q_{i}$ and $p_{i}$ are commuting coordinates and momenta respectively obeying the usual Heisenberg algebra:
\begin{equation}
[q_{i},q_{j}]=0=[p_{i},p_{j}];[q_{i},p_{j}]=i\hslash\delta_{ij}
\end{equation}
and are distinguished by the absence of overhead hats. Obviously, this transformation (\ref{bpshift}) is not an unitary one as it changes the basic
commutation relations, but  
it nevertheless enables us to convert the Hamiltonian in the non-commutative phase-space into a modified Hamiltonian in the usual canonical phase-space, where NC effects will manifest through an explicit dependency of the deformation parameters $\theta$ and $\eta$.
Substituting (\ref{bpshift}) in (\ref{nch}) we can obtain the following form of the Hamiltonian:

\begin{equation}
\mathcal{H}(t)=\alpha(t)(p_{1}^{2}+p_{2}^{2})+\beta(t)(q_{1}^{2}+q_{2}^{2})+\delta(t)(p_{i}q_{i}+q_{i}p_{i})-\gamma(t)(q_{1}p_{2}-q_{2}p_{1});\label{comh}
\end{equation}


where the time-dependent effective coefficients $\alpha,\beta,\gamma,\delta$
are given by,

\begin{equation}
\begin{alignedat}{1}\alpha(t) & =P(t)\left\{ 1-\frac{\theta\eta}{4\hslash^{2}}\right\} +Q(t)\left(\frac{\theta}{2\hslash}\right)^{2}\\
\beta(t) & =Q(t)\left\{ 1-\frac{\theta\eta}{4\hslash^{2}}\right\} +P(t)\left(\frac{\eta}{2\hslash}\right)^{2}\\
\gamma(t) & =\frac{1}{\hslash}\left(\eta P(t)+\theta Q(t)\right)\\
\delta(t) & =\left(\frac{\sqrt{-\theta\eta}}{4\hslash^2}\right)\left(\eta P(t)-\theta Q(t)\right)
\end{alignedat}
\label{eq:alpha exp}
\end{equation}

At this stage, we can recognize that our system Hamiltonian living  effectively an enlarged parameter space and can be written as a combination of three terms, 
\begin{equation}
\mathcal{H}(t)=\mathcal{H}_{gho,1}(t)+\mathcal{H}_{gho,2}(t)+\mathcal{H}_{\boldsymbol{L}}(t)
\label{hsep}
\end{equation}

where $\mathcal{H}_{gho,i}(t)'s$ \textit{(i=1 or 2)} are like a generalised
time dependent harmonic oscillator Hamiltonian along $i^{th}-direction$,

\begin{equation}
\mathcal{H}_{gho,i}(t)=\alpha(t)(p_{i}^{2})+\beta(t)(q_{i}^{2})+\delta(t)(p_{i}q_{i}+q_{i}p_{i})\text{ }\text{ }\text{ }\text{ }\text{(no sum on \ensuremath{i})}\label{hghoi}
\end{equation}

and

\begin{equation}
\mathcal{H}_{\boldsymbol{L}}(t)=-\gamma(t)(q_{1}p_{2}-q_{2}p_{1})
\label{Zeem}
\end{equation}
which is effectively a Zeeman term in Landau like problem, where a magnetic field is present perpendicular to the $ q_{1}-q_{2}$ plane.  In order to diagonalize the system Hamiltonian, at first we introduce the  annihilation (and corresponding creation) operators
\begin{equation}
\boldsymbol{a}_{j}=\left(\frac{\beta}{2\hslash\sqrt{\alpha\beta-\delta^{2}}}\right)^{1/2}\left[q_{j}+\left(\frac{\delta}{\beta}+i\frac{\sqrt{\alpha\beta-\delta^{2}}}{\beta}\right)p_{j}\right];\text{ }j=1,2\label{ol}
\end{equation}

satisfying the commutation relations
\begin{equation}
[\boldsymbol{a}_{i},\boldsymbol{a}_{j}^{\dagger}]=\delta_{ij},
\end{equation}
with $\beta>0$ and $\alpha\beta-\delta^{2}=\left(\frac{P\eta}{2\hslash}-\frac{Q\theta}{2\hslash}\right)^{2}+PQ>0$,
as follows from (\ref{eq:alpha exp}) and from the fact that $PQ>0$.
The Hamiltonian then reduces to the form
\begin{equation}
\mathcal{H}(t)=\hslash\omega\left(\sum_{j=1,2}\boldsymbol{a_{j}^{\dagger}a_{j}}+1\right)+i\hslash\gamma\left(\boldsymbol{a_{1}^{\dagger}a_{2}-a_{2}^{\dagger}a_{1}}\right);\omega=2\sqrt{\alpha\beta-\delta^{2}}.
\label{ttt}
\end{equation}

Observe that the non-diagonal second term, is like the Schwinger representation of $J_{2}$ angular momentum
operator ($\overrightarrow{J}=\boldsymbol{a_{i}}^{\dagger}(\overrightarrow{\sigma})_{ij}\boldsymbol{a_{j}}$). If we now perform yet another unitary transformation,
\begin{equation}
\left[\begin{array}{c}
\boldsymbol{a_{1}}\\
\boldsymbol{a_{2}}
\end{array}\right]\text{ }\rightarrow\text{ }\left[\begin{array}{c}
\boldsymbol{a_{+}}\\
\boldsymbol{a_{-}}
\end{array}\right]=M\left[\begin{array}{c}
\boldsymbol{a_{1}}\\
\boldsymbol{a_{2}}
\end{array}\right]=\frac{1}{\sqrt{2}}\left[\begin{array}{cc}
1 & -i\\
i & -1
\end{array}\right]\left[\begin{array}{c}
\boldsymbol{a_{1}}\\
\boldsymbol{a_{2}}
\end{array}\right]\label{eq:J2 to J3}
\end{equation}
\begin{equation}
\begin{array}{cc}
[\boldsymbol{a_{i},a_{j}^{\dagger}}]=\delta_{ij};[\boldsymbol{a_{i},a_{j}}]=0 & \text{ }(i,j\in\{+,-\})\end{array}
\end{equation}
with $M$ a unitary matrix, which make the term \ref{Zeem} into the exact diagonal form of $J_{3}$, while retaining
the diagonal form of the first term. Accordingly the system Hamiltonian may be rewritten as

\begin{equation}
\begin{alignedat}{1}\mathcal{H}(t) & =\hslash\sum_{j=+,-}\omega_{j}\boldsymbol{a_{j}}^{\dagger}\boldsymbol{a_{j}}+\hslash\omega;\text{ }\omega_{\pm}=\omega\mp\gamma\end{alignedat}
\label{dh}.
\end{equation}

It is useful to define instantaneous eigenvalue equation of this Hamiltonian, viz.,

\begin{equation}
\mathcal{H}(t)\left|n_{1},n_{2};t\right\rangle =E_{n_{1}n_{2}}(t)\left|n_{1},n_{2}; t\right\rangle ,
\end{equation}

and hence instantaneous eigenvalues and eigenstates can virtually be read-off from (\ref{dh}) as, 

\begin{equation}
\begin{array}{c}
E_{n_{1}n_{2}}(t)=\hslash\omega\left(n_{1}+n_{2}+1\right)-\hslash\gamma\left(n_{1}-n_{2}\right);\\
\left|n_{1},n_{2}; t \right\rangle =\frac{\left(\boldsymbol{a_{+}^{\dagger}}\right)^{n_{1}}\left(\boldsymbol{a_{-}^{\dagger}}\right)^{n_{2}}}{\sqrt{n_{1}!}\sqrt{n_{2}!}}\left|0,0;t\right\rangle,~~\boldsymbol{a}_{\boldsymbol{\pm}}(t)\left|0,0;t\right\rangle =0,
\label{ene}
\end{array}
\end{equation}

where $n_{1},n_{2}$ are nonnegative integers. This reproduces the spectra which was also obtained in \cite{gi} as well. This spectrum is clearly non-degenerate and this ensures that there is no crossing of energy levels during the adiabatic process.\\

In this light, it's worth noting that \cite{Jb} looked at essentially the same system, but with Bartelomi's realisation. By carrying out a simple scaling in (\ref{ncms}) and (\ref{bps1}).

\begin{equation}
\theta\rightarrow\xi^{-1}\theta ;~ \eta\rightarrow\xi^{-1}\eta ;
~ \hbar\rightarrow\hbar_{eff}=\hbar\xi^{-1},
\label{scale}
\end{equation}
with $\xi=\xi_{c}$ (\ref{bd}), one can readily verify that the phase-space algebra and spectrum in \cite{Jb} agree with  (\ref{non-com}) and (\ref{ene}), respectively.\\

Finally, we rewrite our primitive operators $\boldsymbol{a}_{1},\boldsymbol{a}_{2}$ of (\ref{ol}) as follows:

\begin{equation}
\begin{alignedat}{1} & \boldsymbol{a}_{i}=A(t)[q_{i}+\left(B(t)+iC(t)\right)p_{i}]\text{ ( 
	{i} \ensuremath{\in}\{1,2\})}\\
\text{where \text{ } } & A(t)=\left(\frac{\beta}{\hslash\omega}\right)^{1/2};~B(t)=\frac{\delta}{\beta};~C(t)=\frac{\omega}{2\beta}.
\end{alignedat}
\label{eq:a in ABC}
\end{equation}

Through the time dependence of $A,B,C,$ we can show that $\boldsymbol{a_{\pm},a_{\pm}^{\dagger}}$ also have explicit time dependence.

\section{Evolution of Ladder operators in Heisenberg picture}

In this section we will study the adiabatic evolution of the system in Heisenberg
picture to obtain the geometric phase shift. The Heisenberg  equation of motion for a generic operator $\hat {O} $ are given by  $\frac{d\hat{O}}{dt}=\frac{1}{i\hslash}[\hat{O},\mathcal{H}]+\frac{\partial\hat{O}}{\partial t}$. These equations of motion are
formally identical to Hamilton's equations of classical
dynamics so that the connection between quantum adiabatic phase and
the classical anholonomy associated with the classical Hannay angle shifts will also be quite transparent in this framework.

 Working in the Heisenberg representation for the ladder operators which are explicitly time dependent, relevant Heisenberg equations of motion take the following forms:

\begin{equation}
\begin{array}{c}
\dfrac{d}{dt}\left[\begin{array}{c}
a_{+}\\
a_{-}\\
a_{+}^{\dagger}\\
a_{-}^{\dagger}
\end{array}\right]=\left[\begin{array}{cccc}
X_{+} & 0 & 0 & Y\\
0 & X_{-} & Y & 0\\
0 & Y^{*} & X_{+}^{*} & 0\\
Y^{*} & 0 & 0 & X^{*}
\end{array}\right]\left[\begin{array}{c}
a_{+}\\
a_{-}\\
a_{+}^{\dagger}\\
a_{-}^{\dagger}
\end{array}\right]\\
X_{\pm}=\frac{\dot{A}}{A}\pm i\left(\gamma\mp2C\beta\mp\frac{(\dot{B}+i\dot{C})}{2C}\right),Y=-\frac{(\dot{B}+i\dot{C})}{2C}.
\end{array}\label{eq:ladder H evol}
\end{equation}
Until now, we've found all of the expressions without any approximation. However, we shall begin to investigate the adiabatic variation of  $P(t)$ and $Q(t)$ from now on. Note that, due to their explicit dependency, $A,B,C,\alpha,\beta,\gamma,\delta$ follow the same order of adiabaticity as P and Q, which is a measure of the parameter's slow time evolution. If $\dot{P},\dot{Q}\approx\epsilon$
, $\ddot{P},\ddot{Q}\approx\epsilon^{2}\ldots$, then $\dot{F}\approx\epsilon$
, $\ddot{F}\approx\epsilon^{2}\ldots$ , where $F$ collectively stands
for $A,B,C,\alpha,\beta,\gamma,\delta$. We won't overlook any terms that are now within the adiabatic limit, but
will merely keep track of adiabaticity order. It will eventually become evident that the geometric phase \cite{dj,ac} can be calculated by ignoring the second or higher order term  in adiabaticity order.\\

We are now able to disentangle the four interrelated differential equations caused by the evolution of annihilation and creation operators in (\ref{eq:ladder H evol}) by first differentiating w.r.t. time and rearranging them appropriately, yielding $\boldsymbol{a}_{+}$ and $\boldsymbol{a}_{-}$ second-order differential equations:

\begin{equation}
\begin{array}{c}
\frac{d^{2}\boldsymbol{a}_{+}}{dt^{2}}=\frac{d\boldsymbol{a}_{+}}{dt}\left(X_{+}+\frac{\dot{Y}}{Y}+X_{-}^{*}\right)+\boldsymbol{a}_{+}\left(\dot{X}_{+}-\frac{\dot{Y}}{Y}X_{+}+YY^{*}-X_{+}X_{-}^{*}\right)\\
\frac{d^{2}\boldsymbol{a}_{-}}{dt^{2}}=\frac{d\boldsymbol{a}_{-}}{dt}\left(X_{-}+\frac{\dot{Y}}{Y}+X_{+}^{*}\right)+\boldsymbol{a}_{-}\left(\dot{X}_{-}-\frac{\dot{Y}}{Y}X_{-}+YY^{*}-X_{-}X_{+}^{*}\right)
\end{array}\label{eq:ladder 2nd diff}
\end{equation}

 Now we observe that if $Y\approx\epsilon$,
then $\dot{Y}\approx\epsilon^{2}$ , and so $\frac{\dot{Y}}{Y}\approx\epsilon$.

Hence substituting $X_{+},X_{-},Y$ from (\ref{eq:ladder H evol}) and
only retaining terms involving first order of adiabaticity $\frac{\dot{Y}}{Y}$, we deduce

\begin{equation}
\frac{d^{2}\boldsymbol{a}_{+}}{dt^{2}}=\frac{d\boldsymbol{a}_{+}}{dt}\left(\mathfrak{P}+\frac{\dot{Y}}{Y}\right)+\boldsymbol{a}_{+}\left(\mathfrak{Q}-\frac{\dot{Y}}{Y}X_{+}\right),\label{eq:ldev22}
\end{equation}\\
where

\begin{equation}
\begin{aligned}\mathfrak{P}= & \left(2\frac{\dot{A}}{A}+2i\gamma+\frac{\dot{C}}{C}\right)\\
\mathfrak{Q}= & i\left(\dot{\gamma}-2\frac{d}{dt}\left(C\beta\right)-\gamma\frac{\dot{C}}{C}-2\frac{\dot{A}}{A}\gamma\right)+\left\{ \gamma^{2}-4C^{2}\beta^{2}-2\dot{B}\beta\right\} +\mathscr{O}\left(\epsilon^{2}\right)
\end{aligned}
\label{gopq}
\end{equation}

As can be seen, the differential equation fulfilled by $\boldsymbol{a}_{-}$ has a similar structure.\\

To continue, we must write (\ref{eq:ldev22}) in its normal form. To do so, we'll need a new time-dependent operator $\boldsymbol{b}(t)$ defined as,

\begin{equation}
\boldsymbol{a_{+}}(t)=\boldsymbol{b}(t)e^{\frac{1}{2}\int_{t}\left(\mathfrak{P}+\frac{\dot{Y}}{Y}\right)d\tau}.\label{eq:normaling}
\end{equation}

Replacing (\ref{eq:normaling}) in (\ref{eq:ldev22}) we obtain

\begin{equation}
\boldsymbol{\ddot{b}}+\boldsymbol{b}\left(\frac{\dot{\mathfrak{P}}}{2}-\frac{\mathfrak{P}^{2}}{4}-\mathfrak{Q}\right)+\boldsymbol{b}\left(\frac{\dot{Y}}{Y}X_{+}-\frac{\mathfrak{P}}{2}\frac{\dot{Y}}{Y}+\mathscr{O}\left(\epsilon^{2}\right)\right)=0,\label{beq}
\end{equation}

where
\begin{equation}
\frac{\dot{Y}}{Y}=\frac{\ddot{B}+i\ddot{C}}{\dot{B}+i\dot{C}}-\frac{\dot{C}}{C}=Z+i\tilde{Z}-\frac{\dot{C}}{C},
\end{equation}
with  both $Z$ and $\tilde{Z}$ corresponds respectively to the real and imaginary part of $\frac{\ddot{B}+i\ddot{C}}{\dot{B}+i\dot{C}}$.

On using the expressions of $\mathfrak{P}$ and $\mathfrak{Q}$
from (\ref{gopq}) we can be recast \ref{beq} in the following form: 

\begin{equation}
\boldsymbol{\ddot{b}}+\boldsymbol{b}(U+iV)=0
\end{equation}

where,

\begin{equation}
\begin{alignedat}{1}U= & 4C^{2}\beta^{2}+2\dot{B}\beta+2\tilde{Z}C\beta+\mathscr{O}\left(\epsilon^{2}\right)\approx\mathscr{O}\left(\epsilon^{0}\right)\\
V= & 2\frac{d}{dt}\left(C\beta\right)-2C\beta \left(Z-\frac{\dot{C}}{C}\right)+\mathscr{O}\left(\epsilon^{2}\right)\approx\mathscr{O}\left(\epsilon\right)
\end{alignedat}
\end{equation}

Note that, since we are working in adiabatic limit, the functions
$U$ and $V$ are very slowly varying function with time. Hence, we can apply the formula for WKB approximation for complex potential \cite{df} given by

\begin{normalsize}
	\begin{equation}
	\boldsymbol{b}(t)=\boldsymbol{b}(0)\left[\frac{C_{1}}{\sqrt{\left|\xi(t)\right|}}\exp\left(\int_{0}^{t}\left(i\xi(\tau)-\phi(\tau)\right)d\tau\right)+\frac{C_{2}}{\sqrt{\left|\xi(t)\right|}}\exp\left(\int_{0}^{t}\left(-i\xi(\tau)+\phi(\tau)\right)d\tau\right)\right]\label{eq:cwkbsol}
	\end{equation}
\end{normalsize}
where, $\sqrt{U+iV}=\xi+i\phi$ and ($C_{1}$, $C_{2}$) are arbitrary coefficients that can be used to find the general solution of the differential equation \ref{eq:cwkbsol} in our case:

\begin{normalsize}
	\begin{equation}
	\begin{alignedat}{1}\xi=\sqrt{\frac{\sqrt{U^{2}+V^{2}}+U}{2}} & \approx\sqrt{U+\frac{V^{2}}{4U}}\approx\sqrt{U}\approx2C\beta+\frac{\dot{B}\beta+C\beta\tilde{Z}}{2C\beta}\\
	\phi=\sqrt{\frac{\sqrt{U^{2}+V^{2}}-U}{2}} & \approx\sqrt{\frac{V^{2}}{4U}}\approx\frac{2\frac{d}{dt}\left(C\beta\right)-2C\beta\left(Z-\frac{\dot{C}}{C}\right)}{4C\beta}
	\end{alignedat}
	\end{equation}
\end{normalsize}

In our adiabatic approximation, we've left out terms in the second order (or beyond) of adiabaticity. We can now see that the solution must adhere to the following boundary condition: $\boldsymbol{b}(t=0)=\boldsymbol{b}(0)$. The cyclicity of the parameters also implies that $\sqrt{\left|\xi(0)\right|}=\sqrt{\left|\xi(T)\right|}$. Finally, only the phase factor of the second component in the solution (\ref{eq:cwkbsol}) produces the dynamical phase of $\boldsymbol{a}_{+}$ with the appropriate sign. This will become clear as we compute $\boldsymbol{a}_{+}(T)$.  As a result, we set $ C_{1}=0$ in (\ref{eq:cwkbsol}). By combining all of these expressions, the specific solution of (\ref{beq}) is obtained as follows:

 \begin{equation}
\begin{aligned}\boldsymbol{b}(T) & \approx\boldsymbol{b}(0)\exp\left(\int_{0}^{T}\left\{ -i\left(2C\beta+\frac{\dot{B}\beta+C\beta\tilde{Z}}{2C\beta}\right)+\phi\right\} d\tau\right)\end{aligned}
\end{equation}

 It may also be noted that $\frac{\dot{Y}}{Y}=Z+i\tilde{Z}-\frac{\dot{C}}{C}$
is an exact differential. Hence, we can write,

\begin{equation}
\int_{0}^{T}\frac{\dot{Y}}{Y}d\tau=\int_{0}^{T}\left(Z+i\tilde{Z}-\frac{\dot{C}}{C}\right)d\tau=\int_{0}^{T}Zd\tau+i\int_{0}^{T}\tilde{Z}d\tau=0
\label{lk}
\end{equation}

Clearly, the above expression (\ref{lk}) indicates that $\phi$ is an exact differential.\\

As a result of (\ref{eq:normaling}), we can effectively delete the term involving only the total derivatives, and the corresponding dynamical and geometric phase shifts emerge in a factorised form as,

\begin{normalsize}
	\begin{equation}
	\begin{alignedat}{1}\boldsymbol{a}_{+}(T) & =\boldsymbol{a}_{+}(0)\exp\left(-i\int_{0}^{T}\left(2C\beta+\frac{\dot{B}\beta+C\beta\tilde{Z}}{2C\beta}\right)d\tau+\frac{1}{2}\int_{0}^{T}\left(\frac{\dot{A}}{A}+2i\gamma+\frac{\dot{C}}{C}+\frac{\dot{Y}}{Y}\right)d\tau\right)\\
	& =\boldsymbol{a}_{+}(0)\exp\left(-i\int_{0}^{T}\left[\left(2C\beta-\gamma\right)+\left(\frac{\dot{B}\beta+C\beta\tilde{Z}}{2C\beta}\right)\right]d\tau\right)
	\end{alignedat}
	\end{equation}
\end{normalsize}

Therefore, we arrive at the following approximate solution,
\begin{equation}
\begin{aligned}
\boldsymbol{a}_{+}(T) & =\boldsymbol{a}_{+}(0)\exp\left(-i\int_{0}^{T}\left[\left(2C\beta-\gamma\right)+\left(\frac{\dot{B}}{2C}\right)\right]d\tau\right)\\
& =\boldsymbol{a}_{+}(0)\exp\left(-\frac{i}{\hslash}\int_{0}^{T}\left(\hslash\omega-\gamma\hslash\right)d\tau-i\int_{0}^{T}\frac{\beta}{\omega}\frac{d}{d\tau}\left(\frac{\delta}{\beta}\right)d\tau\right).\label{eq:a+sol}
\end{aligned}
\end{equation}

The two objects in the exponent reflect the dynamical and geometrical phases, respectively.

Finally, a critical analysis of the solution of the evolution equation of $\boldsymbol{a}_{-}$ in (\ref{eq:ladder 2nd diff}) reveals that a similar approximate solution can be obtained for $\boldsymbol{a}_{-}$, with the exception that $(+\gamma)$ is replaced by $(-\gamma)$, and $\gamma$ enters the solution only via the substitution of (\ref{eq:normaling}). As a result, we have

\begin{equation}
\begin{aligned}
\boldsymbol{a}_{-}(T) & =\boldsymbol{a}_{-}(0)\exp\left(-\frac{i}{\hslash}\int_{0}^{T}\left(\hslash\omega+\gamma\hslash\right)d\tau-i\int_{0}^{T}\frac{\beta}{\omega}\frac{d}{d\tau}\left(\frac{\delta}{\beta}\right)d\tau\right),\label{eq:a-sol}
\end{aligned}
\end{equation}

which gives the correct dynamical phase associated with $\boldsymbol{a}_{-}$.\\

\section{Geometric phases}
The structure of the geometric phase in the Heisenberg picture is examined here. Hence observing the excess phase factor over and above the dynamical phase in the expression of both the creation and annihilation operators $\boldsymbol{a}_{\pm}(T)$ in (\ref{eq:a+sol}) and (\ref{eq:a-sol}), obtained through the coupling between annihilation and creation operators in the above Heisenberg equation of motion (\ref{eq:ladder H evol}) whose leading behavior for adiabatic transport around a closed-loop $ \Gamma$ in time $ T $ can be identified with the geometric phase or Berry phase ( more precisely a geometric phase shift). As will be demonstrated very soon that this phase depends only on the geometry of the circuit $\Gamma$ in parameter space. It is interesting to note that the second phase factor(above) obtained here can readily be converted to the usual expression in terms of the phase gathered by the state vector by going over from the Heisenberg to the Sch\"odinger picture.\\

Moreover, the geometric phase factor $\Phi_{G}$ (over and above the dynamical phase $\frac{i}{\hbar}\int dt (\hbar \omega \pm \gamma \hbar)$) can be rewritten as a line integral in parameter space
 by using the transformation
\begin{equation}
\frac{d}{d\tau}=\frac{d\boldsymbol{R}}{d\tau}\cdot\nabla_{\boldsymbol{R}}
\end{equation}
where \textbf{$\boldsymbol{R}$} represents a vector in the parameter-space and $\nabla_{\boldsymbol{R}}$ denoting the the gradient operator in parameter space.\\

As the system Hamiltonian changes via the parameters in such a way that it makes a circuit $\Gamma$ in the parameter space and returns to its original values, the additional phase  $\Phi_{G}$ can be integrated over this closed-loop $\Gamma$ traversed in time $T$ so that it can also be written as a functional of $\Gamma$ as,

\begin{equation}
\Phi_{G}[\Gamma]=\oint_{\Gamma=\partial S}\frac{\beta}{\omega}\nabla_{\boldsymbol{R}}\left(\frac{\delta}{\beta}\right)\cdot d\boldsymbol{R}=\iint_{S}\nabla_{\boldsymbol{R}}\left(\frac{\beta}{\omega}\right)\times\nabla_{\boldsymbol{R}}\left(\frac{\delta}{\beta}\right)\cdot d\boldsymbol{S}\label{eq:ladderphase},
\end{equation}

where we have made use of Stoke's theorem in the second equality
relating this line integral over a closed curve in parameter space to a
surface integral over $S$, where $S$ is any surface in the parameter space that has the closed loop $\Gamma$ as its boundary. Note that in the final expression all references to the time $\tau$ has disappeared showing that the phase $\Phi_{G}$ depends only on the curve traced out in parameter space and not on how the curve is traversed. Thus $\Phi_{G}$ is a geometric object which is independent of the dynamics in contrast with the dynamical phase.\\

Now substituting $\alpha,\beta,\gamma,\delta$ from (\ref{eq:alpha exp}),
the geometric phase can now be expressed in terms of our 
time dependent primitive parameters $P(t)$, $Q(t)$ and the constant  noncommutative  parameters  $\theta$ and $\eta$ as,

\begin{equation}
\begin{aligned}\Phi_{G}[\Gamma]= & \left( \frac{\sqrt{-\theta \eta}}{4\hbar}\right) \iint_{ S}\nabla_{\boldsymbol{R}}\left(\frac{Q\left(1-\frac{\theta\eta}{4\hslash^2}\right)+P\left(\frac{\eta}{2\hslash}\right)^{2}}{\sqrt{\left(P\left(\frac{\eta}{2\hslash}\right)-Q\left(\frac{\theta}{2\hslash}\right)\right)^{2}+PQ}}\right)\times\\
& \nabla_{\boldsymbol{R}}\left(\frac{P\left(\frac{\eta}{2\hslash}\right)-Q\left(\frac{\theta}{2\hslash}\right)}{Q\left(1-\frac{\theta\eta}{4\hslash^2}\right)+P\left(\frac{\eta}{2\hslash}\right)^{2}}\right)\cdot d\boldsymbol{S}
\end{aligned}
\label{gphase in noncommute}
\end{equation}
Let us introduce, at this stage, the classical non-commutative parameters  $\tilde{\theta}$ and $\tilde{\eta}$ as,
\begin{equation}
\theta=\hbar\tilde{\theta}~~~~~\eta=\hbar\tilde{\eta},
\end{equation}
Accordingly the geometrical phase factor may be rewritten as
\begin{equation}
\begin{aligned}\Phi_{G}[\Gamma]= & \left( \frac{\sqrt{-\tilde{\theta} \tilde{\eta}}}{4}\right) \iint_{ S}\nabla_{\boldsymbol{R}}\left(\frac{Q\left(4-\tilde{\theta}\tilde{\eta}\right)+P\tilde{\eta}^{2}}{\sqrt{\left(P\tilde{\eta}-Q\tilde{\theta}\right)^{2}+4PQ}}\right)\times\\
& \nabla_{\boldsymbol{R}}\left(\frac{P\tilde{\eta}-Q\tilde{\theta}}{Q\left(4-\tilde{\theta}\tilde{\eta}\right)+P\tilde{\eta}^{2}}\right)\cdot d\boldsymbol{S}
\end{aligned}
\label{mute}
\end{equation}
Note at this stage that $\tilde{\theta}\tilde{\eta}< 0$is the dimension-less constant. Also, as desired the denominator of this expression with $PQ>0$, will never vanish. Further, if one of the two types of non-commutativity is absent, i.e. if $\theta$ or $\eta=0$, the geometric phase vanishes. So, for this $2D$ parametric oscillator system, the non-commutative character of phase space geometry as a whole, as well as the geometry of the manifold corresponding to the effective parameter space, plays a vital role in giving rise to the   geometric phase shift.\\

Before we progress further, let us pause for a while and make some pertinent remarks:

	$\bullet$ 
	The appearance of this nonzero geometric phase factor in this case is attributable to the breaking of the discrete time-reversal symmetry of the system Hamiltonian (\ref{nch}),~(\ref{comh}) \cite{j1,j2,j3,j4}. In order to emphasize this matter, we need to explain the deeper meaning of the time-reversal symmetry of time dependent Hamiltonian $\mathcal{H}(t) $ which is determined by a set of time-varying parameters. To begin with, consider this system Hamiltonian $\mathcal{H}(t)$ as a sequence of instantaneous time-independent Hamiltonians associated with each instant $t$, each of which is a distinct Hermitian matrix (finite or infinite) with complex off-diagonal and real diagonal elements. Time reversal symmetry now refers to instantaneous Hamiltonians $ \mathcal{H}(t_{0})$, i.e. the parameter's time dependency is frozen at their values corresponding to that moment say at $t=t_{0}$, which is no longer time-dependent. Thus a time-dependent system Hamiltonian being invariant under time-reversal refers to the fact that, each such instantaneous Hamiltonians in the sequence described by a real symmetric matrix, not just a complex Hermitian.
	
	In other words, if we consider a system, the time evolution of which is governed by this instantaneous Hamiltonian $\mathcal{H}(t_0)$ for some
	finite time interval after $t_0$ and then time reversed at any subsequent time $t > t_0 $, then the corresponding wave-function is obtained simply by performing complex conjugation, without touching the set of parameters occurring there. This is due to the fact that the properties of the time dependent parameters are now kept fixed to their respective values at time $t=t_{0}$ and are thus not affected by neither the continuous time translation nor the discrete time-reversal transformation, i.e. at the instantaneous limit the parameters remain untouched under the operation $t\rightarrow -t$, which we can more correctly refer to as a "quasi-time reversal" transformation. Consequently, under this `` quasi-time reversal" anti-linear and anti-unitary transformation,
	if the system retraces its own history, regardless of which instantaneous Hamiltonian of the original time-dependent system is chosen, we say $\mathcal{H}(t) $ is invariant under time-reversal,
	then we say $\mathcal{H}(t) $ is invariant under time-reversal.
	A precise mathematical definition of time-reversal discrete symmetry \cite{j1,j2} would be:
	
	\begin{equation*}
	\begin{aligned}
	\hat{\Xi}~\hat{H}(t)~\hat{\Xi}^{-1} &= \hat{H}(t)  &&  \text{(without any change in the sign of $t$)}
	\end{aligned}
	\end{equation*}
	where the instantaneous time-reversal operator (anti-linear) $\hat{\Xi}$ preserves all real time-dependent parameter(s). The concept of instantaneous time-reversal symmetry was borrowed from \cite{j5}, where closely related situations were observed in a topological insulator model.\\

	Now, in the usual commutative space: $\mathcal{H}_{c}(t)=P(t)(\hat{p}_{1}^{2}+\hat{p}_{2}^{2})+Q(t)(\hat{x}_{1}^{2}+\hat{x}_{2}^{2})$;
	with $\hat{p}_{1},\hat{p}_{2},\hat{x}_{1},\hat{x}_{2}$ satisfying
	usual Heisenberg algebra. (Instantaneous or quasi) time reversal operator acting as: $\hat{p}_{i}\rightarrow\hat{p}'_{i}=\hat{\Xi}~\hat{p}_{i}~\hat{\Xi}^{-1}=-\hat{p}_{i}$
	and $\hat{x}_{i}\rightarrow\hat{x}'_{i}=\hat{\Xi}~ \hat{x}_{i}~\hat{\Xi}^{-1}=\hat{x}_{i}$
	, shows that the instantaneous Hamiltonian is symmetric under time reversal: $\hat{\Xi}~\mathcal{H}_{c}(t)~\hat{\Xi}^{-1}=\mathcal{H}_{c}(t)$, as the parameters $P(t)$ and $Q(t)$ are not touched at instantaneous limit.\\
	
	However, in the quantum (non-commutative) plane, the time evolution is governed by the Hamiltonian (\ref{nch}) with non-commutative coordinates and momenta satisfying  algebra (\ref{ncms}) or equivalently by the Hamiltonian
	(\ref{comh}) with the usual commutative coordinates and canonical momenta,
	transforming like, $p_{i}\rightarrow-p_{i}$ , $q_{i}\rightarrow q_{i}$,
	under discrete time reversal. Hence, the instantaneous system Hamiltonian $\mathcal{H}(t)=\alpha(t)(p_{1}^{2}+p_{2}^{2})+\beta(t)(q_{1}^{2}+q_{2}^{2})+\delta(t)(p_{i}q_{i}+q_{i}p_{i})-\gamma(t)(q_{1}p_{2}-q_{2}p_{1})$
	is not time reversal symmetric: $\hat{\Xi}~\mathcal{H}(t)~\hat{\Xi}^{-1}\neq\mathcal{H}(t)$;
	the presence of the dilatation operator as well as the angular momentum operator term breaks this symmetry. Particularly, the breaking due to dilatation term is essential for the existence of non-vanishing Berry phase in our case. Actually, whenever a system Hamiltonian continues to support a discrete non-degenerate spectrum and if it is invariant under time reversal, then instantaneous energy eigenfunctions will be real which does not exhibit any geometric phase under adiabatic approximation. In fact, it
	has been shown in \cite{j1,j6}
	that this time-reversal symmetry breaking of the instantaneous Hamiltonian is a necessary, but not sufficient,
	condition for the existence of non-vanishing Berry's  phase\cite{Nd}. And it
	is because of this broken time reversal symmetry, while considering non-commutative phase space, that there arises a natural possibility of obtaining a non-vanishing additional geometrical phase shift in our planar system of simple isotropic oscillator in non-commutative phase space.\\

	$\bullet$ A real valued connection one-form $\boldsymbol{A}$ on effective parameter space ($\mathcal{M}$) can be identified from the above expression \ref{eq:ladderphase} of line integral around a closed path  $\Gamma $ as,

	\begin{equation}
	\boldsymbol{ A}=\frac{\beta}{\omega}d(\frac{\delta}{\beta}) =-\frac{\alpha}{\omega}d(\frac{\delta}{\alpha})-d[tan^{-1}(\sqrt{\frac{\alpha \beta}{\delta^{2}} -1})],
	\label{phg}
	\end{equation}
	
	showing that, upto the total derivative i.e. upto an exact form  the Berry connection one-form can also be written as 
	\begin{equation}
	\boldsymbol{ A}:=-\frac{\alpha}{\omega}d(\frac{\delta}{\alpha}) 
	\label{connec}
	\end{equation}
	This is exactly like a nonsingular gauge transformation, the one-form behaves as a geometric vector potential in this parameter space. The exchange symmetry between $\alpha$ and $\beta$, the coefficients of $\vec{p}^{2}$ and $\vec{q}^{2}$ in the system Hamiltonian (\ref{comh}), is manifested in some way with this property of this one-form. In turn, this structure of the connection one form (\ref{connec}) was also observed in \cite{G,G1}, where a Hamiltonian of the system with the same form as in (\ref{comh}) was previously used to describe a 1D parametric generalised harmonic oscillator. Except that their time-dependent parameters $\alpha, \beta, \delta$ were fundamental in nature, whereas in our model, effective parameters $\alpha, \beta, \delta$ are not fundamental and are instead given in terms of fundamental $P$ and $Q$ via a differentiable \cite{diff} linear mapping (\ref{eq:alpha exp}. Also, none of our time-dependent parameters $\alpha(t), \beta(t), \delta(t), \gamma(t)$ can go to zero during the evolution of the system; otherwise, we'd have $P(t)\propto Q(t)$\footnote{For example, $\gamma(t)=0~~\forall t$, implies from (\ref{eq:alpha exp}), that $P(t)\propto Q(t) ~~\forall t$. In fact, by writing more specifically, we arrive at $\frac{P}{Q}=-\frac{\theta}{\eta}$. } as is clear from (\ref{eq:alpha exp}). In the primitive parameter space ($P,Q$), the closed-circuit $\Gamma$ will then collapse to a 1D axis, generating a vanishing Berry's geometric phase. This is especially true for the parameter $\delta(t)$, which clearly plays an important role here and whose origin can be traced back to (\ref{ol}), where ($\frac{\delta}{\beta}$) occurs as the real part of the coefficient $p _j$; without it, we cannot get any geometrical phase, as (\ref{eq:a+sol},\ref{eq:a-sol}) demonstrates.\\


	 In addition to the above reason, the $\gamma (t)$ occurring in the Zeeman like term (\ref{comh}) must be non-zero for an yet another reason to obtain a non-vanishing geometric phase; it serves an important purpose by allowing us to avoid the crossing of energy levels by lifting the degeneracy, as we have discussed in Section 2. Despite this, it does not have an explicit existence in the form of the Berry connection (\ref{eq:ladderphase},\ref{connec})(if we ignore the linear relations in (\ref{eq:alpha exp}) for now); it rather shows in the dynamical phases of (\ref{eq:a+sol},\ref{eq:a-sol}), correlating with J. Anandan et al.'s observation \cite{ja} where $ U(2)$ was the dynamical group. To uncover the deeper reason behind
	 all this, observe that, although the Hamiltonian $\mathcal{H} (t)$ (\ref{comh},\ref{hsep}) at different times do not commute with each other: $[\mathcal{H}(t), \mathcal{H}(t')]\ne 0$ for $t \ne t'$, it nevertheless can be regarded as an element of the Lie algebra $ su(1,1)\oplus u(1) $, which splits into two commuting factors  as in (\ref{gro}) (see Appendix B). It's also worth noting that these two terms in (\ref{gro}) commute at distinct times. As a result, the time evolution operator (in the Sch\"odinger image) factorizes as,

	\begin{equation}
	\mathcal{U}=\hat{\mathcal{T}}(e^{-\frac{i}{\hslash}\int dt[\alpha(t)(p_{1}^{2}+p_{2}^{2})+\beta(t)(q_{1}^{2}+q_{2}^{2})+\delta(t)(p_{i}q_{i}+q_{i}p_{i})]})\hat{\mathcal{T}}(e^{\frac{i}{\hslash}\int dt\gamma(t)(q_{1}p_{2}-q_{2}p_{1})}).
	\label{angul}
	\end{equation}
	where the time-odering or chronological operator is $\hat{\mathcal{T}}$. After all, $\gamma (t)$, like $\omega (t)$, appears in the integral as $\int_0^T \gamma (t) dt $, which clearly depends on the evolution parameter $T$ and cannot be represented as a functional of the closed-loop $\Gamma$, the latter being the telltale indicator of Berry's geometrical phases (\ref{eq:ladderphase},\ref{phg}): $\Phi_G[\Gamma]=\int_{\Gamma} \boldsymbol{A}$.\\

$\bullet$ Finally, in our case, we would like to emphasise the unitary equivalence of the two alternative Bopp-shifts (\ref{bps1} and \ref{bps3}), as proved in Appendix A. We can see that the geometrical phase $\Phi_{G}$ in (\ref{gphase in noncommute}) was calculated for the scale parameter $\xi$, which can be taken as $\xi=1$ without losing any generality. The appropriate geometric phase $\Phi_{G}$ for any other value of $\xi$ may be easily found by replacing $\theta\rightarrow\xi\theta$  and $\eta\rightarrow\xi\eta$, i.e. using the realization (\ref{bps3}). It's worth, however, noting that if  $\hbar$ is  also scaled likewise to $\hbar\rightarrow\xi\hbar$, $\Phi_{G}$ remains invariant, as the realization (\ref{bps3}) will reduce to the case for $\xi=1$, undoing the scaling transformation exactly like (\ref{scale}).\\

 If one uses realisation (\ref{bps1}), the occurrence of the counterpart of the expression (\ref{gphase in noncommute}) will be absent at the critical value $\xi=\xi_{c}$ (\ref{bd}). On the other hand, the counterpart of (\ref{gphase in noncommute}) will be present in the corresponding realisation (\ref{bps3}) with $ \xi=\xi_{c}$, but it can be proved that it loses its geometrical relevance, and in any case, the Berry's geometrical phase $\Phi_{G}$ vanishes: $\Phi_{G}=0$. The task of demonstrating this can be accomplished by two methods, both of which are basically equivalent, by taking advantage of the unitary equivalence (described above) between two types of Bopp shifts (\ref{bps1},\ref{bps3}) that holds only for $ \xi=\xi_{c}$ (\ref{bd}. To do so, we first apply the relation (\ref{bps1}) in (\ref{nch}) to get the Hamiltonian in the form,

	\begin{equation}
	\mathcal{H}^{1}(t)=\alpha^{(1)}(t)p^{2}_{i}+\beta^{(1)}(t)q^{2}_{i}-\gamma^{(1)}(t)\epsilon_{ij}q_{i}p_{j}
	\label{nh}
	\end{equation}
	where,
	\begin{equation}
	\begin{alignedat}{1}\alpha^{(1)}(t) &
	=\xi_{c}\left(P(t)+\frac{\theta^{2}}{4\hbar^{2}}Q(t)\right)\\
	\beta^{(1)}(t) & =\xi_{c}\left(Q(t)+\frac{\eta^{2}}{4\hbar^{2}}P(t)\right)\\
	\gamma^{(1)}(t) & =\frac{\xi_{c}}{\hbar}\left(\theta Q(t)+\eta P(t)\right)
	\end{alignedat}
	\label{eq:alpha exp2}
	\end{equation}\\
	The above pair (\ref{nh},\ref{eq:alpha exp2}) of equations,  are the counterparts of
	(\ref{comh}) obtained by using (\ref{bpshift}) in
	(\ref{nch}). Not only do we have $\delta^{1}(t)=0 $ $\forall t$ (which is the counterpart of $\delta(t)$ in (\ref{eq:alpha exp})), but we also have the absence of any linear equation relating $\delta^{(1)}(t)$ with fundamental parameters $P(t)$ and $Q(t)$, indicating the absence of a counterpart of (\ref{gphase in noncommute})following from (\ref{eq:ladderphase}). In this instance, one must use (\ref{phg},\ref{connec}), or the second phase factor in the exponents of (\ref{eq:a+sol}) and (\ref{eq:a-sol}), to see that the geometric phase vanishes: $ \Phi_{G}=0$.\\

	On the other hand, to arrive at the same conclusion using the Bopp Shift (\ref{bps3}), we must write down the exact expressions for $\alpha$,$\beta$,$\gamma$,$\delta$ for $\xi=\xi_{c}$, which can be obtained by simply replacing $\theta\rightarrow\xi_{c}\theta;\eta\rightarrow\xi_{c}\eta$ in
	(\ref{eq:alpha exp}), as already noted, to get

	\begin{equation}
	\mathcal{H}^{(2)}(t)=\alpha^{(2)}(t)p^{2}_{i}+\beta^{(2)}(t)q^{2}_{i}-\gamma^{(2)}(t)\epsilon_{ij}q_{i}p_{j}+\delta^{(2)}(t)(q_{i}p_{i}+p_{i}q_{i}), 
	\label{dc}
	\end{equation}
	
	with
	
	\begin{equation}
	\begin{alignedat}{1}\alpha^{(2)}(t)=\alpha(t;\xi_{c}) & =P(t)\left\{ 1-\frac{\xi_{c}^{2}
		\theta\eta}{4\hslash^{2}}\right\} +Q(t)\left(\frac{\xi_{c}\theta}{2\hslash}\right)^{2}\\
	\beta^{(2)}(t)=\beta(t;\xi_{c}) & =Q(t)\left\{ 1-\frac{\xi^{2}_{c}\theta\eta}{4\hslash^{2}}\right\} +P(t)\left(\frac{\xi_{c}\eta}{2\hslash}\right)^{2}\\
	\gamma^{(2)}(t)=\gamma(t;\xi_{c}) & =\frac{\xi_{c}}{\hbar}\left(\eta P(t)+\theta Q(t)\right)\\
	\delta^{(2)}(t)=\delta(t;\xi_{c}) & =\left(\frac{\xi^{2}_{c}\sqrt{-\theta\eta}}{4\hslash^2}\right)\left(\eta P(t)-\theta Q(t)\right)
	\end{alignedat}
	\label{eq:alpha exp3} 
	\end{equation} \\
	
	The non-vanishing character of $\delta^{(2)}(t)$ in this case appears to indicate a non-trivial $\Phi_{G}$. As a result, demonstrating that $\Phi_{G}=0$ in this example by just using (\ref{gphase in noncommute}) may be challenging. However, it turns out that by employing a time-independent unitary transformation $\boldsymbol{U}\in SU(1,1)\otimes U(1)$, this non-zero $\delta^{(2)}(t)$ can be eliminated. Indeed, by making use of the global (time-independent) unitary transformation
	(\ref{unitary}) and the relations in (\ref{equival}), we can easily retrieve from (\ref{nch}) that
	\begin{equation}
	\mathcal{H}^{(2)}(t)=\boldsymbol{U}\mathcal{H}^{(1)}(t)\boldsymbol{U}^{\dagger}
	\label{funda}
	\end{equation}
where the relevant group parameters are provided in (\ref{alph}) and $\beta_{0}=\beta_{1} $ in (\ref{beta1}). This indicates that $U(1)$ sector of the total dynamical symmetry group $SU(1,1)\otimes U(1)$ (\ref{algcom}) (See Appendix-B) is also set by this finely tuned value of group parameter $\beta_{0}=\beta_{1}$ in (\ref{beta1}); it is not arbitrary. This, in turn, fixes all other parameters of $SU(1,1)$ in (\ref{alph}). And, as previously indicated, this feature makes it impossible to establish the vanishing of $ \Phi_{G}$ simply by using (\ref{gphase in noncommute}).
In any case, we can see that the dilatation term in (\ref{dc}) may be removed for any time $t$, showing that the geometric phase factor $\Phi_{G}=0$. As a result, the phase turns out to be integrable for this critical value of $\xi=\xi_{c}$. To show this, through a more conceptually transparent method, let us consider the following identity as,

\begin{equation}
	\alpha^{(2)}(t)\beta^{(2)}(t)-\left(\delta^{(2)}(t)\right)^{2}=\alpha^{(1)}(t)\beta^{(1)}(t)~~~~ \forall t
	\label{SUP}
	\end{equation}
	
	which can be easily shown to follow trivially from (\ref{eq:alpha exp2},\ref{eq:alpha exp3}). This demonstration of the global invariance of characteristic frequency (\ref{su}) under the $SU(1,1)$ or rather  $SO(2,1)=SU(1,1)/\mathbb{Z}_2$ subgroup of $ SU(1,1)\otimes U(1)$ has been shown in Appendix-B i.e 
	\begin{equation}
	\omega(t;\xi_{c})=2\sqrt{\alpha^{(1)}(t)\beta^{(1)}(t)}=2\sqrt{\alpha^{(2)}(t)\beta^{(2)}(t)-(\delta^{(2)}(t))^2}
	\end{equation}
	expressed in terms of one or the other set of the effective parameters. This, as a result, indicates that the parameters are actually linked via the $SO(2,1)$ transformation (\ref{sop},\ref{su}), and as has been discussed in Appendix B, this
	$\delta^{(2)}(t) $ can be regarded as the time component of a space-like
	3-vector. However, a global (time-independent) "Lorentz transformation" in (2+1)D can now eliminate it for all time.. And finally when this $SO(2,1) $ operation is lifted to its covering group $SU(1,1)$ ( See Appendix-B), it gives
	the $ SU(1,1)$ part of the transformation matrix $\textbf{U}\in SU(1,1)\otimes
	U(1)$ (\ref{unitary}) (see Appendix-A).\\
	
	
	In the time-dependent Sch\"odinger equation, no extra connection term (\ref{connection} in Appendix-B) will develop because this $\boldsymbol{U}$ is time-independent, and
	for any reference state $\left|\Psi(t)\right\rangle $ whose time evolution is  guided
	by $ \mathcal{H}^{(2)}(t)$ as
	$i\hbar\partial_{t}\left|\Psi(t)\right\rangle=\mathcal{H}^{(2)}(t)\left|\Psi(t)\right\rangle$,
	we've got a corresponding state $
	\left(\boldsymbol{U}^{\dagger}\left|\Psi(t)\right\rangle\right)$, which evolve in time by the Hamiltonian $\mathcal{H}^{(1)}(t)$, as follows:
	\begin{equation}
	i\hbar\frac{\partial\left(\boldsymbol{U}^{\dagger}\left|\Psi(t)\right\rangle\right)}{\partial t}=\mathcal{H}^{(1)}(t)\left(\boldsymbol{U}^{\dagger}\left|\Psi(t)\right\rangle\right)
	\end{equation}
As a result, we're back to $\mathcal{H}^{(1)}(t)$, where the equivalent  $\delta^{(1)}(t)=0$ and no geometric vector potential develops as a result. In other words, for the scale parameter $\xi$ at this particular value $\xi_{c}$, it is possible to eliminate the dilatation term by applying a time-independent unitary transformation (\ref{funda}) to the system Hamiltonian $\mathcal{H}^{(2)}(t)$ :
\begin{equation}
\mathcal{H}^{(2)}(t)\rightarrow\boldsymbol{U}^{\dagger}\mathcal{H}^{(2)}(t)\boldsymbol{U}
\end{equation}
As a result, the geometric phase factor vanishes.\\

Of course, this current analysis will not hold true for any other values of $\xi$ other than $\xi_{c}$. Finally, a time dependent unitary transformation $\mathcal{W}(t) \in SU(1,1)$ (\ref{W})  can be used to eliminate the dilatation part completely. Unlike the previous situation, however, it is not necessary for $\mathcal{W}(t)$ to belong to the complete product group $SU(1,1)\otimes U(1)$; keeping the $ U(1)$ part becomes unnecessary. As a result, the disappearence of the geometric phase is only apparent in 
nature \cite{g5}, and it reappears in the dynamical part of the total phase, while still keeping its geometrical character \cite{G}. This has been elaborated in Appendix-B.\\

Returning to our original goal, let's see if we can connect the extra geometric phase shift gained in the Heisenberg picture with the more common  approach of computing Berry phases using state vectors. We'll go to the Schr\"odinger picture for that. Let's start by rewriting (\ref{eq:a+sol}) and (\ref{eq:a-sol}) as,

\begin{equation}
\boldsymbol{a}_{\pm}(T)=\boldsymbol{a}_{\pm}(0)\exp\left(-i\Theta_{\pm,d}-i\Phi_{G}\right)
\label{apmmt}
\end{equation}

where

\begin{equation}
\Theta_{\pm,d}=\int_{0}^{T}\left(\hslash\omega\mp\gamma\hslash\right);\Phi_{G}=\int_{0}^{T}\frac{\beta}{\omega}\frac{d}{d\tau}\left(\frac{\delta}{\beta}\right)d\tau\label{fp}
\end{equation}

are the dynamical and geometrical phases respectively.

Let $\mathcal{U}(0,t)$ be our system's Schr\"odinger time evolution operator, as generated by the Hamiltonian (\ref{comh}). Then, $\boldsymbol{a}_{\pm}(t)=\mathcal{U}^{\dagger}(0,t)\boldsymbol{a}_{S\pm}(t)\mathcal{U}(0,t)$,
where $\boldsymbol{a}_{S\pm}(t)$ are the annihilation (corresponding creation) operators in Schr\"odinger
picture. Note that even in the Schr\"odinger  picture, the time dependency of the ladder operators are not totally frozen; its time dependence now stems from a collection of time dependent external parameters.\\

However, in Schr\"odinger picture, Berry's geometric phase was derived under circuital adiabatic condition to ensure that the transported state returns to itself up to a phase factor so that we can write,

\begin{equation}
\begin{split}
&\frac{\left(\boldsymbol{a}_{+}^{\dagger}(T)\right)^{n_{1}}\left(\boldsymbol{a}_{-}^{\dagger}(T)\right)^{n_{2}}}{\sqrt{n_{1}!}\sqrt{n_{2}!}} \left|0,0;t=0\right\rangle _{S}\\
&=\mathcal{U}^{\dagger}(0,T)\frac{\left(\boldsymbol{a}_{S+}^{\dagger}(T)\right)^{n_{1}}\left(\boldsymbol{a}_{S-}^{\dagger}(T)\right)^{n_{2}}}{\sqrt{n_{1}!}\sqrt{n_{2}!}}\mathcal{U}(0,T)\left|0,0;t=0\right\rangle _{S}\\
&=\mathcal{U}^{\dagger}(0,T)\frac{\left(\boldsymbol{a}_{S+}^{\dagger}(T)\right)^{n_{1}}\left(\boldsymbol{a}_{S-}^{\dagger}(T)\right)^{n_{2}}}{\sqrt{n_{1}!}\sqrt{n_{2}!}}e^{-i\phi_{0,0}}\left|0,0;t=T\right\rangle _{S}\\
&=\mathcal{U}^{\dagger}(0,T)\left|n_{1},n_{2};t=T\right\rangle _{S}e^{-i\phi_{0,0}}\\
&=e^{i(\phi_{n_{1},n_{2}}-\phi_{0,0})}\left|n_{1},n_{2};t=0\right\rangle _{S}
\end{split}
\label{cb1}
\end{equation}

where  $\phi_{n_{1},n_{2}}$ represents the total adiabatic phase obtained by the states $\left|n_{1},n_{2};t=0\right\rangle _{S}$ after evolving by $\mathcal{H}(t)$ over its whole period $T$. We also discover, when we use (\ref{apmmt}), that


\begin{equation}
\begin{split}
&\frac{\left(\boldsymbol{a}_{+}^{\dagger}(T)\right)^{n_{1}}\left(\boldsymbol{a}_{-}^{\dagger}(T)\right)^{n_{2}}}{\sqrt{n_{1}!}\sqrt{n_{2}!}} \left|0,0;t=0\right\rangle _{S}\\
& =e^{in_{1}\left(\Theta_{+,d}+\Phi_{G}\right)}e^{in_{2}\left(\Theta_{-,d}+\Phi_{G}\right)}
\frac{\left(\boldsymbol{a}_{+}^{\dagger}(0)\right)^{n_{1}}\left(\boldsymbol{a}_{-}^{\dagger}(0)\right)^{n_{2}}}{\sqrt{n_{1}!}\sqrt{n_{2}!}}\left|0,0;t=0\right\rangle _{S}\\
& =e^{i n_{1}\left(\Theta_{+,d}+\Phi_{G}\right)}e^{i n_{2}\left(\Theta_{-,d}+\Phi_{G}\right)}\left|n_{1},n_{2};t=0\right\rangle _{S}.
\end{split}
\label{cb2}
\end{equation}

It's worth noting that we've used the fact that $\boldsymbol{a}_{\pm}(t=0)=\boldsymbol{a}_{S\pm}(t=0).$
When we compare the two equations (\ref{cb1}) and (\ref{cb2}) above, we get

\begin{equation}
\phi_{n_{1},n_{2}}=\phi_{0,0}+\left[n_{1}(\Theta_{+,d}+\Phi_{G})+n_{2}(\Theta_{-,d}+\Phi_{G})\right]\label{eq:adiaphase}
\end{equation}

As a result, the Berry phase gained by the state vector $\left|n_{1},n_{2};t=0\right\rangle _{S}$ is,

\begin{equation}
\phi_{B}^{(n_{1},n_{2})}=\phi_{B}^{(0,0)}+(n_{1}+n_{2})\Phi_{G}
\end{equation}

The linear character of  Berry phases of different eigenstates is a general conclusion \cite{ss} for any Hamiltonian with equally spaced discrete non-degenerate spectrum. In our case, the total Hamiltonian (\ref{dh}) is divided into two commuting parts corresponding to $\boldsymbol{a}_{+}$ and $\boldsymbol{a}_{-}$, each of which produces its own equispaced energy spectrum in their respective sub-Hilbert spaces  $\mathbb{H}_{\pm}$, the tensor product of which forms the total Hilbert space: $\mathbb{H}=\mathbb{H}_{+}\otimes\mathbb{H}_{-}.$\\

Importantly, the expected value of any operator at time $t$ in a state produced from any initial state and evolving under an adiabatic system Hamiltonian is influenced by the difference in Berry phases of distinct eigenstates, where the ground state contribution $\phi_{B}^{(0,0)}$ cancels out. And this is the only concept used in Berry's phase \cite{fu} experiments. As a result, our derivation provides comprehensive information that may be quite useful in the prediction of such events.

 
 \section{Classical analogue: Hannay angles}
 
  This section is devoted to studying the classical analog of this Berry's quantal geometric
  phase. It has been shown in the literature that a classical system exhibiting
  Hannay's angle \cite{jh,mv} must corresponds to the Berry's geometric phase at the quantum level \cite{M.V}. To clinch this
  the correspondence we will exploit a semi-classical approach, using coherent states \cite{mm}
  and some suitably chosen quantum operators that represent the classical
  action and the conjugate angle variables. It will be useful to recall the notion of Hamiltonians at the instantaneous limit presented in the preceding section to discuss the idea of time-reversal symmetry that has been broken in our situation and to decipher the Hamiltonian of a time-varying system $\mathcal{H}(t)$ (\ref{comh}),  whose time dependency originates  from the  the occurrence of a collection of effective parameters $\alpha(t),\beta(t),\delta(t)$ and $\gamma(t)$  that are time-dependent, as a series of the infinite number of time-independent Hamiltonians associated with the planar system labeled by time say $t_{0}$ and the set of parameter's values are held constant by their respective values at time $t_{0}$ as $\alpha(t_{0}),\beta(t_{0})
  ,\gamma(t_{0}),\delta(t_{0})$. Individual dynamical systems evolving by instantaneous Hamiltonians like $\mathcal{H}(t_{0})$, if left to their own parametric values, will produce time evolution in their own phase-spaces at the classical level, as shown below. Clearly a proper unitary (canonical) transformations of the corresponding quantum (resp. classical) systems can bring each of these instantaneous Hamiltonians $\mathcal {H}(t_0)$ to the standard form of a pair of de-coupled oscillators. And, for each of these classical systems corresponding to instantaneous quantum Hamiltonians, this ensures the occurrence of periodic motion in the classical phase-spaces, allowing us to introduce canonical action and conjugate angle variables. Starting with the annihilation and  creation operators $\boldsymbol{a}_{+}$ and $\boldsymbol{a}^{\dagger}_{-}$, we can easily construct, by a linear transformation, a set of canonically conjugate coordinate and momentum operators, called as quadrature variables as,
  
  
  
  \begin{equation}
  \begin{alignedat}{1}\hat{q}_{\pm}= & \sqrt{\frac{\hslash\alpha}{2\omega_{\pm}}}(\boldsymbol{a}_{\pm}^{\dagger}+\boldsymbol{a}_{\pm})\\
  \hat{p}_{\pm}= & i\sqrt{\frac{\hslash\omega_{\pm}}{2\alpha}}(\boldsymbol{a}_{\pm}^{\dagger}-\boldsymbol{a}_{\pm})
  \end{alignedat}
  \label{quadr}
  \end{equation}
  
  On using (\ref{ol},\ref{eq:J2 to J3}), one can easily see that both $\hat{q}_{\pm}$ and $\hat{p}_{\pm}$ has linear functional dependence on the original phase-space variables $q_{i},p_{i}$:
  \begin{equation}
  \hat{q}_\pm=\hat{q}_\pm(q_{1},q_{2},p_{1},p_{2}),~ \hat{p}_\pm=\hat{p}_\pm(q_{1},q_{2},p_{1},p_{2})
  \end{equation}
  
  At the quantum level, this will correspond, at each one of these instants $t_0$, to a suitable unitary transformation $\mathcal{V}(t_0)\in SU(1,1)\otimes U(1)$. This is the same situation as with time-dependent unitary transformation $\mathcal{W} (t_0)$ as in (\ref{W}) (see Appendix B), which helped us to remove just the crucial dilatation term, without affecting the Zeeman-like expression in  $\mathcal{H}(t)$ (\ref{comh}), except that we are now eliminating the Zeeman like term also. However, because we are dealing with instantaneous classical integrable systems, the explicit form of such a unitary operation $\mathcal{V}(t _0)$ is neither very straightforward to obtain nor essential in our current scenario. In fact, an analogues linear canonical transformation with coefficients determined by the values of $\alpha,\beta,\gamma,\delta$ at $t=t_0$ canonically transforms instantaneous classical systems from \{$q_1,q_2;p_1,p_2$\} canonical pairs to \{$q_{+},q_{-};p_{+},p_{-}$\} canonical pairs at the classical level (where the phase-space variables are nothing more than c-numbers). As a result, we extract the 2D decoupled harmonic oscillator type Hamiltonian, defined only in terms of new phase-space variables, from the classical Hamiltonian in old phase-space variables, which is the classical counterpart of our quantum system Hamiltonian (\ref{comh}) at the moment $t_0$ : 
 
 \begin{equation}
 \mathcal{H}_{Cla} (t_0)=\frac{\omega_{+}^{2}}{2\alpha}q_{+}^{2}+\frac{\alpha}{2}p_{+}^2+\frac{\omega_{-}^{2}}{2\alpha}q_{-}^{2}+\frac{\alpha}{2}p_{-}^2.
 \label{cla}
 \end{equation}

  This is the classical equivalent of the transformed Hamiltonian  (\ref{comh}) under a unitary transformation : $\mathcal{V}(t_0)\mathcal{H}(t_0)\mathcal{V}^{\dagger}(t_0)$. Now, instead of only the time-independent instantaneous Hamiltonians $\mathcal{H}(t_0)$ with parameters frozen at fixed values, we might investigate time-evolution in the whole time-dependent system governed by $\mathcal{H}(t)$. But in that case, we would have needed to add a suitable connection one form \cite{jq} like term $i\hbar\mathcal{V}(t)\partial_{t}\mathcal{V}^{\dagger}(t) $ to the unitary transformed Hamiltonian, such that, as in (\ref{connection}), the total Hamiltonian  $\mathcal{H}_{total}(t)$ can drive the temporal evolution of the unitary transformed states ($\mathcal{V}(t)\left|\Psi(t)\right\rangle $). And as shown in the case of $\mathcal{W} (t)$ (\ref{W}) in Appendix-B, here too we can show that the geometrical phase simply reappears as part of the dynamical phase obtained through $\mathcal{H}_{total}(t)$, but will retain its geometrical nature. Of course, the classical counterpart of this $\mathcal{H}_{total}(t)$ can be obtained by simply adding a term  $\frac{\partial F}{\partial t}$ to the classical Hamiltonian (\ref{cla}), where $F$ is a suitable generating function \cite{sn}.
 Here, of course, this extra time-derivative term in $\mathcal{H}_{total}(t)$ need not bother us, because it suffices to work with the instantaneous Hamiltonian, whereby the external parameters $\alpha$, $\beta$, $\gamma$, and $\delta$ are kept frozen at their values $t=t_{0}$. Furthermore, in the classical case, each of these instantaneous time-independent Hamiltonians $\mathcal{H}_{Cla}(t_0)$ (\ref{cla}) gives rise to circuital motion in phase-space when considered as a separate system, allowing us to introduce corresponding action and its conjugate angle variables, as mentioned above.\\\\
 
Let $\{C(I,\boldsymbol{R})\}$ signify a family of continuous periodic trajectories in the phase space associated with the classical system Hamiltonians $\mathcal{H}_{Cla}(\boldsymbol{R})$ in this classical example \cite{mm} and suppose $\omega(I,\boldsymbol{R})$ is the angular velocity on $C(I,\boldsymbol{R})$, and each curve has a fixed origin for the action variables $I$  having a conjugate angle $v$.
Now, a point in phase space follows a trajectory, corresponding to the duration of adiabatic motion in parameter space, and only the shift in angular variable $v(t)$ evolves in time, and its value at time $t$ is given as,

 \begin{equation}
 v(t)=v(0)+\int_{0}^{t}\omega(I,\boldsymbol{R}(s))\mathrm{d}s+\Delta v_{I}^{\mathrm{H}}(t).
 \end{equation}
 
 This is shown as an integration along the curve $C(I,\boldsymbol{R}(t))$, which includes an additional geometrical contribution, the so-called Hannay's angle $\Delta v_{I}^{\mathrm{H}}(t)$, in addition to the normal dynamical component. We should expect two sets of action-angle coordinates $\{\{I_{i},v_{i}\}:i=1,2\}$ because our classical Hamiltonian $\mathcal {H}_{Cla}(\boldsymbol{R})$ has two degrees of freedom.\\
 
 Now let's consider a coherent states (wave packet)\cite{rj}
 of our two-dimensional harmonic oscillator (\ref{dh}), which are
 supposed to be the best approximations to a classical state and  produce a non-spreading wave packet. The coherent
 states analogous to \cite{lu} in this case are the tensor product
 of two independent Glauber coherent states
 , which are simultaneous (normalised) eigen states
 of the two mutually commuting annihilation operators: 
 
 \begin{equation}
 \begin{alignedat}{1}\left|z_{1},z_{2};\boldsymbol{R}\right\rangle = & \left|z_{1}(\boldsymbol{R})\right\rangle \otimes\left|z_{2}(\boldsymbol{R})\right\rangle \\
 \boldsymbol{a}_{+}\left|z_{1}(\boldsymbol{R})\right\rangle = & z_{1}\left|z_{1}(\boldsymbol{R})\right\rangle \\
 \boldsymbol{a}_{-}\left|z_{2}(\boldsymbol{R})\right\rangle = & z_{2}\left|z_{2}(\boldsymbol{R})\right\rangle \\
 \left|z_{1},z_{2};\boldsymbol{R}\right\rangle = & \mathrm{e}^{-\left(\left|z_{1}\right|^{2}+\left|z_{2}\right|^{2}\right)/2}\sum_{n_{1}=0}^{\infty}\sum_{n_{2}=0}^{\infty}\frac{z_{1}^{n_{1}}}{\sqrt{n_{1}!}}\frac{z_{2}^{n_{2}}}{\sqrt{n_{2}!}}\left|n_{1},n_{2};\boldsymbol{R}\right\rangle 
 \end{alignedat}
 \end{equation}
 
 In addition, $\hat{I}_{i}(\boldsymbol{R})=\hslash\hat{N}_{i}(\boldsymbol{R})$,
 where $\hat{N}_{i}(\boldsymbol{R})$ has been demonstrated in \cite{mm} to be an appropriate quantum operator version for classical action variable $I_{i}$, where $\hat{N}_{i}(\boldsymbol{R})$ is the $i$-th number operator, with $i\in\{+,-\}$ in our instance. Let $\hat{U}_{i}(\boldsymbol{R})$ be a unitary operator described by their action, $\hat{U}_{1}(\boldsymbol{R})\left|n_{1},n_{2};\boldsymbol{R}\right\rangle =\left|n_{1}-1,n_{2};\boldsymbol{R}\right\rangle $;
 $\hat{U}_{1}(\boldsymbol{R})\left|0,n_{2};\boldsymbol{R}\right\rangle =0$ and like-wise for $\hat{U}_{2}(\boldsymbol{R})$.\\
 
 They are basically the same as the well-known polar decompositions of Ladder operators such as $\boldsymbol{a}$ into the so-called number $\hat{N}$ and phase operators $\hat{\theta}$:

 \begin{equation}
 \boldsymbol{a}_{\pm}=\sqrt{\hat{N}_{\pm}}e^{i\hat{\theta}_{\pm}},~~[\hat{N}_{\pm},\hat{\theta}_{\pm}]=i\mathbb{I}
 \end{equation}

 The unitary operator $\hat{U}_{i}(\boldsymbol{R})$ can be thought of as the equivalent of $e^{-i\hat{\theta_{i}}}.$ The expectation values of these operators in the state $\left|z_{1},z_{2};\boldsymbol{R}\right\rangle $ can be easily computed to get, $I_{i}=\left\langle \hat{I}_{i}(\boldsymbol{R})\right\rangle =\left|z_{i}\right|^{2}\hslash$
 and $\left\langle \hat{U}_{i}(\boldsymbol{R})\right\rangle =e^{i\times arg(z_{i})}$,
 so that we can identify in the classical limit $z_{j}=\sqrt{\frac{I_{j}}{\hslash}}e^{-iv_{j}}$.\\

 And this natural connection between the ladder operators of the quantum system and the corresponding action and angle-like operators was the main motivation behind our unconventional approach to determine the geometric phases by solving the evolution-equations of $\boldsymbol{a}_{\pm}$ in Heisenberg picture, which also provides a natural background for semi-classical correspondence. In fact, the additional geometrical part of the phases acquired by $\boldsymbol{a}_{\pm}$ over a complete period of adiabatic-excursion, as found in (\ref{eq:a+sol},\ref{eq:a-sol}), is precisely the expression for Hannay angle shifts obtained due to the corresponding classical adiabatic evolution, as we identify below.
 As a result, we can compute Berry's quantal phases as well as Hannay's angle from (\ref{eq:a+sol},\ref{eq:a-sol}) in one go. Furthermore, if we had gone the usual route, we would have had to determine the exact energy eigenfunctions of the quantum dynamical system in order to obtain the necessary geometric phases, which is a difficult task for a generalised two-dimensional Harmonic oscillator like (\ref{comh}). As a result, while our overall procedure was not standard, it was more tailored to our desired goals.\\
 
 Returning to our original topic, we now take the wave packet below as the initial state, which best approximates the initial conditions of the equivalent classical adiabatic transport,
 
 \begin{equation}
 \left|z_{1},z_{2};\boldsymbol{R}(0)\right\rangle =\mathrm{e}^{-\left(\left|z_{1}\right|^{2}+\left|z_{2}\right|^{2}\right)/2}\sum_{n_{1}=0}^{\infty}\sum_{n_{2}=0}^{\infty}\frac{z_{1}^{n_{1}}}{\sqrt{n_{1}!}}\frac{z_{2}^{n_{2}}}{\sqrt{n_{2}!}}\left|n_{1},n_{2};\boldsymbol{R}(0)\right\rangle 
 \end{equation}
 and evolve it adiabatically over a complete period, to get, using (\ref{eq:adiaphase})
 
 \begin{equation}
 \begin{aligned}\mathcal{U}(0,T)\left|z_{1},z_{2};\boldsymbol{R}(0)\right\rangle = & \mathrm{e}^{-\left(\left|z_{1}\right|^{2}+\left|z_{2}\right|^{2}\right)/2}\sum_{n_{1}=0}^{\infty}\sum_{n_{2}=0}^{\infty}\frac{z_{1}^{n_{1}}}{\sqrt{n_{1}!}}\frac{z_{2}^{n_{2}}}{\sqrt{n_{2}!}}e^{-i\phi_{0,0}}\left|n_{1},n_{2};\boldsymbol{R}(T)\right\rangle \times\\
 & e^{-in_{1}(\Theta_{+,d}+\Phi_{G})}e^{-in_{2}(\Theta_{-,d}+\Phi_{G})}\\
 = & \mathrm{e}^{-\left(\left|z_{1}\right|^{2}+\left|z_{2}\right|^{2}\right)/2}\times e^{-i\phi_{0,0}}\sum_{n_{1}=0}^{\infty}\sum_{n_{2}=0}^{\infty}\frac{\left(z_{1}e^{-i(\Theta_{+,d}+\Phi_{G})}\right)^{n_{1}}}{\sqrt{n_{1}!}}\times\\
 & \frac{\left(z_{2}e^{-i(\Theta_{-,d}+\Phi_{G})}\right)^{n_{1}}}{\sqrt{n_{2}!}}\left|n_{1},n_{2};\boldsymbol{R}(T)\right\rangle \\
 = & e^{-i\phi_{0,0}}\left|z_{1}e^{-i(\Theta_{+,d}+\Phi_{G})},z_{2}e^{-i(\Theta_{-,d}+\Phi_{G})};\boldsymbol{R}(T)\right\rangle 
 \end{aligned}
 \label{eq:coherent evolution}
 \end{equation}

 As a result, the coherent state associated with the initial system Hamiltonian $\mathcal{H}(0)$ evolves to another coherent state $\left|z_{1}(T),z_{2}(T);\boldsymbol{R}(T)\right\rangle $ at time T, where the vectors are specified modulo overall phases, where
 $z_{1}(T)=z_{1}e^{-i(\Theta_{+,d}+\Phi_{G})}$ and $z_{2}(T)=z_{2}e^{-i(\Theta_{-,d}+\Phi_{G})}$.
  Thus, from the evolution of $z_{1}$ and $z_{2}$ in (\ref{eq:coherent evolution}),
 through a complete period of the adiabatic Hamiltonian, we can identify
 $\Theta_{\pm,d}=\int_{0}^{T}\omega_{\pm}(t')dt'$, where $\omega_{i}=\frac{1}{\hslash}\frac{\partial E_{n_{1},n_{2}}}{\partial n_{i}}$(Like
 $\frac{\partial\mathcal{H}_{Cla}}{\partial I_{i}}$), with the dynamical
 phases and $\Phi_{G}$ (from (\ref{eq:ladderphase},\ref{mute}))
 with the angular shift which was obtained classically by Hannay.\\\\
 
 The expectation values of the new set of time-dependent phase-space operators, namely the quadrature operators (\ref{quadr}), are given by \cite{rj},
 
 \begin{equation}
 \begin{alignedat}{1}\left\langle \hat{q}_{\pm}\right\rangle = & \sqrt{2\hslash/\omega_{\pm}}Re(z_{i})\\
 \left\langle \hat{p}_{\pm}\right\rangle = & \sqrt{2\hslash\omega_{\pm}}Im(z_{i})\\
 & \text{(i=1,2 respectively)}
 \end{alignedat}
 \label{scrodinger}
 \end{equation}
 
 The mean values of these phase space operators in the transported state are derived by utilizing the above-mentioned parametrization of $z_{1}$ and $z_{2}$ as follows:

 \begin{equation}
 \begin{alignedat}{1}\left\langle \hat{q}_{\pm}\right\rangle _{T}= & \sqrt{2I_{i}/\omega_{\pm}}Cos(v_{i}(0)+\Theta_{\pm,d}+\Phi_{G})\\
 \left\langle \hat{p}_{\pm}\right\rangle _{T}= & -\sqrt{2I_{i}\omega_{\pm}}Sin(v_{i}(0)+\Theta_{\pm,d}+\Phi_{G}),
 \label{qhan}
 \end{alignedat}
 \end{equation}
 demonstrating that the classically canonical phase space variables, i.e. the classical counterparts of quadrature operators  $\hat{q}_{\pm},\hat{p}_{\pm}$ (\ref{quadr}) , follow a classical trajectory. As a result, the geometric phase factor $\Phi_{G}$ enters the non-stationary coherent-state through all of its stationary components, namely the energy eigenstates and the classical correspondence limit, as follows:
 
  \begin{equation}
 \hbar \rightarrow 0, ~\lvert z_{i} \lvert \rightarrow \infty,~\sqrt{I_{i}} =\sqrt{\hbar} \lvert z_{i} \lvert \rightarrow~finite,
 \end{equation}
 the phase component of $ z_{i} $ determines the angle variable which is conjugate to the action $I_{i} $. As a result, the extra phase of $ z_{i} $, i.e. one above and beyond the dynamical phase, can be identified with Hannay's angle, which can be fully understood using classical considerations.\\\\

 \section{Discussion}
We have studied a model of a time-dependent parametric oscillator system, defined in a planar phase-space with non-commuting
coordinates and momenta, wherein the external parameters are slowly varying periodic function of time. Although Aharanov and Anandan \cite{ya} show that periodicity, not adiabaticity, is more important in the computation of the geometrical phase, we still find Berry's original adiabatic approach to be easier to implement. By considering a novel phase space transformation (Bopp shift) we reduced the problem on the effective commutative space. We have shown that in this process it is not mandatory to modify the Planck constant as is generally believed \cite{bao,Jz}.
We may introduce a dilation operator in the system Hamiltonian involving commutative phase space variable ($[q_i,q_j]=0=[p_i,p_j];[q_i,p_j]=i \hbar \delta_{ij}$) by this Bopp shift, which is crucial in creating this Berry's geometrical phase. The instantaneous energy eigenstates and spectrum in this model were computed analytically by exploiting this non-canonical phase-space transformation. We have also provided an unconventional but novel methodology to compute this additional geometrical phase shift over and above the dynamical phase initially in Heisenberg's picture and then connection with the conventional Berry phase in the Sch\"odinger picture is beautifully exhibited in this approach.  Finally, the classical counterpart of Berry's phase i.e the Hannay angle was also computed using two-mode oscillator coherent states. Furthermore, the emergent geometrical phase shift is shown to be dependent on both kind of non-commutative parameters ($\theta$ and $\eta$) or equivalently the classical deformation ($\tilde{\theta}$ and $\tilde {\eta}$) parameters, and it vanishes if any one of them vanishes. As a result, when considering circuital adiabatic excursion in the parameter space, we can conclude that the quantum phase space geometry induces a suitable geometry on the circuit $\Gamma $ in the effective parameter space of the system, which manifests in the emergence of the associated geometrical phase shift.\\

We'd also like to point out that the effective commutative Hamiltonian $\mathcal{H}(t)$ (\ref{comh}), obtained by using the representation (\ref{bpshift}), takes its value in $su(1,1)\oplus u(1)$ Lie algebra (\ref{gro}), and  the explicit breaking of time-reversal symmetry of the family of instantaneous Hamiltonians $\mathcal{H}(t)$'s relies on this time-dependent Lie algebra element, which is a requirement for obtaining Berry's phase.  As a result, the eigenspace of the instantaneous Hamiltonians $\mathcal{H}(t)$ can be viewed as a representation space for the group $SU(1,1)\otimes U(1)$, and geometrical phase appears naturally in this situation, as shown in \cite{BDR,zh}. Our result, clearly corroborates this general observation. For our purposes, the deformation parameters $\theta$ and $\eta$ can be treated as fundamental parameters in some appropriate energy scale, and the resulting geometrical phase then can also be considered fundamental as well.\\


Finally, for the case of more realistic models, a planar system of charged non-relativistic anyons with fractional spin (related to $\theta$) and an additional harmonic trap subjected to a normal uniform magnetic field $B$ (related to $\eta$) can be expected to exhibit Berry phase with quantum phase space non-commutativity \cite{VN,Balu,Ams}, wherein the mass and spring constant parameters are assumed to vary adiabatically as periodic functions of time.\\

\section { Appendix A}

As we have already pointed out that there exists another realization of the  phase-space non-commutative algebra
given in \cite{bao,Jz}, unlike the one which is used by us (\ref{bpshift}). 
And Berry phase too was investigated by using that realization in non-commutative phase-space, albeit
in a completely different phenomenological model involving gravitational
potential well \cite{AP}, but the Berry's phase were found to vanish, where a scaled equivalent version of the realization (\ref{bps1}) \cite{bao} given below, was used.  
What we would like to emphasize here is that, our realization (\ref{bps3}) of the one-parameter ($\xi$) family of non-commutative algebra (\ref{ncms}) given below is more general and
the realization (\ref{bps1}) arising in \cite{Jz, bao} are unitarily related to our proposed realization
(\ref{bps3}), only for a particular value of the scale factor $\xi=\xi_c$ (\ref{bd}) and hence at this critical value one can expect to get the
same physical results, but not for any other $\xi\neq\xi_{c}$. And for this case (\ref{bps1}) will not hold.
In contrast, our realization (\ref{bps3}) will persists to hold for any other values of $\xi$, i.e. for $\xi\neq\xi_c$ also.\\

To draw the equivalence between the above mentioned realizations (\ref{bps1}), (\ref{bps3}) holding only for $\xi=\xi_{c}$ (\ref{bd}), let's consider
the following structure of noncommutativity among the phase space
variables: 
\begin{equation}
\begin{array}{c}
[\hat{x}_{i},\hat{x}_{j}]=i\xi\theta\epsilon_{ij};[\hat{p}_{i},\hat{p}_{j}]=i\xi\eta\epsilon_{ij};[\hat{x}_{i},\hat{p}_{j}]=i\hslash\delta_{ij}\end{array};\theta\eta<0;\label{ncms}
\end{equation}
where $\theta$ and $\eta$ are non-commutative constant type parameters, respectively; $\epsilon_{ij}$ is an anti-symmetric tensor(constant), and $\xi$ is a scaling factor. The normal (commuting) coordinates $q_{i}$ and conjugate momenta $p_{i}$ are then introduced, which satisfy the Heisenberg algebra:  $[q_{i},q_{j}]=0=[p_{i},p_{j}];[q_{i},p_{j}]$ $=i\hslash\delta_{ij}$.
In contrast to their non-commutative counterparts ($\hat{x_{i}}$'s
and $\hat{p_{i}}$ 's), these phase-space variables $q_{i}$'s and $p_{i}$'s have no over-head hats for brevity, in the notations we have adopted.\\

\begin{doublespace}
	In \cite{Jz}, the consistent realization in terms of the  undeformed
	variables $(q_{i},p_{i})$ are given through the following linear mapping :
	
	\begin{equation}
	\begin{alignedat}{1}\hat{x}_{i}^{(1)} & =\sqrt{\xi}(q_{i}-\frac{\theta}{2\hslash}\epsilon_{ij}p_{j})\\
	\hat{p}_{i}^{(1)} & =\sqrt{\xi}(p_{i}+\frac{\eta}{2\hslash}\epsilon_{ij}q_{j}),
	\end{alignedat}
	\label{bps1}
	\end{equation}
	
\end{doublespace}
which holds only if 
\begin{equation}
\xi=\xi_{c}:=(1+\frac{\theta\eta}{4\hbar^{2}})^{-1};4\hbar^{2}+\theta\eta>0
\label{bd}
\end{equation}

But this representation of the deformed algebra (\ref{ncms}) is not unique as has been pointed out in
\cite{VN}. 
Indeed, here we provide below another possible representation (realization) of (\ref{ncms}) in terms
of another new linear transformation, as
\begin{equation}
\begin{alignedat}{1}\hat{x}_{i}^{(2)} &
=q_{i}-\frac{\xi\theta}{2\hslash}\epsilon_{ij}p_{j}+\frac{\xi\sqrt{-\theta\eta}}{2\hslash}\epsilon_{ij}q_{j}\\
\hat{p}_{i}^{(2)} &
=p_{i}+\frac{\xi\eta}{2\hslash}\epsilon_{ij}q_{j}+\frac{\xi\sqrt{-\theta\eta}}{2\hslash}\epsilon_{ij}p_{j},
\end{alignedat}
\label{bps3}
\end{equation}
We thus have a novel realization that is valid for any value of scale factor
$\xi$ which need not be fixed to the specific value given in (\ref{bd}). This is unlike
the one in (\ref{bps1}). Clearly, neither of the transformations (\ref{bps1}) or
(\ref{bps3}) represent canonical (unitary) transformations, as they can change the basic
commutation brackets. It is, however, quite natural that for the scale factor $\xi$, set to the value in (\ref{bd}), the realizations
should be unitary equivalent. We now find out this unitary transformation
explicitly, which map the equivalent commutative representation   (\ref{bps1}) to the  other one 
(\ref{bps3}).
To that end, let us consider the following unitary operator:

\begin{equation}
\textbf{U}=\exp[-i\frac{\sigma\textbf{D}}{\hbar}]~\exp[-i\frac{\beta_{0} \textbf{L}}{\hbar}]~\exp[-i\alpha_{2}\vec{p}^{2}]~\exp[-i\alpha_{1}\vec{q} ^{2}],
\label{unitary}
\end{equation}
\\
where $\textbf{D}=\vec{q}.\vec{p}+\vec{p}.\vec{q}$ and $\textbf{L}=\vec{q}\wedge\vec{p}$ 
are respectively the crucial dilatation term and angular momentum operators\footnote{In a normal commutative plane, the former represents a scalar operator, whereas the latter represents a pseudo scalar operator and provides appropriate transformations. It's also common knowledge that the three scalar generators ($\textbf{D},\vec{p}^{2},\vec{q}^{2}$)
	form a closed $SO(1,2)$ algebra \cite{g5},
	while $\textbf{L}$ commutes with all of them:$[\textbf{L},\vec{q}^{2}]=[\textbf{L},\vec{p}^{2}]=[\textbf{L},\textbf{D}]=0$. }, and relates these two representations as,

\begin{equation}
\hat{x}_{i}^{(2)}=\textbf{U}\hat{x}_{i}^{(1)}\textbf{U}^{\dagger},\hat{p}_{i}^{(2)}=\textbf{U}\hat{p}_{i}^{(1)}\textbf{U}^{\dagger}
\label{equival}
\end{equation}\\
It's worth noting that the parameters $\sigma$ and $\beta$ are dimensionless in this case, whereas the other parameters like $\alpha_ 1$ and $\alpha_ 2$ are dimensionful. Using Hadamard identity, we can now demonstrate that,

\begin{equation}
\begin{alignedat}{1}\hat{x}_{i}^{(2)} & =Aq_{i}-B\epsilon_{ij}p_{j}+C\epsilon_{ij}q_{j}+Dp_{i}\\
\hat{p}_{i}^{(2)} & =Ep_{i}+F\epsilon_{ij}q_{j}+G\epsilon_{ij}p_{j}-Hq_{i},
\end{alignedat}
\label{bps2l}
\end{equation}

where 

\begin{equation}
\begin{alignedat}{1}A= & \lambda\sqrt{\xi}[cos(\beta_{0})+\alpha_{1}\theta sin(\beta_{0})],\ \ \ \ \ B=\frac{\sqrt{\xi}}{\lambda}[(\frac{\theta}{2\hbar}-2\alpha_{1}\alpha_{2}\theta\hbar)cos(\beta_{0})+2\alpha_{2}\hbar sin(\beta_{0})]\\
C= & \sqrt{\xi}\lambda[sin(\beta_{0})-\alpha_{1}\theta cos(\beta_{0})],\ \ \ \ \ D=\frac{\sqrt{\xi}}{\lambda}[(\frac{\theta}{2\hbar}-2\alpha_{1}\alpha_{2}\hbar\theta)sin(\beta_{0})-2\alpha_{2}\hbar cos(\beta_{0})]\\
E= & \frac{\sqrt{\xi}}{\lambda}[(1-4\alpha_{1}\alpha_{2}\hbar^{2})cos(\beta_{0})+\eta\alpha_{2}sin(\beta_{0})],\ \ \ \ \ F=\lambda\sqrt{\xi}[\frac{\eta}{2\hbar}cos(\beta_{0})+2\alpha_{1}\hbar sin(\beta_{0})]\\
G= & \frac{\sqrt{\xi}}{\lambda}[(1-4\alpha_{1}\alpha_{2}\hbar^{2})sin(\beta_{0})-\eta\alpha_{2} cos(\beta_{0})],\ \ \ \ \ H=\lambda\sqrt{\xi}[\frac{\eta}{2\hbar}sin(\beta_{0})-2\alpha_{1}\hbar  cos(\beta_{0})].\\
\lambda= & \exp(-2\sigma).
\end{alignedat}\label{abcefg}
\end{equation}

We haven't assigned any specific values to these eight coefficients yet. By comparing (\ref{bps2l}) with (\ref{bps3}), $A$ - $H$ in (\ref{abcefg}) can be easily found, and has been displayed below in two segregated clusters:

\ 
\\

\begin{equation}
H=0,\ \ \ \ \ D=0,\ \ \ \ \ A=1,\ \ \ \ \ C=\xi\frac{\sqrt{-\theta\eta}}{2\hbar}\ \ \ \ \ 
\label{ebfgg}
\end{equation}
and,\\
\begin{equation}
B=\xi\frac{\theta}{2\hbar},\ \ \ \ \ E=1,\ \ \ \ \ F=\frac{\xi\eta}{2\hbar},\ \ \ \ \ G=\xi\frac{\sqrt{-\theta\eta}}{2\hbar}\ \ \ \ \ 
\label{evigpg}
\end{equation}

The reason behind for this segregation is that a simple inspection suggests that we can solve  $\alpha_{1}$, $\alpha_{2}$, and $\lambda$ in terms of the single parameter $\beta_{0}$ by making use of first three equations in (\ref{ebfgg}) as,

 \begin{eqnarray}
\begin{array}{rcl}
&  & \alpha_{1}=\frac{\eta}{4\hbar^{2}}tan(\beta_{0})\\
\\
&  & \alpha_{2}=\frac{\theta}{4\hbar^{2}}[\frac{tan(\beta_{0})}{1+\frac{\theta\eta}{4\hbar^{2}}tan^{2}(\beta_{0})}]\\
\\
&  & \lambda=[\sqrt{\xi}(cos(\beta_{0})+\alpha_{1}\theta sin(\beta_{0}))]^{-1}
\end{array}\label{alph}
\end{eqnarray}

and then this parameter $\beta_{0} $ , can be determined by making use of the fourth equation in (\ref{ebfgg}) to get the following quadratic  algebraic equation:

\begin{equation}
\frac{\xi}{4\hbar^{2}}(-\theta\eta)^{\frac{3}{2}}tan^{2}(\beta_{0})-(\frac{\theta\eta}{2\hbar}-2\hbar)tan(\beta_{0})-\xi\sqrt{-\theta\eta}=0,
\end{equation}

yielding the following two roots for $\beta_{0}$:

\begin{equation}
\beta_{1}= tan^{-1}( \sqrt{\frac{-\theta\eta}{4\hbar^{2}}})  \ \ \ \ \ \beta_{2}=-tan^{-1}((\frac{-\theta\eta}{4\hbar^{2}})^{\frac{3}{2}})
\label{beta1}
\end{equation}.

Now, after a lengthy but straightforward computation, it can be
 easily verified that only $\beta_{1}$ from (\ref{beta1}) along with $\alpha_{1}$, $\alpha_{2}$, and $\lambda$ from (\ref{alph}) when substituted to the above mentioned set of expressions of $B$, $E$,$F$ and $G$ in (\ref{abcefg}), they establish compatibility with the corresponding expression given in (\ref{evigpg}). This therefore provides a particular illustration of the unitary equivalence of two different realizations (\ref{bps1},\ref{bps3}) for specific choice of the value of $\xi=\xi_c$ in (\ref{bd}).  The realization (\ref{bps1}) will not hold for other values of $\xi$, in contrast to the representation
(\ref{bps3}) which persists to hold. In-fact, the realization 
(\ref{bps3}) is more general and in our present chapter and we are basically working
with the algebra (\ref{non-com}) and its realization (\ref{bpshift}), which are described by  the equations (\ref{bps3},\ref{ncms}) themselves with $\xi=1$.\\

\section{ Appendix B}
Following the Wei-Norman method \cite{q} , here we  discuss the the Lie-algebraic structure \cite{r} of our system Hamiltonian (\ref{comh}).
 To see this algebraic structure let us introduce the generators: 
\begin{equation}
\boldsymbol{K_{+}}=\frac{iq^{2}_{i}}{2};\boldsymbol{K_{-}}=\frac{ip^{2}_{i}}{2};
\boldsymbol{K_{0}}=\frac{i(p_{i}q_{i}+q_{i}p_{i})}{4};~~~
\boldsymbol{L}=\epsilon_{ij}q_{i}p_{j}
\label{alg}
\end{equation}

It can be shown quite easily that the   $\boldsymbol{K_{\pm}}$,
$\boldsymbol{K_{0}}$ and $\boldsymbol{L}$, satisfy the $su(1,1)\oplus u(1)$ Lie
algebra \cite{Gi}:
\begin{equation}
[\boldsymbol{K_{0}},\boldsymbol{K_{\pm}}]=\pm \hbar \boldsymbol{K_{\pm}};~~
[\boldsymbol{K_{+}},\boldsymbol{K_{-}}]=-2\hbar\boldsymbol{K_{0}};~~[\boldsymbol{K_{\pm}},\boldsymbol{L}]=[\boldsymbol{K_{0}},\boldsymbol{L}]=0
\label{algcom}
\end{equation}

where $\boldsymbol{L}$ is the $u(1)$ group generator which commutes with all $su(1,1)$
generators. The product of exponentials of all generators, like in (\ref{unitary}),
 will generate all the elements of the Lie group $SU(1,1)\otimes U(1)$.\\

It will be useful to introduce the dimensionless group generators
$\boldsymbol{T}_{1} , \boldsymbol{T}_{2}$ $ \boldsymbol{T}_{0}$ defined through
$\boldsymbol{K}_{\pm}$  and $\boldsymbol{K}_{0}$ as,  

\begin{equation}
\boldsymbol{K}_{+}=\theta (\boldsymbol{T}_{1}+i \boldsymbol{T}_{2});
\boldsymbol{K}_{-}=-\eta (\boldsymbol{T}_{1}- i\boldsymbol
{T}_{2});\boldsymbol{T}_{0}=\frac{\boldsymbol{K}_{0}}{\sqrt{-\theta\eta}}; \theta\eta<0
\label{dimalg}
\end{equation}

where $ \theta$ and $\eta$ are the dimension-full non-commutative parameters, which help us to assign a consistent dimension of all the
SU(1,1) generators in (\ref{alg}) and  in terms of the these dimensionless basis
$\boldsymbol{T}_{\mu} $ ( where $\mu=0,1,2 $), the above (\ref{algcom})
Lie algebra take a more suggestive form as,

\begin{equation}
[\boldsymbol{T_{0}},\boldsymbol{T_{i}}]=i \tilde{\hbar}\epsilon_{ij} 
\boldsymbol{T_{j}};~~~~~ [\boldsymbol{T_{i}},\boldsymbol{T_{j}}]=-i\tilde{\hbar}
\epsilon_{ij}\boldsymbol{T_{0}}~~~~(i,j=1,2),
\end{equation}
where $\tilde{\hbar}= \frac{\hbar}{\sqrt{-\theta\eta}}$ is the  modified dimentionless 
Planck's constant. It may be noted at this stage that $\boldsymbol{T_{0}}$ and
$\boldsymbol{T_{1}}$ are anti-hermitian like $\boldsymbol{K_{\pm}}$ and
$\boldsymbol{K_{0}}$, but $\boldsymbol{T_{2}}$ is still hermitian.\\\\
A faithful 2D finite representation
\cite{gd,amo}  "$\Pi$" of this SU(1,1) is
furnished by the hermitian Pauli matrices $\vec{\boldsymbol{\sigma}}$'s 
as\footnote{Observe at this stage that in this finite dimensional representation, it is rather $\Pi({\boldsymbol{T}_{0}})$ which is only hermitian and $\Pi({\boldsymbol{T}_{i}})$'s are anti-hermitian. This is a typical and peculiar feature of finite-dimensional representations of the Lie-algebra corresponding to a non-compact unitary groups like SU(1,1).}

\begin{equation}
\Pi({\boldsymbol{T}_{0}})=\frac{\tilde{\hbar}}{2}
\sigma_{3},\Pi({\boldsymbol{T}_{i}})=- \frac{i\tilde{\hbar}}{2}\sigma_{i}
\end{equation}
On using this representation, one can easily verify that, any trace-less su(1,1) algebra
element $ A^{\mu}  \Pi(\boldsymbol{T}_{\mu})$, with parameters $A^\mu$ and
$\mu=0,1,2$, takes the following form
\begin{equation}
A^{\mu}  \Pi(\boldsymbol{T}_{\mu})=\frac{\tilde{\hbar}}{2}\begin{pmatrix}
A^{0}& -A \\
A^{*} & -A^{0} 
\end{pmatrix} ;~~~~A=A^{1}+iA^{2}
\end{equation}

If this element is now subjected to a adjoint action by $\mathcal {U}
\in SU(1,1)$ as
\begin{equation}
A^{\mu}  \Pi(\boldsymbol{T}_{\mu})\rightarrow \mathcal {U}A^{\mu} 
\Pi(\boldsymbol{T}_{\mu})\mathcal {U}^{\dagger}:=
B^{\mu}\Pi(\boldsymbol{T}_{\mu}), 
\label{Amu}
\end{equation}\\
(where we could easily replace $\mathcal {U}\rightarrow \mathcal{V}\in
SU(1,1)\otimes U(1) $, as $[\boldsymbol{L},\boldsymbol{K_{\mu}}]=0~~ \forall
~\mu$.) the resulting element in (\ref{Amu}) will again be another $su(1,1) $ element with some new set of parameters  $B^{\mu}$, where the trace-less behavior will be maintained along with the determinant.
Particularly, we note that the preservation of determinant indicates that we must have the following identity :
\begin{equation}
(A^{1})^{2}+(A^{2})^{2}-(A^{0})^{2}=(B^{1})^{2}+(B^{2})^{2}-(B^{0})^{2}.
\label{sop}
\end{equation}

We immediately observe that here all the components of $A^{\mu}$ organize themselves to form a Lorentz 3-vector transforming under $SO(2,1)$ Lorentz
transformation in (2+1)D, where  $A^{0}$'s and $A^{i}$ may be thought of representing temporal and spatial components respectively. This connection of the Lorentz group $SO(2,1)$ in $3$ space-time dimensions with its covering group (double cover) $SU(1,1)$ or $SL(2,\mathbb{R})$ is  documented within the literature: $SO(2,1)=SU(1,1)/\mathbb{Z}_2=SL(2,\mathbb{R})/\mathbb{Z}_2$; all of them are locally isomorphic, but not globally.\\

Now, we can clearly write our system Hamiltonian (\ref{comh}) in terms of the linear combination of these $SU(1,1)\otimes U(1)$ group generators (\ref{alg}) as,

\begin{equation}
\mathcal{H}(t)=-2i[\alpha(t)\boldsymbol{K_{-}} +\beta(t)
\boldsymbol{K_{+}}+2\delta(t)\boldsymbol{K_{0}}]-\gamma(t)\boldsymbol{L}=\mathcal{H}_{gho}(t)-\gamma(t)\boldsymbol{L}
\label{gro}
\end{equation}

Of course, a part of Hamiltonian $H_{gho}(t)$ in (\ref{hghoi})
is an $su(1,1)$ Lie algebra element. Re-writing this in 		    	terms of generators 
$\boldsymbol{T}_{\mu}:=(\boldsymbol{T}_{0},\boldsymbol{T}_{1},\boldsymbol{T}_{2})$
introduced in (\ref{dimalg}) we get

\begin{equation}
\mathcal{H}_{gho}(t)=-2i A^{\mu}(t)\boldsymbol{T}_{\mu}
\label{ha},
\end{equation}\\

where

\begin{equation}
A^{\mu}(t)=\begin{pmatrix}
A^{1}(t) \\
A^{2}(t) \\

A^{0}(t) \\
\end{pmatrix}=\begin{pmatrix}
(-\eta\alpha(t)+\theta\beta(t)) \\
i(\eta\alpha(t)+\theta\beta(t)) \\
(2\sqrt{-\theta\eta}\delta(t)) \\
\end{pmatrix}
\label{Atr}
\end{equation}
Note that $A^2(t)$  occurs here as a purely imaginary component, which preserves the self-adjoint property of $\mathcal{H}_{gho}(t)$ (\ref{ha}). At this stage, this 3- Lorentz vector $A^{\mu}(t)$ can be scaled appropriately by $\tilde{\hbar}$ as
\begin{equation}
A^{\mu}(t)\rightarrow \tilde{A}^{\mu}(t):=\tilde{\hbar}A^{\mu}(t)
\end{equation}

Now the invariance of SO(2,1) norm of the vector $\tilde{A}^{\mu}(t)$ readily implies
\begin{equation}
(\tilde{A}^{1}(t))^{2}+(\tilde{A}^{2}(t))^{2}-(\tilde{A}^{0}(t))^{2}=4\hbar^{2}[\alpha(t)\beta(t)-\delta^{2}(t)]=\hbar^{2}
\omega^{2}(t)>0,
\label{su}
\end{equation}

where $\omega(t)>0$ in (\ref{ttt})  is the frequency of the $\mathcal{H}_{gho}(t)$ and is invariant under the local (instantaneous) $SO(2,1)$ Lorentz transformation.\\

Furthermore, $2\delta(t)\propto A^0(t)$ herein (\ref{Atr},\ref{su})   can be  identified with the  temporal component of the space-like 3-vector
$A_{\mu}$. Consequently, imagine the tip of the 3-vector $A^{\mu}(t_{0})$ (at any instant
$t=t_{0}$) will lie on a 2D-hyperboloid
whose tangent plane is orthogonal to $A^{\mu}$ and as time evolves a closed curve $\Gamma$ in parameter space is traced out by the tip of the vector $A^{\mu}$ and the surface can be taken to be a member of one-parameter family of such hyperboloids.


Further, at any particular instant
$t=t_{0}$, the space-like nature of 3-vector $A^{\mu}$ implies that under a suitable local(time-dependent) SO(2,1) transformation we can eliminate $\delta(t_0)$ in this particular Lorentz frame.\\

 On the other hand, to do this, let us transform the Hamiltonian 
$\mathcal{H}_{gho}(t)$ (\ref{gro}), (\ref{ha}) under a unitary transformation, albeit time dependent, which belonging to the non-compact covering group $\mathcal{W}(t)\in SU(1,1)$ in the fashion of
(\ref{Amu}). Of course, here, one can also  consider the more bigger group $SU(1,1)\otimes U(1)$ , but the trivial U(1) element is quite insignificant here and hence it is optional in nature. This has to be contrasted with (\ref{funda}) in section-3.4, where we need to select a particular U(1) element other than identity element.\\

At this point, we should emphasize that it is no more difficult to construct such a unitary operator 
$\mathcal{W}(t)$. To illustrate the point, we consider SO(2,1)
transformation $\Lambda (t_{0})$ transforming the triplet at instantaneous limit:
\begin{equation}
(\alpha(t_{0}),\beta(t_{0}),\delta(t_{0}))\rightarrow(\alpha^{'}(t_{0}),\beta^{'}(t_{0}),\delta^{'}(t_{0})):=(\alpha(t_{0}),\beta^{'}(t_{0}),0)
\label{cotra}
\end{equation}

in such a way that the coefficient of the dilatation term vanishes.
 Using
(\ref{su}), we can easily obtain
\begin{equation}
\beta^{'}(t_{0})=\frac{\alpha(t_{0})\beta(t_{0})-\delta^{2}(t_{0})}{\alpha^{'}(t_{0})}
\end{equation}

It is easily verified that a corresponding unitary
transformation $\mathcal{W}(t)$ at an arbitrary time $t$ can be taken as,
\begin{equation}
\mathcal{W}(t)=\exp[{\frac{i}{\hbar}\frac{\delta(t)}{2\alpha(t)}\vec{q}^{2}}]
\label{W}
\end{equation}
 Accordingly the instantaneous total Hamiltonian $ \mathcal{H}(t)$ would suffer the following unitary transformation 
\begin{equation}
\mathcal{H}(t)\rightarrow\mathcal{W}(t)\mathcal{H}(t)\mathcal{W}^{\dagger}(t)=\alpha(t)p^{2}_{i}+(\frac{\alpha(t)\beta(t)-\delta^{2}(t)}{\alpha(t)})
q^{2}_{i}-\gamma(t)\epsilon_{ij}q_{i}p_{j}
\label{htra}
\end{equation}
 However, recognizing, from the time dependent Sch\"odinger equation,
 that (\ref{htra}) should not be identified as the time evolution  generator of the transformed  state
($\mathcal{W}(t)\left|\Psi(t)\right\rangle $) as $\mathcal{W}(t)$
explicitly time dependent. Indeed, it is easy to see that the time
evolution of the transformed state ($\mathcal{W}(t)\left|\Psi(t)\right\rangle $) is governed by  the modified Hamiltonian
$\tilde{\mathcal{H}}(t)$, obtained through the following transformation 

\begin{equation}
\mathcal{H}(t)\rightarrow\tilde{\mathcal{H}}(t)=\mathcal{W}(t)\mathcal{H}(t)\mathcal{W}^{\dagger}(t)-i\hbar\mathcal{W}(t)\frac{d}{dt}\mathcal{W}^{\dagger}(t)
\label{connection}
\end{equation}
so that $i\hslash \partial_{t}(W(t)\left|\psi(t)\right \rangle)=\tilde{\mathcal{H}}(t)(W(t)\left|\psi(t)\right \rangle)$ satisfied. It has to be contrasted with the previous case involving global (time-independent) unitary transformation $\boldsymbol{U}$ (\ref{unitary}) (See discussion below (\ref{SUP})) connecting two Hamiltonians  $ \mathcal{H}^{(1)}(t)$ and $\mathcal{H}^{(2)}(t)$ (\ref{funda}) respectively.\\

Now, expanding this $\tilde{\mathcal{H}}(t)$ we obtain

\begin{equation}
\tilde{\mathcal{H}}(t)=\alpha(t)p^{2}_{i}+(\frac{\alpha(t)\beta(t)-\delta^{2}(t)-\frac{\alpha}{2}\frac{d}{dt}(\frac{\delta(t)}{\alpha(t)})}{\alpha(t)})q^{2}_{i}-\gamma(t)\boldsymbol{L}=\mathcal{H}_{sho}(t)-\gamma(t)\boldsymbol{L}
\label{d}
\end{equation}

This is similar to a standard bi-harmonic oscillator Hamiltonian with the additional Zeeman-like interaction $\gamma(t)\boldsymbol{L}$. We can also see that $\gamma(t)\boldsymbol{L}$, $\mathcal{H}_{sho}(t)$ and $\tilde{\mathcal{H}}(t)$ commute pair-wise between each other at different times. As a result, they have simultaneous eigenstates that are instantaneous.\\

To find out the eigenstates of the system Hamiltonian (\ref{d}) one may introduce the annihilation operator
 \begin{equation}
 \tilde{\boldsymbol{a}}_{j}=\left(\frac{\tilde{\beta}}{4\alpha\hbar^{2}}\right)^{\frac{1}{4}} \left[
 q_{j}+i\sqrt{\frac{\alpha}{\tilde{\beta}}}p_{j}\right];j=1,2
 \end{equation} 
 with
 $\tilde{\beta}=(\frac{\alpha(t)\beta(t)-\delta^{2}(t)-\frac{\alpha}{2}\frac{d}{dt}(\frac{\delta(t)}{\alpha(t)})}{\alpha(t)})$,
 satisfying the commutation relation
 $[\hat{\tilde{\boldsymbol{a}}}_{j},\hat{\tilde{\boldsymbol{a}}}_{k}^{\dagger}]=\delta_{jk}$

Accordingly, the system Hamiltonian (\ref{d}) may be written in the form 
\begin{equation}
\tilde{\mathcal{H}}(t)=\hbar\tilde{\omega}(t)(\tilde{\boldsymbol{a}}^{\dagger}_{j}\tilde{\boldsymbol{a}}_{j}+1)+i\hbar\gamma(t)\epsilon_{jk}\tilde{\boldsymbol{a}}^{\dagger}_{j}\tilde{\boldsymbol{a}}_{k}~~~~(j,k)\in \{1,2\}
\label{a}
\end{equation}
where
\begin{equation}
\tilde{\omega}(t)=2\sqrt{(\alpha(t)\beta(t)-\delta^{2}(t))-\frac{\alpha(t)}{2}\frac{d}{dt}(\frac{\delta(t)}{\alpha(t)})}.
\label{arxiv}
\end{equation}
Again it is convenient to introduce the operators $\boldsymbol{a}_{\pm}$
through a time
independent canonical transformation (\ref{eq:J2 to J3}), in terms of make the Hamiltonian takes a much simpler form (viz. like number operators):
\begin{equation}
\tilde{\mathcal{H}}=\hbar\tilde{\omega}(\tilde{\boldsymbol{a}}^{\dagger}_{j}\tilde{\boldsymbol{a}}_{j}+1)-\hbar\gamma(t)(\tilde{\boldsymbol{a}}^{\dagger}_{+}\tilde{\boldsymbol{a}}_{+}-\tilde{\boldsymbol{a}}^{\dagger}_{-}\tilde{\boldsymbol{a}}_{-}); ~~~~~~~~~j\in \{+,-\}.
\end{equation}

The usual non-degenerate basis states (instantaneous) considered are

\begin{equation}
\begin{aligned}
&~~~~~~~~~~~~\left|n_{+},n_{-};(t)\right\rangle
=\frac{\left(\tilde{\boldsymbol{a}}^{\dagger}_{+}\right)^{n_{+}}\left(\tilde{\boldsymbol{a}}^{\dagger}_{-}\right)^{n_{-}}}{\sqrt{n_{+}!}\sqrt{n_{-}!}}\left|0,0;(t)\right\rangle;~~\tilde{\boldsymbol{a}}_{\boldsymbol{\pm}}(t)\left|0,0;(t)\right\rangle =0.\\
\label{sp}
\end{aligned}
\end{equation}
Note that here we have inserted the time parameter $t$ within the parenthesis as $(t)$ in order to discriminate these set of eigenstates from (\ref{ene}) which are clearly not the same; they are build upon different instantaneous vacuum states.\\

Correspondingly the instantaneous time dependent eigenvalues are given by,
 
\begin{equation}
\begin{aligned}
E_{n_{+},n_{-}}(t) &
=\hbar\tilde{\omega} (n_{+}+n_{-}+1)-\hbar\gamma(t)(n_{+}-n_{-})\\
&
\simeq\hbar(n_{+}+n_{-}+1)\omega(t)\left[1-\frac{\alpha(t)}{\omega(t)^{2}}\frac{d}{dt}(\frac{\delta(t)}{\alpha(t)})\right]-\hbar\gamma(t)(n_{+}-n_{-})
\end{aligned}
\label{c}
\end{equation}
where $\omega=2\sqrt{\alpha\beta-\delta^{2}}$ . In order to find the geometrical phases it suffices to work in the
first order of adiabaticity.  This is tantamount to neglecting the terms beyond the first order time derivatives of
the slowly varying parameters in (\ref{arxiv},\ref{c}).\\

Now, since $\tilde{\mathcal{H}}(t)$ and $\mathcal{H}(t)_{sho}$ in (\ref{d}) commutes with each other, they have same eigenspaces.
Consequently, we can re-express an eigenstate (\ref{sp}) of $\tilde{\mathcal{H}}(t)$, as
a linear superposition of eigenstates of $\mathcal{H}(t)_{sho}$:
\begin{equation}
\left|n_{+},n_{-};(t)\right\rangle=\sum_{n_1+n_2=n_{+}+n_{-}}
C_{n_1,n_2}^{n_{+},n_{-}} \left|n_1,n_2;(t)\right\rangle_{sho}^{2D}
\end{equation}
where we have defined the eigenstates of $\mathcal{H}(t)_{sho}$ as
\begin{equation}
\left|n_1,n_2;(t)\right\rangle_{sho}^{2D}=\frac{\left(\tilde{\boldsymbol{a}}^{\dagger}_{1}\right)^{n_{1}}\left(\tilde{\boldsymbol{a}}^{\dagger}_{2}\right)^{n_{2}}}{\sqrt{n_{1}!}\sqrt{n_{2}!}}\left|0,0;(t)\right\rangle
\end{equation}

whence $n_1+n_2=n_{+}+n_{-}$. This constraint assures that the eigenstates of $\mathcal{H}(t)_{sho}$ \cite{bhn} come from one eigenspace. Because the annihilation operators diagonalizing $\tilde{\mathcal{H}}(t)$ are derivable from the ladder operators of $\mathcal{H}(t)_{sho}$, using time-independent invertible linear transformation (\ref{eq:J2 to J3}, $C_{n_1,n_2}^{n_{+},n_{-}}$'s are explicitly time independent. This also assures that the identical vacuum state $\left |0,0;(t)\right\rangle$ gets annihilated by both sets of annihilation operators $\{\tilde{\boldsymbol{a}}_{1},\tilde{\boldsymbol{a}}_{2}\}$ or ${\tilde{\boldsymbol{a}}_{\pm}}$  at time $t$.\\
Because we're only interested in the region where the circuital adiabatic theorem holds, the Berry's phase for an eigenstate  $\left|n_{+},n_{-}\right\rangle$, if it exists, is:

\begin{equation}
\begin{aligned}
\Phi^{(G)}_{(n_{+},n_{-})} &=-i \int d t\left\langle n_{+},n_{-};(t)\left|\frac{d}{d
	t}\right| n_{+},n_{-};(t)\right\rangle\\
&=-i \int d t \sum_{\substack{n_1+n_2=m_1+m_2 \\
		=n_{+}+n_{-}
	}}C_{m_1,m_2}^{n_{+},n_{-}\star}C_{n_1,n_2}^{n_{+}n_{-}}~
	{^{2D}_{sho}\left\langle m_1,m_2(t)\right|}\frac{d}{d t}\left|
	n_1,n_2;(t)\right\rangle^{2D}_{sho}
	\end{aligned}
	\end{equation}
	Hence sandwiching a pair of tuples $(m_1,m_2)$ and $(n_1,n_2)$ term inside the
	sum, can be performed in two possible ways: either (i) $m_1=n_1, m_2=n_2$ or (ii) $m_1\neq n_1,
	m_2\neq n_2$.
	Let us focus our attention to the case of second possibility (ii) first as,
	\begin{equation}
	\begin{aligned}
	{^{2D}_{sho}\left\langle m_1,m_2;(t)\right|}\frac{d}{d t}\left|
	n_1,n_2;(t)\right\rangle^{2D}_{sho}
	=\left({^{1D}_{sho}\left\langle m_1(t)\right|}\frac{d}{d t}\left|
	n_1(t)\right\rangle^{1D}_{sho}\right) \times
	\left({^{1D}_{sho}}\left\langle m_2(t)|n_2(t)\right\rangle_{sho}^{1D}\right)\\
	+\left({^{1D}_{sho}\left\langle m_2(t)\right|}\frac{d}{d t}\left|
	n_2(t)\right\rangle^{1D}_{sho}\right) \times
	\left({^{1D}_{sho}}\left\langle m_1(t)|n_1(t)\right\rangle_{sho}^{1D}\right) = 0
	\end{aligned}
	\end{equation}
	On the other hand, for the case of first possibility (i) $(m_1,m_2)=(n_1,n_2)$, also we get,
	\begin{equation}
	\begin{aligned}
	\int dt \left({^{2D}_{sho}\left\langle n_1,n_2;(t)\right|}\frac{d}{d t}\left|
	n_1,n_2;(t)\right\rangle^{2D}_{sho} \right) &=
	\int dt \left({^{1D}_{sho}\left\langle n_1(t)\right|}\frac{d}{d t}\left|
	n_1(t)\right\rangle^{1D}_{sho}\right.\\
	&\left. +{^{1D}_{sho}\left\langle n_2(t)\right|}\frac{d}{d
		t}\left| n_2(t)\right\rangle^{1D}_{sho}\right)=0.
	\end{aligned}
	\end{equation}
	This shows the geometrical phase of a pair of independent 1D harmonic oscillators don't exhibit any Berry's phase:  $\Phi^{(G)}_n=0$, which in turn, apparently implies that the total system
	hamiltonian $\tilde{\mathcal{H}}(t)$ does not produce any Berry's phase by itself, apparently.\\
	
	However, the total dynamical phase factor acquired by state vector
	$\left|n_{+},n_{-};(t)\right\rangle$ after a complete cycle $\Gamma$ of time period $T$ by the transformed Hamitonian $\tilde{\mathcal{H}}(t)$, is obtained by using (\ref{c}), to get:
	\begin{equation}
	\begin{aligned}
	\Phi_{n_{+},n{-}}(T) &=\int_{0}^{T} dt \frac{ E_{n_{+},n_{-}}(t)}{\hbar}\\
	&= \int_{0}^{T} dt
	\left[(n_{+}+\frac{1}{2})(\omega+\gamma)+(n_{-}+\frac{1}{2})(\omega-\gamma)-(n_{+}+n_{-}+1)\frac{\alpha}{\omega}\frac{d}{dt}\left(\frac{\delta}{\alpha}\right)\right],
	\end{aligned}
	\end{equation}

	%
	%
	%
	

	Although the last term is now part of the dynamical phase, it retains its geometric nature because it represents the line integral of a real valued one-form $\boldsymbol{A}$ (\ref{phg}) along the closed circuit $\Gamma$ in parameter space as $\int_{\Gamma}\boldsymbol{A}$, and thus the additional phase is a functional of $\Gamma$ and matches exactly with Berry's phase (\ref{fp}). When analysing the whole adiabatic phase as a whole, this illustration shows how the key dilatation term which is responsible for Berry's phase can  apparently be gotten rid of by a time-dependent unitary (canonical) transformation, it, nevertheless re-emerges in disguise inside the dynamical phase part.

\chapter{Quantum theory on (1+1) dimensional quantum space-time}\label{galvec}

So far, we have dealt with planar systems having noncommutativity between spatial coordinates and also between
momentum components, wherein we have considered the time $t$ to be a simple 
commutative parameter. But, as we have already indicated previously that any attempt
to reconcile the two fundamental pillars of modern physics: quantum mechanics and Einsteins general relativity, should incorporate the impossibility 
of localization of a space-time event in the vicinity of
Planck length scale ($l_{P}=\sqrt{\frac{\hbar G}{c^{3}}}$) within the formalism itself. In other words, the modeling of space-time by a smooth differential
manifold may break down at this scale-the scale of quantum gravity \cite{GT}. This suggests that both space and time coordinates 
together have to become non-commutative operators in a more fundamental and consistent formulation of quantum theory of gravity. And the by-product of such a formalism was expected to
provide a natural way to tame the divergences of quantum field theories. On the other hand, the nature of
time itself is expected to play an important role in the formulation of  quantum theories of gravitational
field for a diffeomorphism (or time re-parametrization) invariant system. In fact, this
has been studied for zero spatial dimensions case in \cite{grqm}.\\

In-fact, the nature of time and its eventual status, at a more fundamental level i.e. at the level of quantum mechanics itself,remains quite unclear even today, and it is an open question whether our concepts of space-time are fully compatible jointly with both quantum mechanics and special relativity itself. This situation is obvious to any practitioners of quantum theory, who must have observed the asymmetrical role played by space and time coordinates, in the sense that time is regarded as a $c$-number evolution parameter and not elevated to the level of operators, unlike the spatial coordinates. In fact, it was argued long back by Pauli \cite{Wpauli} that if time `$t$' is also elevated to the level of quantum operators then the energy spectrum will not be bounded from below any more.In this context, we can mention about some the remarkable features of quantum mechanics with operator-valued time coordinate have been discussed in great detail in \cite{jmo,gm,rk,jbo,ore}.We can also point out to another important work was due to T. D. Lee \cite{TD} who have shown that time can be considered as a dynamical variable to formulate a path-integral approach for non-relativistic field theories.\\


In this chapter, we devoted ourselves to formulate consistent one-dimensional non-relativistic quantum mechanics, where both space and time coordinates have been promoted to the level of operator-valued coordinates. And there is some valid motivation for pursuing such an investigation. First of all, the question of this quantum nature of space-time is interesting in its
own right, as it would be interesting to see how quantum mechanics and space-time
are intimately tied up, although the dynamical nature of space-time may become a sort of `liability', as far as the interpretation \cite{ci,k.v,ea} of canonical quantum gravity is concerned. In this context, one may recall the work of Unruh and
Wald \cite{Wun} , who have  proposed a formulation of canonical quantum gravity using Sch\"odinger-like quantum mechanical equation in which time coordinate (t) appears as a
non-dynamical parameter. On the other hand, considerations of quantum gravity and
black hole physics strongly suggest that in the vicinity of Planck scale, space-time
coordinates become quite fuzzy and non-commutative in nature. The geometry and dynamics of such a space-time can perhaps be analyzed by
Connes non-commutative geometry and his famous spectral action principle, paving the way towards a consistent theory of quantum gravity. This is of course a long and ambitious project and we have no intension to study it here. Rather, our aim remains quite modest here, in the sense that we would like to simply 
understand the basic quantum mechanics with built-in time-space noncommutativity. This will simply serve as a toy
model,  where it would be worthwhile to study the basic Sch\"odinger quantum mechanics \cite{Bhg}
for a canonical (constant) non-commutative Moyal type space-time (($[\hat{x}^{\mu},\hat{x}^{\nu}]=i\theta^{\mu\nu}$) ). There are a few interesting papers in the literature, dealing with some phenomenological \cite{mli,NCt} investigations, which reveal
various interesting points of this problem. Very recently, some of us have analyzed the
dynamics of a non-relativistic, as well as a relativistic dynamical system in Lie algebraic
type ($\kappa$ deformed) space-time \cite{spv,spvi}. Moreover, it is important to mention that the
case of relativistic quantum space-time, say in $3+1$ dimensional space-time, with constant matrix-valued non-commutative parameter ($\theta$) 
violates Lorentz invariance, although it can, however, be restored in some sense using twisted Hopf-algebra \cite{mcp}. Furthermore, it was claimed in certain string theoretical analysis [\cite{ms}  that the noncommutativity of space-time coordinates may give rise violation of both causality \cite{nss,agl} and unitarity \cite{gom}. Subsequently,  Doplicher \cite{Db} and his collaborators have shown quite rigorously that
it is indeed possible to construct non-commutative field theories which are ultraviolet finite to
all orders, and unitarity is preserved. They pointed out that there were certain conceptual shortcoming in the previous papers claiming such violations of unitarity. 
 They, in particular, emphasized that the evolution parameter should never  be identified
with the eigenvalues of the operator valued time coordinate ($\hat{t}$). In the backdrop of all these studies, we consider the simplest $(1+1)$ D Moyal space-time and try to develop quantum mechanics on it using the Hilbert Schmidt (HS) operators. Although quantum mechanics in $1+1$ D Moyal space-time was also investigated in \cite{apG}, our analysis is will employ a novel approach involving the powerful HS operators, which, as we have seen already in chapter 2 that it provides an operatorial framework to describe quantum mechanics on the quantum (Moyal) plane with time being an ordinary $c$ number.\\ 

 The chapter is organized as follows: In section 4.1, we start from a time re-parametrization invariant toy model of a
non-relativistic system which however, does not have a built-in and manifest with time re-parametrization symmetry. But, this symmetry can be introduced by
simply enlarging the configuration space, where we treat both space and
time as configuration space variables that evolve with respect to some new evolution parameter ($\tau$). We then show how to obtain the Sch\"odinger equation in quantum mechanics, where time behaves as the evolution parameter. This was basically
inspired by a previous work by Deriglazov \cite{Adez}. In section 4.2, we introduce Moyal
like non-commutative quantum space-time for the time re-parametrization invariant
system, where we adopt the Hilbert-Schmidt operator formulation to study quantum
mechanics on this space-time. Then, we provide a coherent state-based approach to
extract an equivalent commutative Sch\"odinger equation in section 4.3. Section 4.4
has a detailed discussion on the time evolution of the Gaussian wave packet for a free
particle and in section 4.5, we proceed on to study the possible implication of quantum space-time noncommutativity in the harmonic oscillator model. Section 4.6 deals
with the modified rate of transition probability in time-dependent perturbation theory.
Finally, concluding remarks are made in Sec.4.7. There are two appendices; where in appendix A we have carried out a  Dirac analysis to obtain a canonical type deformed space-time algebra and in appendix B, we derive the self adjoint property of quantum space-time operators.

\section{Time re-parametrization symmetry and quantum mechanics in (1+1) dimension }

Let us start with a brief review of the time re-parametrization invariant form of the action integral \cite{Aht} for the ($1+1$) dimensional non-relativistic system, where both space and time are treated as a configuration space variables. Following Deriglazov \emph{et al} \cite{Adez}, we first consider the action of a non-relativistic particle in the presence of the external potential (in general time-dependent) $V(x,t)$ as
\begin{equation}
S[x(t)]=\int_{t_1}^{t_2} dt L_{0}^{(t)}\left(x,\frac{dx}{dt}\right);~~~L^{(t)}_{0}=\frac{1}{2}m\left(\frac{dx}{dt}\right)^2 - V(x,t),
\label{n1}
\end{equation}
where time ($t$) is the evolution parameter. However, this description is not symmetric in space and time as is evident from the structure of the action  (\ref{n1}) itself and is also not  invariant under time re-parametrization. But, we can introduce a new evolution parameter $\tau$ so that both the time and space coordinates can be parametrized as $t=t(\tau)$ and $x=x(\tau)$ and then treat both of them as configuration space variables \cite{H.C.Co} in an enlarged configuration space. As such the parameter $\tau$ can be taken to be an arbitrary one, except that the function $t(\tau)$ is just  required to be a monotonically increasing function of new evolution parameter $\tau$. With this consideration, the above action (integrated over the line element $d\tau$ with Lagrangian $L^{\tau}_{0}$) can be re-written as  
\begin{equation}
S[x(\tau),t(\tau)]=\int_{\tau_1}^{\tau_2} d\tau L^{\tau}_{0}(x,\dot{x},t,\dot{t}),~~~L^{(\tau)}_{0}(x,\dot{x},t,\dot{t})=\frac{1}{2}m\frac{\dot{x}^2}{\dot{t}} -\dot{t}V(x(\tau),t(\tau)),
\label{react}
\end{equation}
where the over-head dot denotes a differentiation with respect to  $\tau$ i.e. $\dot{t}=\frac{dt}{d\tau}$, $\dot{x}=\frac{dx}{d\tau}$. It can be easily verified that the above action (\ref{react}) is manifestly invariant under the finite reparametrisations,:
\begin{equation}
\tau\rightarrow \tau^{'}=\tau^{'}(\tau);~~~~~ x^{\mu}(\tau)\rightarrow x^{'\mu}(\tau^{'})=x^{\mu}(\tau).
\end{equation}
Here we immediately note that $x^{\mu}$ transforms as scalars under one dimensional diffeomorphisms ( re-parametrization).\\

Therefore, the canonical conjugate momenta $p_{t}(\tau)$ and $p_{x}(\tau)$, corresponding to the dynamical variables $t(\tau)$ and $x(\tau)$ are now given by
\begin{equation}
p_x =\frac{\partial L^{(\tau)}_{0}}{\partial\dot{x}}=m\frac{\dot{x}}{\dot{t}}=m\left(\frac{dx}{dt}\right)
\end{equation}
and
\begin{equation} \label{n1}
\begin{split}
p_t & = \frac{\partial L^{(\tau)}_{0}}{\partial \dot{t}}=-\frac{1}{2} m\frac{\dot{x}^2}{\dot{t}^2}-V(x,t)=-\frac{1}{2} m\left(\frac{dx}{dt}\right)^2-V(x,t) \\
& =-\frac{p_x^2}{2m}-V(x,t)=-H \
\end{split}
\end{equation}
where $H = \frac{p_x^2}{2m} + V(x,t)$  is the Hamiltonian associated with the primitive Lagrangian $L^{(t)}_{0}$.\\

This (\ref{n1}) indicates that this system is endowed with a primary constraint
\begin{equation}
\phi=p_t + H\approx 0.\label{5}
\end{equation}
Following the nomenclature of Dirac \cite{Aht,PMD}, here $\approx $ refers to the weak equality.\\

Expectedly, the Legendre transformed canonical Hamiltonian $H^{(\tau)}_{0}$ corresponding to Lagrangian $L^{\tau}_{0}$
vanishes,
 
\begin{equation}
H^{(\tau)}_{0} = p_t\dot{t}+p_x\dot{x}-L^{(\tau)}_{0}=0.\label{nn}
\end{equation}
which is nothing but a manifestation of the re-parametrization invariance of (\ref{react}).

\subsection{Canonical Formulation: Dirac's constraint analysis}
At this stage, by exploiting the Legendre transformation we can rewrite the Lagrangian in the first-order \cite{Rma} canonical form as,
\begin{equation}
L^{(\tau)}_{f}=p_{t}\dot{t}+p_{x}\dot{x}-e(\tau)\phi=p_{\mu}\dot{x}^{\mu}-e(\tau) (p_{t}+H), ~~\mu=0,1,
\label{symf}
\end{equation}
where $e(\tau)$ is an arbitrary Lagrange multiplier which enforces the constraint (\ref{5}) and $p_{\mu}=(p_{t},p_{x})$. It may be noted that we have used greek indices in both the supercripts and subscripts to denote the components of position and momentum respectively. This, however, should not give rise to any confusion, as there is no non-trivial space-time metric. In contrast to the Lagrangian $L^{(\tau)}_{0}$, the first order form of the Lagrangian $L^{(\tau)}_{f}$ is simpler in the sense that it does not contain any variables and its derivative(s) with respect to $\tau$ in the denominator. Furthermore, here the field $e(\tau)$ transforms as a scalar density under $\tau$ re-parametrization ($\tau\rightarrow \tau^{'}(\tau)$) as,
 \begin{equation}
  e(\tau)\rightarrow e^{'}(\tau^{'})=(\frac{d\tau}{d\tau^{'}}) e(\tau),
 \end{equation}
 so that we can interpret $e(\tau)$ as the einbein field on the world line.\\

 Further, we can reinterpret the dynamical variables  $x^{\mu}$ and $p_{\mu}$ of the first order action (\ref{symf})as the configuration variables in an extended phase-space. The canonical momentum conjugate to $x^{\mu}$, $p_{\mu}$ and $e$ are, 
 \begin{eqnarray}
 \pi_{\mu}^{x}&=&\frac{\partial L^{(\tau)}_{f}}{\partial \dot x^{\mu}}=p_{\mu}\nonumber\\
 \pi_{p}^{\mu}&=&\frac{\partial L^{(\tau)}_{f}}{\partial \dot p_{\mu}}=0\nonumber\\
 \pi_{e}&=&\frac{\partial L^{(\tau)}_{f}}{\partial \dot e}=0.\nonumber
 \end{eqnarray}
In the first-order formulation, it may be noted that none of the canonical momenta involve velocities, and these have to be interpreted as primary constraints. These are given by
 \begin{eqnarray}
 &&\Phi=\pi_{e}\approx0\label{3.8}\\
 &&\Phi_{1,\mu}=\pi_{\mu}^{x}-p_{\mu}\approx0\label{3.9}\\
 &&\Phi_{2}^{\mu}=\pi_{p}^{\mu}\approx0.\label{3.10}
 \end{eqnarray}
 
 Since non zero Poisson bracket exists only for canonical pairs
 \begin{equation}
 \{x^{\mu},\pi_{\nu}^{x}\}=\delta^{\mu}_{~\nu},~\{p_{\mu},\pi_{p}^{\nu}\}=\delta_{\mu}^{~\nu}
 \end{equation}
 the Poisson brackets between among the constraints are given by
 \begin{eqnarray}
 &&\{\Phi,\Phi\}=\{\Phi,\Phi_{1,\mu}\}=\{\Phi,\Phi_{2,\mu}\}=0\label{3.11}\\
 &&\{\Phi_{1,\mu},\Phi_{1,\nu}\}=0\label{3.12}\\
 &&\{\Phi_{1,\mu},\Phi_{2}^{\nu}\}=-\delta_{\mu}^{~\nu}\label{3.13} \\
 &&\{\Phi_{2,\mu},\Phi_{2,\nu}\}=0\label{3.14}. 
 \end{eqnarray}
 Because the Poisson algebra of the primary constraints $\Phi_{1,\mu}$ and $\Phi_{2}^{\mu}$ does not close, they form a set of second class constraint. Accordingly, the other left-over constraint $\Phi$ is considered as a first-class constraint. However, the set of second-class constraints can be eliminated as shown later, through the introduction of Dirac brackets.\\

 Since our system described in terms of a first-order Lagrangian (\ref{symf}), the canonical Hamiltonian $H_{C}$ of the system can also be written as,
 \begin{eqnarray}
 H_C=e(\tau)(p_{t}+H).\label{3.15}
 \end{eqnarray}
 
 Following Dirac, the total Hamiltonian is a sum of canonical Hamiltonian and a suitable linear combination of primary constraints with suitable Lagrange multipliers:
 \begin{eqnarray}
 H_T=e(\tau)(p_{t}+H)+\lambda \Phi+\lambda_{1,\mu}\Phi_{1,\mu}+\lambda_{2,\mu}\Phi_{2}^{\mu}.\label{3.17}
 \end{eqnarray}
 Now we need to ensure that all the constraints must hold for all times. Hence, one has to impose the time consistency of the constraint (\ref{3.8}) that leads to the following secondary constraint

 \begin{eqnarray}
 \phi=\{H_T,\pi_e\}=p_{t}+H\approx0.\label{3.18}
 \end{eqnarray}
 
 Clearly, there are no other secondary constraints. The second class constraint sector $\Phi_1,\Phi_2$ is next eliminated by using Dirac brackets. The first step is to compute the constraint matrix,
\begin{eqnarray} 
 \Lambda_{ab}&=&\begin{pmatrix}
 \{\Phi_{1,\mu},\Phi_{1,\nu}\}	 &  \{\Phi_{1,\mu},\Phi_{2}^{\nu}\}\\
 	 \{\Phi_{2}^{\mu},\Phi_{1,\nu}\} &  \{\Phi_{2}^{\mu},\Phi_{2}^{\nu}\}
 \end{pmatrix}\\
  &=&\begin{pmatrix}
  0 & -\delta_{\mu}^{~\nu}\\
  \delta^{\mu}_{~\nu} &0
  \end{pmatrix}
 \end{eqnarray}
 
 Now we can write the inverse of  $\Lambda_{ab}$ as $(\Lambda^{-1})_{ab}$ such that
 \begin{equation}
 (\Lambda^{-1})_{ab}=\begin{pmatrix}
 0 & \delta_{\mu}^{~\nu}\\
- \delta^{\mu}_{~\nu} &0
 \end{pmatrix}
 \end{equation}
 satisfying $\Lambda_{ab}(\Lambda^{-1})_{bc}=\delta_{ac}$.\\
 At this stage, we can compute the various Dirac brackets using the definition:
 \begin{eqnarray}
 \{f,g\}_{DB}=\{f,g\}-\{f,\Phi_{a}\}(\Lambda^{-1})_{ab}\{\Phi_{b},g\},~ a,b=1,2.\label{3.22}
 \end{eqnarray}
 A straightforward computation then shows that the Dirac brackets among the dynamical variables are satisfy the following algebra
 \begin{eqnarray}
 \begin{array}{rcl}
 &&\{x^{\mu},x^{\nu}\}_{DB}=0\\
 &&\{p_{\mu},p_{\nu}\}_{DB}=0\\
 &&\{x^{\mu},p_{\nu}\}_{DB}=\delta^{\mu}_{~\nu},~~ \mu,\nu=0,1.
 \end{array}
 \label{3.23}
 \end{eqnarray}
 
  Since the primary first class constraint $\Phi$ is the conjugate momentum corresponding to the Lagrange multiplier $e(\tau)$ (\ref{3.8}), it is not physically relevant. On the other hand, the secondary constraint $\phi$ (\ref{3.18}) satisfy vanishing Dirac brackets with all primary constraints as,
  \begin{eqnarray}
  &&\{\phi,\phi\}_{DB}=\{\phi,\Phi\}_{DB}=0\label{3.24}\\
  &&\{\phi,\Phi_{1,\mu}\}_{DB}=\{\phi,\Phi_{2}^{\mu}\}_{DB}=0.\label{3.25}
  \end{eqnarray}
Accordingly the secondary constraint $\phi$ is a first-class and hence should act as a generator of $\tau$ re-parametrization transformation. For instance, this $\tau$-evolution can be identified with an unfolding "gauge transformation" \cite{bisho}. It will be worth emphasizing at this stage that the analysis in this section is quite independent of the Hamiltonian/ Lagrangian one starts with. Particularly, the set of equations (\ref{3.23}) is not sensitive to the structure of the Hamiltonian occurring in (\ref{symf}) and also in the first class constraint (\ref{3.18}). \\\\\
  
  \subsection{Quantum theory: Commutative space-time picture} 
  
Now to formulate the quantum theory of the above model we have to promote all the phase space variables ($t,x,p_t,p_x$) to the level of phase-space operators satisfying Heisenberg algebra (in the unit $\hbar = 1,c=1$):
\begin{equation}
[\hat{t},\hat{x}]=0=[\hat{p_t},\hat{p_x}],~~[\hat{t},\hat{p_t}]=i=[\hat{x},\hat{p_x}].\label{A2}
\end{equation}
We then look for an appropriate Hilbert space on which these operators act and furnish a unitary representation of this algebra. The quantum counterpart of the 2D configuration space is now the Hilbert space $L^2(\mathcal{R}^2)$. We now introduce the simultaneous ``spatio-temporal" eigenbasis $\left|t,x\right\rangle$ of the commutating  operators $\hat{t}$ and $\hat{x}$ satisfying
\begin{equation}
\hat{t}\left|t,x\right\rangle=t\left|t,x\right\rangle,~~~\hat{x}\left|t,x\right\rangle=x\left|t,x\right\rangle.\label{A3}
\end{equation}
 A completeness and orthonormality relations read:
\begin{equation}
\int dt dx \left|t,x \right\rangle \langle t,x | =\mathbb{I}, ~~~\left\langle t,x | t^{'},x^{'} \right\rangle=\delta(t-t^{'})\delta(x-x^{'}).
\end{equation}

The coordinate representations of phase space operators are given by
\begin{equation} \label{n2}
\begin{split}
\left\langle x,t | \hat{x} | \psi \right\rangle = x\left\langle x,t |\psi \right\rangle , & \left\langle x,t |\hat{t} |\psi \right\rangle = t \left\langle x,t |\psi \right\rangle\\
\left\langle x,t |\hat{p}_x |\psi \right\rangle = -i\partial_x \left\langle x,t |\psi \right\rangle , & \left\langle x,t |\hat{p}_t |\psi \right\rangle = -i\partial_t \left\langle x,t |\psi \right\rangle \
\end{split}
\end{equation}
where $\psi (x,t)=\left\langle t,x|\psi\right\rangle \in L^2(\mathcal{R}^2)$ and can be formally identified with the  state function satisfying the following norm:
\begin{equation}
\left\langle\psi |\psi\right\rangle =\int dtdx~ \psi^\ast(x,t)\psi(x,t) < \infty .\label{nn2}
\end{equation}
The associated inner product is given by
\begin{equation}
<\psi|\phi>_=\int dt dx \psi^{\star}(t,x)\phi(t,x)
\label{in2}
\end{equation}

In order to obtain an appropriate probabilistic interpretation of non-relativistic quantum mechanics, however, we need to consider a physical Hilbert space which is the linear span of all possible physical states. Mathematically, this can be identified by projecting out physical states ($\left|\psi;phy\right\rangle$) by imposing the subsidiary condition
\begin{equation}
\hat{\phi}\left|\psi;phy\right\rangle=(\hat{p_t} + \hat{H})\left|\psi;phy\right\rangle=0.\label{nn1}
\end{equation}
where $\hat{\phi}$ is the operator form of the first class constraint (\ref{3.18}). This is tantamount to demanding the gauge invariance nature of physical states.\\

 
The coordinate representation of quantum constraint equation (\ref{nn1})  readily yields the time evolution of the physical states or wave-functions ($<t,x|\psi;phy >:=\psi_{phy}(x,t)$) as
\begin{equation}
i\frac{\partial}{\partial t} \psi_{phy}(x,t)=\left( -\frac{1}{2m} \frac{\partial^2}{\partial x^2} + V(x,t)\right) \psi_{phy}(x,t)\label{A7}
\end{equation}
which will be recognized as the time-dependent Schr\"odinger equation.  
Note that it is independent of the parameter $\tau$, as its $\tau$-evolution is frozen, as can be easily observed by using (\ref{nn},\ref{nn1}).\\

 Now to look at the probabilistic interpretation we recall that the continuity equation corresponding to the Sch\"odinger equation is given by
\begin{equation}
\frac{\partial \rho}{\partial t}+\frac{\partial J_{x}}{\partial x}=0
\label{conti}
\end{equation}
with $\rho=\psi^{\star}_{phy}(x,t)\psi_{phy}(x,t)$ and $J_{x}=\frac{1}{m}Im(\psi^{\star}_{phy}(x,t)\partial_{x}\psi_{phy}(x,t))$. Correspondingly, after performing a spatial integration in both side of (\ref{conti}) ranging from -$\infty$ to +$\infty$ we can write,
\begin{equation}
\partial_t\int_{-\infty}^{\infty} \rho dx =-\int_{-\infty}^{\infty} (\partial_xJ_x) dx= 0
\end{equation}
where we have used the fact that our physical wave-function $\psi_{ph}(x,t)\to 0$ as $x\to \pm\infty$. Thereby the nonnegative $\rho$ is interpreted as a probability density and this notion of physical state on a "space-like surface" indicates the conservation of the total probability $\int_{-\infty}^{\infty} \rho dx$. From the expression of the continuity equation (\ref{conti}), the variable $t$ can be identified as the evolution parameter and we argue that  for $\psi_{ph}(x,t)$ to be well behaved it has to satisfy the following condition at a constant time slice as 
\begin{equation}
\langle \psi; phy |\psi;phy \rangle_{t} =	\int_{-\infty}^{\infty} \,dx\,\, \psi_{phy}^*(x,t)\psi_{phy}(x,t) < \infty\label{e44}
\end{equation}

Thus the physical states satisfying (\ref{A7}) cannot be elements of the Hilbert space of  square integrable wave-functions ($\psi_{phy}(x,t)$)  belonging to $L^2(\mathcal{R}^2)$. We shall, however, restrict our attention to physical states and we shall henceforth  simply write it as $\psi(x,t)$ for brevity and put in the subscript $p$ as and when necessary. More precisely, the probabilistic notion in quantum mechanics is then recovered by replacing the inner-product (\ref{in2}) by the one, which involves only a spatial integration at a constant time slice:
\begin{equation}
\left\langle\psi |\phi\right\rangle_t : =\int_{t} dx ~\psi^\ast(x,t)\phi(x,t).\label{in1}
\end{equation}

We shall refer this as ``induced inner product". Clearly, any normalizable states with $L^2(\mathcal{R}^1)$ inner product (\ref{in1}) may not be so with respect to that of $L^2(\mathcal{R}^2)$ (\ref{in2}): $L^2(\mathcal{R}^2)\subset L^2(\mathcal{R}^1)$. As an illustration, we may consider a stationary state like $\psi (x,t) = e^{-iEt} \phi (x)$. Finally, note that the self-adjoint property of the derivative representation of $\hat{p}_t=-i\partial_t$ in (\ref{n2}) does not hold anymore in the Hilbert space $L^2(\mathcal{R}^1)$ with associated inner product (\ref{in1}), as it is impossible to demand that $\mid\psi(x,t)\mid \to 0$ as $\mid t\mid \to \infty$. In contrast, in $L^2(\mathcal{R}^2)$, this would have allowed one to carry out integration by parts and drop boundary contribution. Indeed, this is deeply related to the original Pauli's objection \cite{Wpauli} in regard to the elevation of $\left( \hat{t}, \hat{p}_t \right)$ to the level of operators. His arguments were very simple, which we can recall here very briefly. Considering an energy eigenstate $\left|E\right\rangle$ satisfying $\hat{p}_t \left|E\right\rangle = - \hat{H} \left|E\right\rangle = - E \left|E\right\rangle$ , the state $e^{i \alpha \hat{t}} \left|E\right\rangle$ too will be an eigenstate $\left|E-\alpha\right\rangle$ with energy eigenvalue $(E - \alpha)$, where $\alpha$ is an arbitrary real parameter, allowing the spectrum of the system Hamiltonian $H$ to have the support which forms a   continuum with values in the entire range $\left( -\infty, \infty \right)$. Particularly, this is in direct contradiction with the existence of systems where energy is positive definite or at least bounded from below. Of course, there were some investigation to evade this problem \cite{ore}, but, we are not going to pursue our analysis in those directions. Rather, we shall follow the conventional method, where $\hat{p}_t$ is now excluded from the dynamical phase-space variables, along with $\hat{t}$. The latter, when `demoted' to a $c$-number parameter, is now identified with the new evolution parameter with $\left( -i\partial_t \right)$ having no association with $\hat{p}_t$ anymore, so that (\ref{A7}) has now the status of a postulate.\\

\section{Quantum space-time: Non-commutative picture }

In section 4.1, we observed that the first-order action is endowed with many variables ($x,p_{x},t,p_{t},e$) that allow more freedom for theoretical interest. Also, it has been pointed out in chapter 2 that the symplectic first-order Lagrangian (\ref{action}) with Chern-Simon term in momentum space can generate a non-canonical symplectic structure between spatial components. Therefore, from this experience we can add a momentum space Chern-Simon like term in (\ref{symf}) to obtain a non-commutative generalization of the space-time structure as following:
\begin{equation}
L_\theta^{(\tau)}= p_{\mu}\dot{x}^{\mu}+\frac{\theta}{2}\epsilon^{\mu\nu}p_{\mu}\dot{p}_{\nu}-e(\tau)(p_{t}+H)
\label{Nc1}
\end{equation}
where $\theta$ is turned out to be the constant non-commutative parameter for the variables $x^{\mu}$. Namely, the system $ L_{\theta}^{(\tau)}$
has the same number of physical degrees of freedom as the
initial system $L_{f}^{(\tau)}$ and quasi invariant under re-parametrization in the sense that $L^{\tau}_{\theta}$ is now invariant up-to a total derivative only. Since the Lagrangian is first order (\ref{Nc1}), following Dirac's algorithm (see Appendix A), one can easily observe that there are only two first-class constraints as $\pi_{e}\approx 0$ and $\Phi=p_{t}+H\approx0$. However, the second class sector can be taken into account by the use of Dirac brackets, just as in the previous section. Thus, one finds, the following relevant Dirac brackets among the configuration space variables   

\begin{equation}
\{t,x\}_{DB}=\theta,\,\,\,\,\,\,\{t,p_t\}_{DB}=1=\{x,p_x\}_{DB}.
\label{dbp}
\end{equation} 
So we can see that the non-commutative nature of configuration space emerges very naturally from a constrained classical system. Note that this is shown just to motivate the appearance of non-commutativity between space-time coordinates naturally at the classical level before setting up a quantum mechanical description of the quantum space-time.\\

\subsection{Quantum theory: The quantum Hilbert space}

In order to provide a description for the non-relativistic quantum mechanics in quantum space-time, we will elevate the classical Dirac brackets (\ref{dbp}) to the level of commutation brackets as given below, 
\begin{equation}
[\hat{t},\hat{x}]=i\theta\label{e1}
\end{equation} with $\theta$ being the non-commutative parameter, along with 
\begin{equation}
[\hat{p}_t,\hat{p}_x]=0,\,\,[\hat{t},\hat{p}_t]=i=[\hat{x},\hat{p}_x].\label{e2}
\end{equation} We have considered $\hbar= 1,c=1$ throughout this chapter. (\ref{e1}) and (\ref{e2}) as a whole represents the non-commutative Heisenberg algebra (NCHA).\\
 
The representation non-commutative space-time algebra in (\ref{e1}) is furnished by a Hilbert space $\mathcal{H}_{c}$:
\begin{equation}
\mathcal{H}_c=Span \left\{|n\rangle = \frac{(\tilde{b}^{\dagger})^n}{\sqrt{n!}}|0\rangle;\,\,\hat{b}=\frac{\hat{t}+i\hat{x}}{\sqrt{2\theta}},~[\hat{b},\hat{b}^{\dagger}]=\mathbb{I}_{c} \right\}.\label{e3}
\end{equation}
This is just the counterpart of (\ref{hc}) of section 2 and can be regarded as the non-commutative configuration space. Let us now introduce an arbitrary associative non-commutative operator algebra ($\hat{\mathcal{A}}_{\theta}$) acting on this configuration space $\mathcal{H}_c$ (\ref{e3}) as
\begin{equation}
\hat{\mathcal{A}}_{\theta}=\{|\Psi)= \sum_{m,n} c_{n,m}|m\rangle\langle n|=\Psi (\hat{t},\hat{x})\}\label{e59}
\end{equation}
which is basically the set of all polynomial in $\hat{t}$ and $\hat{x}$ i.e. all the algebra elements generated by $\hat{t}$ and $\hat{x}$ and also can be thought of as the universal enveloping algebra corresponding to (\ref{e1}). At this stage one should note that $\hat{\mathcal{A}}_{\theta}$ has not yet been endowed with any inner product structure and consequently can't be identified with a Hilbert space. As far as the notations are concerned, like in section 2, the elements of $\mathcal{H}_c$ and $\hat{\mathcal{A}}_{\theta}$ are denoted by the angular ket $| .\rangle$ and round ket $| . )$ respectively.\\

 Again as in chapter 2, we can introduce a subset of $\hat{\mathcal{A}}_{\theta}$ (\ref{e59})as the set of `Hilbert Schmidt' (HS) operators, with finite HS norm, acting on $\mathcal{H}_c$ (\ref{e3}), given by,
\begin{equation}
\mathcal{H}_q= Span\left\{\psi(\hat{t},\hat{x})\equiv |\psi)\in\mathcal{B}(\mathcal{H}_{c}) ; \, \Vert\psi\Vert_{HS}:= \sqrt{tr_{c}(\psi^{\dagger}\psi)} <\infty\right\} \subset \hat{\mathcal{A}}_{\theta},\label{e4}
\end{equation}
where $tr_{c}$ indicates the trace over non commutative space-time configuration space and $\mathcal{B}(\mathcal{H}_{c})\subset \hat{\mathcal{A}}_{\theta}$ is a set of bounded operators on $\mathcal{H}_{c}$.  Now this set is equipped with the inner product
\begin{equation}
	(\psi(\hat{t},\hat{x}),\phi(\hat{t},\hat{x})):=tr_{c}(\psi^{\dagger}(\hat{t},\hat{x})\phi(\hat{t},\hat{x}))
	\label{iop}
\end{equation}
and forms a Hilbert space on its own and is a dense subspace of $\mathcal{B}(\mathcal{H}_{c})$.  We now define the quantum space-time coordinates ($\hat{T},\hat{X}$) and corresponding conjugate momentum operators ($\hat{P_{t}},\hat{P}_{x}$)  by their actions on a typical element $|\psi(\hat{t},\hat{x}))\in \mathcal{H}_{q}$ as,
\begin{align}
&\hat{T}|\psi)=\hat{t}\psi(\hat{t},\hat{x}),\,\,\,\,\,\hat{X}|\psi)=\hat{x}\psi(\hat{t},\hat{x}),\nonumber\\
&\hat{P}_t|\psi)=\frac{1}{\theta}[\hat{x},\psi(\hat{t},\hat{x})],\,\,\,\hat{P}_x|\psi)=-\frac{1}{\theta}[\hat{t},\psi(\hat{t},\hat{x})] \label{e5}
\end{align}
Note that here upper case letters $\hat{T}$ and $\hat{X}$ are used to distinguish them their lower case counterparts $\hat{t}$ and $\hat{x}$ as their domain of actions are different; while $(\hat{t},\hat{x})$ act on $\mathcal{H}_{c}$, the $(\hat{T},\hat{X})$ act on $\mathcal{H}_{q}$. But since they satisfy isomorphic commutator algebra, $(\hat{T},\hat{X})$ can be regarded as the representation of ($\hat{t}, \hat{x}$). Further observe here that the  quantum space-time operators have been taken to act by left multiplication and the corresponding momentum operators adjointly on $\mathcal{H}_{q}$. It is easily verified using (\ref{e5}) that ($ \hat{T},\hat{X}, \hat{P}_{t},\hat{P}_{x}$) satisfies the same non-commutative Heisenberg algebra (\ref{e1},\ref{e2}). These new-fashioned operators are known as super operators as they operate on space of operators. In fact, they can act on the entire $\hat{\mathcal{A}}_{\theta}$. \\

However, we can also define a set of right-acting operators $\{\hat{X}^{\mu}_{R}=(\hat{T}_{R},\hat{X}_{R})\}$ on $\mathcal{H}_{q}$:
\begin{equation}
\hat{T}_{R}\psi(\hat{t},\hat{x})=\psi(\hat{t},\hat{x})\hat{t},~\hat{X}_{R}\psi(\hat{t},\hat{x})=\psi(\hat{t},\hat{x})\hat{x}, 
\end{equation}
satisfying the so-called ``opposite" commutation relation: $[\hat{T}_{R},\hat{X}_{R}]=-i\theta$. Accordingly the action of the momentum operators may be rewritten as
\begin{equation}
\hat{P}_{\mu}\psi(\hat{t},\hat{x})=\frac{1}{\theta}\epsilon_{\mu\nu}[\hat{X}^{\nu}_{L}-\hat{X}^{\nu}_{R}]\psi(\hat{t},\hat{x});~~\mu,\nu=0,1
\label{adjp}
\end{equation} 
where $\epsilon_{\mu\nu}$ is the antisymmetric Levi-Civita symbol,with $\epsilon_{01}$. Upto this, all the constructions are completely in parallel with the ones provided in Chapter-2
. However, it is clear from the discussion of the previous section that HS norm (\ref{e4}) has its commutative counterparts in (\ref{nn2}). Clearly, the associated inner-product (\ref{iop}), like the counterpart (\ref{in2}), needs also to be modified to some form analogous to (\ref{in1}) if we want to have a proper probabilistic interpretation. We take it up in the next section.

\section{Recovery of effective commutative theory} \label{sec4}

Now it is clear that in view of $\theta \neq 0$, we cannot find a counterpart of the common space-time eigenstate $|x,t\rangle $ (\ref{A3}). However, we can recover an effective commutative theory using the coherent state approach as we have already discussed in chap.2. In terms of this Fock basis, we may define a set of coherent states belonging to $\mathcal{H}_c$ (\ref{e3}) as

\begin{equation}
|z\rangle = e^{-\bar{z}\hat{b}+z\hat{b}^{\dagger}}|0\rangle\,\in\mathcal{H}_c\label{e43}
\end{equation}
 
which is an eigen state of the anihilation operator : $\hat{b}|z\rangle=z|z\rangle$, where $\hat{b}=\frac{\hat{t}+i\hat{x}}{\sqrt{2\theta}}$ and $z$ is dimensionless complex number which is given by,
\begin{equation}
z=\frac{t+ix}{\sqrt{2\theta}}
\label{hjl}
\end{equation}

Here $t$ and $x$ are effective commutative coordinate variables. We can now construct another coherent state (operator) in $\mathcal{H}_q$ (\ref{e4}) made out of the bases $|z\rangle \equiv |t,x\rangle$ by taking their outer product as
\begin{equation}
|z,\bar{z})\equiv |z)=|z\rangle\langle z|=\sqrt{2\pi \theta}\,\,|x,t)\,\in \mathcal{H}_q ;\,\,\,\,\,\,\,\,\,B|z)=z|z)
\label{cohe}
\end{equation}
where $|x,t)$ is a dimension-full basis, and the  annihilation operator $B=\frac{\hat{T}+i\hat{X}}{\sqrt{2\theta}}$ is a representation of the operator $\hat{b}$ in $\mathcal{H}_q$ (\ref{e4}). It can also be checked that the basis $|z,\bar{z})\equiv |z)$ satisfies the over-completeness property:
\begin{equation}
\int \,\,\frac{d^2z}{\pi}|z,\bar{z})\,\star_{V}\,(z,\bar{z}|= \int dtdx \, |x,t) \star_{V} (x,t| = \textbf{1}_q,
\label{vnc}
\end{equation} \\
where the product $*_V$ is given by,
\begin{equation}
*_V=e^{\overleftarrow{\partial_z}\overrightarrow{\partial_{\bar{z}}}}=e^{\frac{i\theta}{2}(-i\delta_{ij}+\epsilon_{ij})\overleftarrow{\partial_i}\overrightarrow{\partial_j}};\,\,\,\,i,j=0,1;\,\,\,x^0=t,x^1=x;\,\,\,\,\,\epsilon_{01}=1
\end{equation}


Then the coherent state representation of an abstract state $\psi(\hat{t},\hat{x})$ gives the usual coordinate representation of a state just like formal quantum mechanics.
\begin{equation}
\psi(x,t)=\frac{1}{\sqrt{2\pi\theta}}(z,\bar{z}|\psi(\hat{x},\hat{t}))=\frac{1}{\sqrt{2\pi\theta}}tr_{c}[|z\rangle\langle z|\psi(\hat{x},\hat{t})]=\frac{1}{\sqrt{2\pi\theta}}\langle z|\psi(\hat{x},\hat{t})|z\rangle\label{e45}
\end{equation}
and is called the symbol of the HS operator $\psi(\hat{x},\hat{t})$.
The corresponding representation of a composite operator say $\psi(\hat{x},\hat{t}) \phi(\hat{x},\hat{t})$ is  given by composing the corresponding symbols through Voros star product given as 
\begin{equation}
(z|\psi(\hat{x},\hat{t})\phi(\hat{x},\hat{t}))=(z|\psi(\hat{x},\hat{t})) \,*_V\,(z|\phi(\hat{x},\hat{t}))
\label{como}
\end{equation} 
We shall demonstrate that with this Voros star product probability density is positive definite. The basis $|z,\bar{z})$ referred as Voros basis in the literature \cite{basup} and is compatible with POVM (positive operator-valued measure) \cite{jab} in contrast to Moyal basis which is associated with a similar kind of star product named as Moyal star product \cite{basup}.\\

 However, since the momenta components $\hat{P}_x$ and $\hat{P}_t$ still commutes, it is possible to define joint eigenstate $|p,E)$, satisfying,
\begin{equation} 
\hat{P}_x|p,E)=p|p,E),~~\hat{P}_t|p,E)=-E|p,E).\label{A15}
\end{equation}
 Indeed, it can be easily checked that the following state in the quantum Hilbert space satisfies this equation.  
\begin{equation}
|p,E)=\sqrt{\frac{\theta}{2\pi}}e^{-i(E\hat{t}-p\hat{x})}\label{23}
\end{equation}
It is now quite straight forward to see the orthonormality and completeness relation takes the following form:
\begin{equation} \label{24}
(p,E|p^\prime ,E^\prime)=\delta (p-p^\prime) \delta (E-E^\prime);~~~\int dp dE~ |p,E)(p,E|=\textbf{1}_q.
\end{equation}

The overlap of a states $|x,t)$ with this momentum state is

\begin{equation} \label{eq42}
(x,t|p,E)=\frac{1}{\sqrt{2\pi\theta}} (z|p,E)= \frac{1}{2\pi} e^{-\frac{\theta}{4}(E^2 + p^2)} e^{-i(Et-px)}.
\end{equation}


Now,  using (\ref{vnc})  the overlap of two arbitrary states ($|\psi),|\phi)$) in the quantum Hilbert space can be written in the form
\begin{equation} \label{innpro_voros}
(\psi|\phi) = \int dtdx ~ \psi^\ast(x,t) \star_{V} \phi(x,t)
\end{equation}

Also note that the overlap of the basis $|x,t)$ (\ref{cohe}) and its primed counterpart is given by

\begin{equation} \label{a3}
(x^\prime, t^\prime | x, t ) = \frac{(z^\prime|z)}{2\pi\theta} = \frac{\mid\langle z^{'} |z\rangle\mid^{2}}{2\pi\theta} = \delta_{\sqrt{\theta}} (t^\prime - t) \delta_{\sqrt{\theta}} (x^\prime - x)
\end{equation}
where
\begin{equation} \label{a4}
\delta_\sigma (x) = \frac{1}{\sigma\sqrt{2\pi}} e^{-\frac{x^2}{2\sigma^2}} ~;~~ \int dx \, \delta_\sigma (x) = 1
\end{equation}

It may be noted that the $\left( \delta_{\sqrt{\theta}} (t^\prime - t)
\delta_{\sqrt{\theta}} (x^\prime - x) \right)$ - as a whole, plays the role of Dirac's $\delta$-distribution in our non-commutative  space-time, provided they are composed with the Voros star product. This can be seen quite transparently from the derivation of the following identity, by making use of \ref{24} and \ref{eq42},
\begin{equation} \label{5a1}
\int dt^\prime dx^\prime \, \left( \delta_{\sqrt{\theta}} (t - t^\prime) \, \delta_{\sqrt{\theta}} (x - x^\prime) \right) \star_{V}^\prime \psi \left( x^\prime, t^\prime \right) = \psi \left( x, t \right), 
\end{equation}
with
\begin{equation}
\psi(x,t) = (x,t|\psi) = \int dE dp \, (x,t|E,p)(E,p|\psi)
\end{equation}
where it is essential to retain both $\delta_{\sqrt{\theta}} (t)$ and $\delta_{\sqrt{\theta}} (x)$ together. This, in turn, can be seen easily by making use of the identity :
\begin{equation} \label{7b1}
\int dt' dx' \, \delta_{\sqrt{\theta}} (t - t^\prime) \, \delta_{\sqrt{\theta}} (x - x^\prime) \star_{V}^\prime e^{-i \left( Et^\prime - px^\prime \right)} = e^{-i \left( Et - px \right)}.
\end{equation}

Here $\star_{V}^\prime$ indicates that the relevant higher-order derivatives involve $t^\prime$ and $x^\prime$. Besides, to recover effective usual commutative theory with proper probabilistic interpretations of quantum mechanics we recall Pauli's objection and exclude `$t$' and `$p_t$' from the dynamical variables. This, however, does not indicate that $\hat{t}$ is no longer an operator; it still satisfies the commutation $[\hat{t}, \hat{x}] = i \theta$, but the pair of canonical commutators involving time and its translation generator $\hat{p}_t$ in (\ref{e2}) are disregarded. Particularly, the operator $\frac{1}{\theta} \, ad\,\hat{x}$ can not be identified with the representation of time translation generator $\hat{P}_t$. On the other hand, $e^{i\alpha\hat{t}}$ generates space translation in $\mathcal{H}_c$ in the sense that its action on an eigenstate $|a\rangle$ of $\hat{x}$, satisfying $\hat{x}|a\rangle = a|a\rangle$ yields a shifted eigenstate of the position operator $\hat{x}$ as $\hat{x} \, \left( e^{i\alpha\hat{t}} |a\rangle \right) = \left( a + \alpha \theta \right) \left( e^{i\alpha\hat{t}} |a\rangle \right)$, so that we may with impunity write $e^{i\alpha\hat{t}} |a\rangle =|a + \alpha\theta\rangle$. Thus its outer product will be $|a \rangle\langle a | = e^{i\alpha\hat{t}} |a + \alpha \theta\rangle\langle a + \alpha \theta| e^{-i\alpha\hat{t}}$. In its infinitesimal version, we will be recognized as the $\hat{P}_x$ is equivalent to $\left(-\frac{1}{\theta}\right) ad\,\hat{t}$, as it occurs in \ref{e5}. Therefore, finally again in non-commutative quantum mechanics, the Schr\"odinger equation, to be introduced below, will have the status of a postulate. We have more to say on this point in the sequel. Note that we will regard the coherent state basis $|x,t)$ as ``quasi-orthonormal bases", as Gaussian function (\ref{a4}) are can be considered some kind of ``regularised Dirac's $\delta$-distribution function", in the sense of the identity (\ref{5a1}) and also for the fact that $\delta_\sigma (x) \to \delta (x)$ as $\sigma \to 0$. It is quite clear at this stage that all these expressions of the previous section i.e. their commutative counterparts are reproduced in the limit $\theta \to 0.$\\

\subsection{Sch\"odinger equation and an induced inner product}

To obtain the effective commutative Sch\"odinger equation, we introduce coordinate (coherent state) representation of the phase space operators, which will be useful to construct this. As we know that for space-time super operators operators have both left and right actions on $\mathcal{H}_{q}$ as well as on $\mathcal{A}_{\theta}$.  Thus the coherent state representation of space-time operators  $\{\hat{X}_{L},\hat{T}_{L}\}$, acting on any arbitrary state $|\Psi)\in\mathcal{A}_{\theta}$, can be written as,
 
\begin{equation}
( x,t|\hat{X}_L \, \Psi(\hat{x},\hat{t})) =\frac{1}{\sqrt{2\pi\theta}}(z,\bar{z}|\hat{x}\Psi) = \frac{1}{\sqrt{2\pi\theta}} \, \left\langle z|\hat{x}|z\right\rangle \star_{V} (z,\bar{z}|\Psi(\hat{x},\hat{t}))
\end{equation}
Finally making use the fact of \eqref{e45} this readily yields
\begin{equation}
( x,t|\hat{X}_L \, \Psi(\hat{x},\hat{t})) = X_\theta \, (x,t|\Psi(\hat{x},\hat{t})) \equiv X_\theta \, \Psi(x,t)
\end{equation}
with
\begin{equation}
X_\theta^L \equiv X_\theta =  \left[x+\frac{\theta}{2}(\partial_x-i\partial_t)\right].\label{47}
\end{equation}
Proceeding exactly in the same way, we obtain the representation of $\hat{T}_{L}$ as
\begin{equation}
T_\theta^L \equiv T_\theta =  \left[t+\frac{\theta}{2}(\partial_t+i\partial_x)\right].\label{48}
\end{equation}
 Now, it is trivial to prove the self-adjointness property of both $X_\theta^L$ and $T_\theta^L$, w.r.t. the inner product \eqref{innpro_voros}. In $\mathcal{H}_{q}$ by considering an arbitrary pair of different states $|\psi_1)$, $|\psi_2) \in \mathcal{H}_q$ and their associated symbols by exploiting associativity of Voros star product. We also note that since this analysis will not involve any integration by parts i.e. it is not sensitive to the integration measure, this self-adjointness property of $X_\theta^L$ and $T_\theta^L$ will persist to hold for the `induced' inner product \eqref{7a2} below, the counterpart of (\ref{in1}), (see Appendix B) as well.\\

The corresponding expressions for right acting operators $\{\hat{X}_{R},\hat{T}_{R}\}$  are obtained as follows
\begin{equation}
X_\theta^R \equiv \left[x+\frac{\theta}{2}(\partial_x+i\partial_t)\right];~T_\theta^R \equiv \left[t+\frac{\theta}{2}(\partial_t-i\partial_x)\right].\label{49}
\end{equation}
We, however, proceed to develop the coherent state representational adjoint action of momenta operators in (\ref{adjp}) which are essentially given by the difference of the left and right actions as,

\begin{equation} \label{mom_op_act}
\hat{P}_t \Psi(\hat{x},\hat{t})=\frac{1}{\theta}(\hat{X}_L - \hat{X}_R)\Psi(\hat{x},\hat{t});~ \hat{P}_x \Psi(\hat{x},\hat{t})=-\frac{1}{\theta}(\hat{T}_L - \hat{T}_R)\Psi(\hat{x},\hat{t}).
\end{equation}
The overlap with $|x,t)$ allows us to write 
\begin{equation}
(x,t|\hat{P}_t \psi(\hat{x},\hat{t}))= -i \partial_t\psi(x,t)~;~~ (x,t|\hat{P}_x \psi(\hat{x},\hat{t}))=-i\partial_x\psi(x,t)
\label{25}
\end{equation}

 Since our non-commutative theory (\ref{Nc1}) is endowed with a secondary first-class constraint (see Appendix A) $\Phi\approx 0$ (\ref{e87}), we must consider the role of this first-class constraint as an operator representation to complete our construction. Thus, in order to obtain an effective commutative Sch\"odinger equation in non-commutative space-time we must project onto the set of physical states $|\Psi,phy)=\Psi_{phy}(\hat{x},\hat{t})$ for which
 
 \begin{equation}
 (\hat{P}_t+\hat{H})|\Psi,phy)=0;~\Psi_{phy}(\hat{x},\hat{t})\in \hat{\mathcal{A}}_{\theta} \label{e37}
 \end{equation} 
 where $\hat{H}=\frac{\hat{P}_x^2}{2m}+V(\hat{X},\hat{T})$.
    We are now ready to write down the time dependent Schr\"{o}dinger equation in quantum space-time by taking the representation in $|x,t)$ basis as
    \begin{equation}
    (x,t|\hat{P}_t+\hat{H}|\Psi;phy)=0
    \label{physical}
    \end{equation}
    By making use of (\ref{25}), this readily yields,
    \begin{equation}
    i\partial_t \Psi_{phy}(x,t)= \left[-\frac{1}{2m}\partial_x^2+ V(x,t)\, \star_{V}\right] \Psi_{phy}(x,t)\label{e70}
    \end{equation}
Taking complex conjugate of the equation we get
\begin{equation}
-i\partial_t \Psi_{phy}^*(t,x)= -\frac{1}{2m}\partial_x^2 \Psi_{phy}^*(x,t) + \Psi_{phy}^{*}(x,t)\ \star_{V} V(t,x).\label{e71}
\end{equation}
Now using (\ref{e70}) and (\ref{e71}) one therefore obtain the continuity equation
\begin{equation}
\partial_t \rho_{\theta} =-\partial_x J_{\theta}^{x}\label{e72}
\end{equation}
where 
\begin{align}
\rho_{\theta} &= \Psi_{phy}^*(x,t)\,\star_{V}\,\Psi_{phy}(x,t),\nonumber\\
J_{\theta}^{x}&= \frac{1}{2im}[\Psi_{phy}^*\star_{V}(\partial_x \Psi_{phy})-(\partial_x\Psi_{phy}^*)\star_{V}\Psi_{phy}].
\label{prob}
\end{align}
 After taking a spatial integration in both side of (\ref{e72}) ranging from -$\infty$ to +$\infty$ we can write,
\begin{equation}
\partial_t\int_{-\infty}^{\infty} \rho_{\theta} dx =-\int_{-\infty}^{\infty} (\partial_xJ_\theta^{\theta}) dx=0
\end{equation}
where we have used the fact of $\Psi_{phy}(x,t) ,\to 0$ as $x\to \pm\infty$ i.e. $\Psi_{phy}(x,t)$ is well behaved, then right hand side of the above equation becomes zero giving the conservation of the total probability $\int_{-\infty}^{\infty} \rho_{\theta} dx$ whereby we can write $\rho_{\theta}(x,t)$ in a manifestly positive definite form
\begin{equation} \label{e40}
\rho_{\theta}(x,t) = \Psi_{phy}^*(x,t)\,\star_{V}\,\Psi_{phy}(x,t)= \frac{1}{2\pi\theta} \Psi_{Ph}^\ast(z,\bar{z})\star_V \Psi_{Ph}(z,\bar{z}) = \frac{1}{2\pi\theta} \sum_{n=0}^\infty \frac{1}{n!}|\partial_z^n \psi_{ph}(z,\bar{z})|^2 \geqslant 0.
\end{equation} 
In view of the non-negative condition \eqref{e40} $\rho_{\theta}(x,t)$ can indeed be interpreted as probability density for a particular time.
From the expression of the continuity equation (\ref{e72}), the variable $t$ can be identified as the evolution parameter and we argue that for $\Psi_{ph}(x,t)$ to be well behaved it has to satisfy the following condition at a constant time slice.
\begin{equation}
\langle \Psi; phy |\Psi;phy \rangle_{t} =	\int_{-\infty}^{\infty} \,dx\,\, \Psi_{phy}^*(x,t)\star_{V}\Psi_{phy}(x,t) < \infty.
\label{op}
\end{equation}
This implies that $\Psi_{phy}(x,t)\in L_{\star}^2(\mathcal{R}^1)$\footnote{It may be noted that    $L_{\star}^2(\mathcal{R}^1)$ involves $\theta$-deformed multiplication rule.  }  which is a bigger space than $L_{\star}^2(\mathcal{R}^2)$.  Clearly,the corresponding inner-product for the space should also be defined by a spatial i.e. $x$-integration for a fixed time slice as,
\begin{equation}
\langle \Psi; phy |\Phi;phy \rangle_{t} =	\int_{-\infty}^{\infty} \,dx\,\, \Psi_{phy}^*(x,t)\star_{V}\Phi_{phy}(x,t) < \infty.
\label{op1}
\end{equation}   
Here, of course, we need to emphasize that here $t$ should not be identified as coordinate time i.e. as an
the eigenvalue of $\hat{t}$, and more precisely, this is a coherent state expectation value:
\begin{equation}
t=<z|\hat{t}|z>,~~~x=<z|\hat{x}|z>.
\end{equation}

 Since the physical states satisfying the Schr\"{o}dinger's equation are normalizable with respect to the inner product (\ref{op1}) of $L_{\star}^2(\mathcal{R}^1)$. This imposition is clearly indicative of the fact that the physical states $\Psi_{phy}(\hat{x},\hat{t})$ should belong to a suitable subspace  of $\hat{\mathcal{A}}_{\theta}$ (\ref{e59}) which is bigger than its constrained subspace $\mathcal{H}_q$, for which the norm is given by the inner product defined for $L_{\star}^2(\mathcal{R}^2)$ (\ref{innpro_voros}).\\

Further observe that one can introduce
\begin{equation} \label{pi_t}
\pi_t = \int_t dx ~| x,t ) \star_{V} ( x,t |, 
\end{equation}
and the inner-product (\ref{op1}) can be re-written  as
\begin{equation}
(\Psi;phy|\pi_{t}|\Psi;phy)=\int_{-\infty}^{\infty} \,dx\,\, \Psi_{phy}^*(x,t)\star_{V}\Phi_{phy}(x,t).
\end{equation}
This is the appropriate induced inner product \cite{arm,lc} for physical Hilbert space and is the counterpart of (\ref{in1}) for the commutative case, to which it tends to in the commutative ($\theta \rightarrow 0$) limit. 
However, it may be noted that (\ref{pi_t}) satisfies, only approximately, a modified version of projection operator identity for leading approximation of $\theta$ :
\begin{equation} \label{7a2}
\pi_{t^\prime} \pi_t \approx \pi_{t^\prime} \, \delta_{\sqrt{\theta}} (t^\prime - t)
\end{equation}
This indicates that any pair of projection operators $\pi_t$ and $\pi_{t+\delta t}$ separated by a time interval $\delta t$ need not be exactly orthogonal.\\

Now in order to obtain stationary physical states for a time independent one dimensional bound system ($V(\hat{x},\hat{t})=V(\hat{x})$) we consider the completeness relation \eqref{24} satisfied by the basis $|p,E)$, and introduce the projector:

\begin{equation} \label{cal_p_E}
\mathcal{P}_E = \int dp \, |p,E)(p,E| ~~;~~ \mathcal{P}_{E^\prime} \mathcal{P}_E = \mathcal{P}_E \delta(E^\prime - E)
\end{equation}
where $E$ will be identified with the eigenvalue of the Hamiltonian in a moment.

Using this projection operator $\mathcal{P}_E$ we can define the projected state vector $|\psi)_E = \mathcal{P}_E |\psi)$ and eventually its coherent state representation :
\begin{equation} \label{eq50}
\psi_E (x,t) \equiv (x,t|\psi)_E = \int dp \, (x,t|p,E)(p,E|\psi) = \frac{1}{\sqrt{2\pi}} \int dp ~ e^{-i \left( Et - px \right)} \, e^{-\frac{\theta}{4} \left( E^2 + p^2 \right)} \, \psi_E (p)
\end{equation}
where we define $\psi_E(p) \equiv \frac{1}{\sqrt{2\pi}} ( p,E | \psi )$. States like $|\psi)_E=\int dp |p,E)(p,E|\psi)$ will span a subspace of the physical Hilbert space  in (\ref{e59}) $L_{\star}^2(\mathcal{R}^1)$ as $|p,E)$ can be easily shown to satisfy (\ref{e37}) i.e. 
\begin{equation}
(\hat{P}_{t}+\hat{H})|p,E)=0
\end{equation}

We now compute the $\mathcal{H}_{q}$ norm (\ref{e4})  projected states $|\psi)_E$  by making use of \eqref{eq50} to get
\begin{equation} \label{5a2}
\int dt dx \, \psi^\ast_E (x,t) \star_{V} \psi_E (x,t) = \frac{e^{-\frac{\theta}{2}E^2}}{2\pi} \int dt dx dp dp' \, e^{-\frac{\theta}{4}\left(p^2 + p'^2\right)} \, \psi^\ast_E (p) \psi_E (p') \left( e^{i\left(Et - px\right)} \star_{V} e^{-i\left(Et - p'x\right)} \right)
\end{equation}

A short calculation shows that `$t$'-dependence cancels out even in the presence of Voros star product, yielding a divergent integral which can be written in another equivalent form, where the $x$-integration is replaced by $p$-integration :
\begin{equation} \label{5a3}
\int dt dx \, \psi^\ast_E (x,t) \star_{V} \phi_E (x,t) = \int dt dp \, \psi^\ast_E (p) \, \phi_E (p)
\end{equation}
However, since $t$ should no longer be counted as a dynamical variable, thus we should actually compute the norm using the ``induced" inner product (\ref{op1}) only i.e, excluding  the $t$-integration and retain only the convergent integral over $x$ or $p$ :
\begin{equation} \label{5a4}
\left( \psi_E | \psi_E \right)_t := \int_t dx \, \psi^\ast_E (x,t) \star \psi_E (x,t) = \int_t dp \, \psi^\ast_E (p) \, \psi_E (p)
\end{equation}
where the presence of `$t$' at the bottom of the integral sign implies that the integration has to be carried out in a constant $t$-surface. 
We finally remark that \eqref{5a4}, upon normalization, can be considered as the non-commutative extension of Parseval's theorem. And it is also understood that $|\psi)_{E}$ no longer be an element of $\mathcal{H}_{q}$ as for the case of generic physical states. \\

Finally, it can be checked that $\psi_E (x,t)$ \eqref{eq50} satisfies the following effective time-independent Schr\"odinger equation for time-independent potential $V(x)$ :
\begin{equation} \label{eqb2}
E \, \psi_E (x, t) = - \frac{1}{2m} \frac{\partial^2 \psi_E (x,t)}{\partial x^2} ~+~ V(x) \star_{V} \psi_E (x,t)
\end{equation}
where $E$ can be identified with the energy eigenvalues of the bound states. With all these formal aspects of our formalism in place, we can now investigate its application to some dynamical systems. In the next section, we begin with an investigation of free particle and later we study the behavior of a particle under the simple harmonic potential.\\
\section{Free particle wave packet in the noncommutative space-time} \label{sec5}

In this section our intention is to introduce a Gaussian wave packet for a free particle moving under the hamiltonian,
\begin{equation} \label{H_free}
\hat{H} = \frac{\hat{P}_{x}^2}{2m}
\end{equation}
and study the possible effect of noncommutativity that can be observed in its time evolution.
Let us define a projection operator $\hat{\rho}_{free}$ as

\begin{equation}
\hat{\rho}_{free}=\int dE \delta(E-E_{p})\mathcal{P}_E;~~E_{p}=\frac{p^{2}}{2m}
\end{equation}
where we have used the fact of ``on-shell'' implementation of a non-relativistic free particle dispersion relation for each momentum component, through the inclusion of an appropriate delta function in \eqref{24}. This allows us to ‘project’ states (at least those lying in $\hat{\mathcal{A}}_{\theta}$) onto the states lying in the space of physical states $|\Psi;phy)$ (solution of the Sch\"odinger  equation corresponding to the free particle).\\

   
Thus the action of $\hat{\rho}_{free}$ on a generic state $|\Psi)$, such as $|\Psi) \in \hat{\mathcal{A}}_{\theta}$ is then given by,
\begin{equation} \label{rho_action}
|\Psi)\rightarrow |\Psi;phy)=\hat{\rho}_{free} |\Psi)~\equiv \int dp ~ \Psi(p,E_p) ~ |p,E_p) ~~;~ \Psi(p,E_p) = ( p,E_p |\Psi ),
\end{equation}
where the above operator $\hat{\rho}$ \eqref{rho_action} too satisfies the property of a projection operator with respect to the induced inner product (\ref{op1}):
\begin{equation} \label{7a3}
\hat{\rho}_{free} \, \pi_t \, \hat{\rho}_{free} = \hat{\rho}_{free},
\end{equation}
this in turn implies that
\begin{equation}
(\hat{\rho}_{free})\pi_{t}|\Psi;phy)=|\Psi;phy);
\end{equation}
i.e., $\hat{\rho}_{free}$ acts as the identity operator in the space for physical states $\{|\Psi;phy)\}$ equipped with the induced inner product (\ref{op1}).\\

It can be easily checked that the physical states ($|\Psi;phy)$) are projected out by imposing the operatorial counterpart of the constraint ($\Phi=p_{t}+\frac{p_{x}^{2}}{2m}\approx 0$):
\begin{equation}
\hat{\Phi}|\Psi;~~ phy)=0;~~~\hat{\Phi}=\hat{P}_{t}+\frac{\hat{P}^{2}_{x}}{2m},
\end{equation}
 which leads to the Sch\"odinger  equation for free particle in a natural way.\\
 
Now, we can compute the induced overlap between two on-shell states $|p, E_p )$ and $|p^{'},E_{p^{'}})$ as,
\begin{equation} \label{7a1}
\left( p^\prime, E_{p^\prime} | p, E_p \right)_t=(p^{'},E_{p^{'}}|p,E_{p}) = \int_t dx \, ( p^\prime, E_{p^\prime} | x,t ) \star ( x,t | p,E_p ) = \frac{1}{2\pi} \delta(p^\prime - p)
\end{equation}
where we have used the fact of ``quasi-projection operator" \eqref{pi_t} and (\ref{eq42}).  And this indicates that the ``on shell" states form orthonormal basis, provided we make use of the aforesaid induced inner product.\\

The coherent state representation of $|\Psi;phy)$ \eqref{rho_action} corresponding to \eqref{gwp_mom} then yields

\begin{equation}
\Psi_{phy} (x,t) = \left( x,t | \Psi;phy \right)=\int dp~(x,t|p,E_{p})~(p,E_{p}|\Psi)
\end{equation}

 We now consider a Gaussian wave function for the free particle in momentum space as:
 \begin{equation} \label{gwp_mom}
 ( p,E_p | \Psi )~\equiv\Psi(p,E_p) = \frac{\sqrt{\sigma}}{\pi^{1/4}} \, e^{-\frac{\sigma^2 p^2}{2}}
 \end{equation}
 The  corresponding coherent state representation of wave-packet yields

\begin{equation}
\Psi (x,t) = \left( x,t | \Psi;phy \right) = \frac{\sigma^{1/2}}{2\pi^{5/4}} \, \int dp \, e^{-\frac{\theta p^4}{16 m^2} - \lambda p^2 + ipx},
\label{gwp}
\end{equation} 
where

\begin{equation}
\lambda= (\frac{\sigma^{2}}{2}+\frac{\theta}{4}+i\frac{t}{2m}).
\end{equation}

If only first order in $\theta$ deformations ($\mathcal{O}(\theta)$) are considered, then one can show that $\Psi (x,t)$ can be recast in the following simple-looking form,
\begin{equation}
\Psi (x, t) \simeq \frac{1}{2\pi^{3/4}} \sqrt{\frac{\sigma}{\lambda}} \,\, \left[ 1 + \theta \, f(x; \lambda) \right] e^{-\frac{x^2}{4\lambda}}
\end{equation}
where, the function $f(x;\lambda)$ is
\begin{equation*}
f(x; \lambda) = \frac{1}{16m^2} \left( - \frac{3}{4\lambda^2} + \frac{3x^2}{4\lambda^3} - \frac{x^4}{16\lambda^4} \right)
\end{equation*}
exhibiting a deviation in the functional form, away from Gaussian shape in coordinate space. The  $\theta$-deformation then appears both in the exponential factor as well as in the amplitude. This  clearly demonstrates that the width $d$ of the deformed Gaussian term at a later time $t$ gets enhanced due to non-commutativity:

\begin{equation} \label{d}
d = \sqrt{2 \mid\lambda\mid} = \left[ \left( \sigma^2 + \frac{\theta}{2} \right)^2 + \left( \frac{t}{m} \right)^2 \right]^{\frac{1}{4}}
\end{equation}
This further indicates that even for an infinite spread in the Gaussian wave packet \eqref{gwp_mom} in the momentum space ($\sigma \to 0$), and the spread in coordinate space $x$ cannot be squeezed bellow the length scale $\sim \sqrt{\frac{\theta}{2}}$.\\

\section{Schr\"odinger equation and energy spectra of harmonic oscillator} \label{sec6}

In this section, we start by considering the operatorial version of the Schr\"odinger equation \eqref{e37} for the time-independent harmonic oscillator potential $V(\hat{X}) = \frac{1}{2} m \omega^2 \hat{X}^2$ :

\begin{equation}
[\hat{P}_{t}+\hat{H}]|\Psi;phy)_{H.O}=0,
\label{scho}
\end{equation}
where  $\hat{H}=\frac{\hat{P}^{2}_{x}}{2m}+\frac{1}{2}m\omega^{2}\hat{X}^{2}$ and let $|\Psi;phy)_{H.O}$ be a physical states corresponding to the harmonic oscillator system.
Now the coherent state representation of the above equation (\ref{scho}) yields

\begin{equation}
i\frac{\partial}{\partial t}\Psi_{H.O}(x,t)=[-\frac{1}{2m}\frac{\partial^{2}}{\partial^{2}x}+V(X_{\theta})]\Psi_{H.O}(x,t)
\label{scho}
\end{equation}
where we have used the simpler notation $(x,t|\Psi;phy)_{H.O}=\Psi_{H.O}(x,t)$ for brevity.\\

Introducing a non-unitary operator 
\begin{equation}
S=e^{\frac{\theta}{4}(\partial^{2}_{t}+\partial^{2}_{x})}e^{-i\frac{\theta}{2}\partial_{t}\partial_{x}},
\label{twist}
\end{equation} 
 and after some algebra, one can easily show that the space-time coordinate representation $X_{\theta}$ (\ref{47}) and $T_{\theta}$ (\ref{48}) can be casted in simpler forms as,
 \begin{equation}
 X_{\theta}=x+\frac{\theta}{2}(\partial_{x}-i\partial_{t})=SxS^{-1};
 \label{reps}
 \end{equation}
 
  \begin{equation}
  T_{\theta}=t+\frac{\theta}{2}(\partial_{t}+i\partial_{x})=S^{\dagger}t(S^{\dagger})^{-1}
  \label{time}
  \end{equation}
respectively.

Now, for time independent potential $V(X_{\theta})$, it is straightforward to show that 
\begin{equation}
V(X_{\theta})=V(SxS^{-1})=SV(x)S^{-1}
\label{tayl}
\end{equation} 

Using the relation (\ref{reps}) and (\ref{tayl}) we can show the above Sch\"odinger  equation (\ref{scho}) can be rewritten as
\begin{equation}
i\partial_{t}\Psi_{c}(x,t)=[-\frac{1}{2m}\frac{\partial^{2}}{\partial x^{2}}+\frac{1}{2}m \omega^{2} x^{2}]\Psi_{c}(x,t),
\label{cho}
\end{equation} 
with 
\begin{equation}
\Psi_{c}(x,t)=S^{-1} \Psi_{H.O}(x,t),
\label{vp}
\end{equation}
where $\Psi_{c}(x,t)$ stands for the wave-function in commutative space-time. Thus (\ref{cho}) is nothing but a time-dependent Sch\"odinger equation for a one-dimensional harmonic oscillator that lives in usual space-time.\\

Of particular interest are the stationary states, viz. eigenstates of the  Hamiltonian operator, which will obey the so called time independent Sch\"odinger  equation:
\begin{equation}
i\partial_{t}\Psi_{c}(x,t)=[-\frac{1}{2m}\frac{\partial^{2}}{\partial x^{2}}+\frac{1}{2}m \omega^{2} x^{2}]\Psi_{c}(x,t)=E\Psi_{c}(x,t)
\label{chl}
\end{equation}

Following the usual approach, we now introduce the  annihilation and creation operators as
\begin{equation}
b =\frac{1}{\sqrt{2m\omega}}(m\omega x+\frac{\partial}{\partial x})  ~;~ b^\dagger = \frac{1}{\sqrt{2m\omega}}(m\omega x-\frac{\partial}{\partial x}) ~;~ [b, b^\dagger] = 1
\end{equation}
In terms of these operators we can re-write the above equation (\ref{chl}) as
\begin{equation}
\omega\left(b^\dagger b + \frac{1}{2}\right) \Psi_{c}(x,t)=E \Psi_{c}(x,t)
\label{huku}
\end{equation}
Since the operator $b^{\dagger}b$ is the  number operator with non-negative integers as its eigenvalues, the corresponding eigenvalues of the operator on the left-hand side of the (\ref{huku}) gives the energy eigenvalues: $E_{n}=\omega (n+\frac{1}{2})$.  Therefore, the exactly solvable harmonic oscillator model (\ref{cho}) in commutative space-time implies that the corresponding model (\ref{scho}) in non-commutative space-time is also exactly solvable, and vice versa. In fact, our models (\ref{scho}) and (\ref{cho}) in
non-commutative and commutative spaces, respectively,
share the same set of energy eigenvalues and their eigenstates
relate to each other by the relation (\ref{vp}).\\

The above discussions demonstrate that as long as the
explicit $\theta$ dependence does not show up in the energy spectrum, it will be encoded only in the stationary state wave-functions of harmonic oscillator \ref{scho} in non-commutative space-time. To illustrate this point by an explicit example, let us start by considering $\Psi^{(0)}_{c} (x,t)$, the ground state wave function of (\ref{chl}) of commutative space-time  with energy $E_0$, satisfying the condition $b \Psi^{(0)}_{c} (x,t) = 0$. Thus the time evolution of the \emph{un-normalized}  ground state wave-function factorizes as
\begin{equation}
\Psi^{(0)}_{c} (x,t)=  e^{-\frac{m\omega x^{2}}{2}} e^{-iE_{0}t}=\int ~dp~ g^{(0)}(p)~ e^{i(px-E_{0}t)}; ~~~ with~~g^{(0)}(p)=e^{-\frac{p^2}{2m\omega}}
\label{gswave}
\end{equation} 
where $g^{0}(p)$ is a corresponding \emph{un-normalized}  wave-function in momentum space.\\

 However, the corresponding ground state wave function in non commutative space time  i.e. in the coherent state representation ($\Psi^{(0)}_{H.O}(x,t)$) can be easily obtained by making use of (\ref{twist}) and (\ref{vp})  to get the following form of \emph{un-normalised} ground state wave function (normalizable with respect to the inner product (\ref{op1}):
 
 \begin{equation}
 \Psi^{(0)}_{H.O}(x,t)=S^{-1}\Psi^{(0)}_{c}(x,t)= e^{-\left[\frac{(x-\frac{\theta E_0}{2})^2}{2\sigma^2_\theta}\right]} \, e^{-iE_0 t} ~~;~ \sigma^2_\theta = \frac{\theta}{2}+\frac{1}{m \omega}.
 \label{ncwh}
 \end{equation}

This clearly displays energy ($E_{0}$) dependent shift in the origin towards the right side and a modified width $\sigma_\theta$ which is a manifestation of a breakdown of parity invariance. Clearly, this bears a resemblance to the form of the ground state wave function in commutative space-time quantum mechanics $\Psi^{0}_{c}(x,t)$, except for the $\theta$-deformation in the width $\sigma_\theta$. With $\theta \to 0$, one gets back the familiar commutative form of the \emph{un-normalized} ground state wave function (\ref{gswave}).\\
 
 Now we can compute the probability density factor $\rho_{\theta}$ using \eqref{prob} and integrate it over only $x$ to set it to unity. Thus, with the correct normalization factor, we obtain
 \begin{equation} \label{gr_ho_pd}
 \rho_{\theta}(x) = \frac{1}{\tilde{\sigma_\theta} \sqrt{2\pi}} e^{-\left[\frac{(x-\theta E_0)^2}{2\tilde{\sigma_\theta}^2}\right]} ~;~ \tilde{\sigma_\theta}^2 = \frac{\sigma_\theta^2}{2} \left[ 1 + \frac{\theta}{2\sigma^2_\theta} \right]
 \end{equation}
 One can recast in a simpler form, using \eqref{a4} as,
 \begin{equation} \label{gr_ho}
 \rho_{\theta}(x) = \delta_{\tilde{\sigma_\theta}} \left( x - \theta E_0 \right)
 \end{equation}
 
 It may be noted the probability density-independent of time and is positive definite. We can now study the effect of infinitely large confining potential, by considering the limit $\omega \to \infty$ in presence of nonzero noncommutativity $(\theta \neq 0)$. In this limit, $\tilde{\sigma_\theta} \to \sqrt{\theta}$, preventing the squeezing of the wave-packet in a region $\lesssim \sqrt{\theta}$. This is a purely a nontrivial effect of noncommutativity as the inherent noncommutativity in space-time provides a natural impenetrable barrier and does not allow for localization to a sharply defined point. Thus the probability density $\rho(x)$ at some point $x$ will have contributions from the vicinity and points from the finite, however small, region around that point. In a certain sense, noncommutativity therefore thus essentially introduces a non-locality within the theory, where the notion of a point particle, it seems, has got to be necessarily replaced by some extended object of a `cloud' A similar situation was observed also in the context of spatial noncommutativity \cite{RR}.\\

Interestingly, if we takes both the limits $\omega \to \infty$ and $\mid \theta\mid \to 0$, in such a way that their product is a finite constant: $\omega\mid\sqrt{\theta}\mid \rightarrow~ finite~constant $, then $\rho(x) \to \delta(x)$, as one can expect in the commutative limit. One should note that the the non-local features appear for both the ground state wave function \eqref{ncwh} and also in the expression of the probability density \eqref{gr_ho_pd}. One can thus expect to see the effect of non-commutativity through non-local behaviours in some measurable quantities or observables .\\

\section{Transition probability in presence of space-time noncommutativity} \label{sec8}

In this section, we intend to investigate the modification due to the presence of space-time noncommutativity, if any, in the rate of transition probability amplitude of a system when it undergoes a transition from a given initial stationary state to a particular final stationary state. For this study let us first recall what we did for the case of time dependent perturbation theory in ordinary quantum mechanics, described by a Hamiltonian $H_0$ and satisfies time independent Schr\"odinger equation as
\begin{equation}
H_0 |\psi\rangle_{n}=E_n |\psi\rangle_{n}.
\end{equation}
These stationary eigenstates $|\psi\rangle_{n}$ evolve in time as
\begin{equation}
|\psi(t)\rangle_{n}=e^{-iE_nt}|\psi\rangle_{n}.
\end{equation}
Now if we switch on a perturbation say $V(t)$ then its time evolution will be governed by the Hamiltonian $H=H_0 + V(t)$ satisfying time dependent Schr\"odinger equation as
\begin{equation}
H|\psi(t)\rangle = i\frac{\partial}{\partial t}|\psi(t)\rangle,
\label{timed}
\end{equation}
We work with the ansatz for the perturbed state-vectors $|\psi(t)\rangle=\sum_{n} C_n(t) e^{-iE_{n}t}|\psi\rangle_{n}$ where the coefficients $C_n(t)$ are themselves explicitly time dependent. \\

In our non-commutative case, for physical states the above equation (\ref{timed}) can be rewritten as
\begin{equation}
H_0|\Psi; phy)+V(\hat{t})|\Psi;phy) = -P_t|\Psi;phy).
\label{pertu}
\end{equation}
Now using coherent state basis \eqref{cohe}, above equation will take the following form as
\begin{equation}
(x,t|H_0|\Psi;phy)+(x,t|V(\hat{t})|\Psi;phy) = i\frac{\partial}{\partial t}(x,t|\Psi;phy),
\end{equation}
or equivalently,
\begin{equation}
H_0(P_x,X_\theta) \, \Psi_{phy}(x,t) + V(T_\theta) \, \Psi_{phy}(x,t) = i \, \partial_t \Psi_{phy}(x,t).
\label{kuj}
\end{equation}
Let us now consider a generic non-stationary physical state which satisfies the above Schr\"odinger equation and may be expanded as
\begin{equation}
\Psi_{phy}(x,t) = \sum\limits_n C_n (t) e^{-iE_n t}\phi^{(n)}(x;\theta)
\end{equation}
with $ H_{0}(P_{x},X_{\theta})\phi^{(n)}(x;\theta)=E_{n}\phi^{(n)}(x;\theta).$ Here the subscript $n$ implies to their association with the $n-th$ energy levels $E_n$ of the unperturbed Hamiltonian $H_{0}$. Now we should insert the expression of $\Psi_{phy}(x,t)$ in the above Schr\"odinger equation (\ref{kuj}) and solve with the initial condition $C_{n}(t=0)=\delta_{ni}$ (as the system is in the initial state  $\phi^{(i)}(x;\theta)$ at $t=0$). Proceeding thus we have
\begin{equation}
\begin{split}
\sum_n i\dot{C}_n (t) \, & e^{-iE_n t}\phi_n(x;\theta) + \sum_n E_n C_n (t) \, e^{-iE_n t}\phi_n(x;\theta) \\
& = \sum_n H_0(P_x,X_\theta) \, C_n (t) \, e^{-iE_n t}\phi_n(x;\theta) + \sum_n V(T_\theta) \, C_n (t) \, e^{-iE_n t}\phi_n(x;\theta) \
\end{split}
\end{equation}
Simplifying,
\begin{equation}
\sum_n i\dot{C}_n (t) \, e^{-iE_n t}\phi_n(x) = \sum_n V(T_\theta) \, C_n (t) \, e^{-iE_n t}\phi_n(x).
\end{equation}

Now using (\ref{time}) we obtain
\begin{equation}
\sum_{n} i \dot{C}_{n} e^{-E_{n}t}\phi^{(n)}(x;\theta)=\sum_{n}  S^{\dagger}V(t)(S^{\dagger})^{-1} C_{n}(t) e^{-iE_{n}t} \phi^{(n)}(x;\theta)
\label{eqi}  
\end{equation}
where we have merely used the fact that $V(S^{\dagger} t(S^{\dagger})^{-1})=S^{\dagger}V(t)(S^{\dagger})^{-1}.$\\

If the time-dependent perturbation $V(t)$ is weak then $C_{n}$for $n\neq i$ will also be a slowly varying function of time and its higher-order time derivative can be disregarded to calculate the rate of transition probability. At this stage, it is important to observe that the coherent state representation of any bound stationary states ($e^{-iE_{n}t}\phi^{(n)}(x;\theta)$) in non-commutative space-time is related to  the stationary states ($e^{-iE_{n}t}\phi_{c}^{(n)}(x)$) of the corresponding
model in commutative space-time through the non-unitary operator (\ref{twist}):
\begin{equation}
e^{-iE_{n}t}\phi_{c}^{(n)}(x)= S^{-1}[ e^{-iE_{n}t}\phi^{(n)}(x)]
\end{equation} 

Now, after some algebra, the above equation (\ref{eqi}) may be rewritten as 
\begin{equation}
\sum_{n} i \dot{C}_{n}~ e^{-E_{n}t}\phi_{c}^{(n)}(x)=\sum_{n}C_{n}(t)~  [UV(t)U^{-1}]~e^{-iE_{n}t} \phi_{c}^{(n)}(x).  
\end{equation}
Therefore
\begin{equation}
\sum_{n} i \dot{C}_{n}(t)~ e^{-E_{n}t}\phi_{c}^{(n)}(x)=\sum_{n}C_{n}(t)~  V(UtU^{-1})~e^{-iE_{n}t} \phi_{c}^{(n)}(x),  
\end{equation}
with $U=e^{i\theta\partial_{x}\partial_{t}}.$  All information about space-time noncommutativity may be encoded into this unitary operator $U$.\\

Finally, after using the Hadamard lemma, we arrive at
\begin{equation}
\sum_{n} i \dot{C}_{n}(t)~ e^{-E_{n}t}\phi_{c}^{(n)}(x)=\sum_{n}C_{n}(t)~  V(t+i\theta \partial_{x})~e^{-iE_{n}t} \phi_{c}^{(n)}(x),
\end{equation}

Using a Taylor expansion up-to first order in $\theta$ (as $\theta$ is taken to be very small and the perturbation is also weak), we can write,
\begin{equation}
\begin{split}
\sum_n i\dot{C}_n (t) \, e^{-iE_n t} \, \phi^{(n)}_{c}(x) & \simeq \sum_n V(t) \, C_n (t)e^{-iE_n t} \phi^{(n)}_c(x) \\
& -\frac{\theta}{2}\sum_n C_n (t)\frac{\partial V(t)}{\partial t} \left(-i \partial_x \right) e^{-iE_n t}\phi_{c}^{(n)}(x) + \mathcal{O} (\theta^2) \
\end{split}
\end{equation}
Now taking the scalar product with the  $e^{iE_mt} \, \phi^{*(m)}_{c}(x)$ we arrive at
\begin{equation}
\begin{split}
 i\dot{C}_m (t)  & \simeq  \sum_n  \, C_n (t) \, e^{i(E_{m}-E_{n})t}\int ~dx~\phi_{c}^{\ast(m)}(x) \,V(t)  \,  \, \phi_{c}^{(n)}(x) \\
& - \frac{\theta}{2} \sum_n C_n (t) \, e^{i(E_{m}-E_{n})t}\int~dx~\phi_{c}^{\ast(m)}(x) \,  \, \frac{\partial V(t)}{\partial t} \, (-i\partial_x) \,  \, \phi_{c}^{(n)}(x) \
\end{split}
\label{sl}
\end{equation}

Since the right-hand side of the above equation (\ref{sl}) has a matrix element of $V$and it suffices to retain the term $C_{i} (t)$  there and  for weak perturbation $V(t)$ the coefficient $C_i (t)$ are approximately same as $C_i (t=0)$ and $C_{i}$ too in this order of approximation is 1, and so lowest contributing order $\theta$ is given by
 \begin{equation}
 i\dot{C}_m (t) \simeq e^{i(E_{m}-E_{n})t}[\, < \phi_{c}^{(m)} |V(t)|\phi_c^{(i)} > - \frac{\theta}{2}  \langle \phi_{c}^{(m)} |\frac{\partial V(t)}{\partial t}\hat{P}_{x}| \phi_{c}^{(i)} \rangle]
 \end{equation}
 Therefore if the system undergoes transition in time $T$, then for initial state $| \phi_{c}^{(i)} \rangle$ and final state $| \phi_{c}^{(f)} \rangle$, we have
 
 \begin{equation}
C_f (T) \simeq -i \int_0^T dt \,e^{i(E_{f}-E_{n})t} \left[ < \phi_{c}^{(f)} |V(t)|\phi_c^{(i)} > - \frac{\theta}{2}  \langle \phi_{c}^{(f)} |\frac{\partial V(t)}{\partial t}\hat{P}_{x}| \phi_{c}^{(i)} \rangle \right]
 \end{equation}
 
 and the transition probability for the transition $i\rightarrow f$ is
 \begin{equation}
 Prob_{f\leftarrow i}=\mid C_{f}(T)\mid^{2}=\mid \int_0^T dt \,e^{i(E_{f}-E_{n})t} \left[ < \phi_{c}^{(f)} |V(t)|\phi_c^{(i)} > - \frac{\theta}{2}  \langle \phi_{c}^{(f)} |\frac{\partial V(t)}{\partial t}\hat{P}_{x}| \phi_{c}^{(i)} \rangle \right]\mid^{2}
 \end{equation}
 
 The transition rate of the system for a total time $T$ is then given by
 \begin{equation}
\boldsymbol{Rate}_{f\leftarrow i}=\frac{1}{T}\mid \int_0^T dt \,e^{i(E_{f}-E_{i})t} \left[ < \phi_{c}^{(f)} |V(t)|\phi_c^{(i)} > - \frac{\theta}{2}  \langle \phi_{c}^{(f)} |\frac{\partial V(t)}{\partial t}\hat{P}_{x}| \phi_{c}^{(i)} \rangle \right]\mid^{2}
 \end{equation}
In the smooth limit $\theta \rightarrow 0$, we get back the commutative result. The explicit presence of $\theta$ dependent correction term in the relevant first-order transition rate is clearly a non-commutative effect for which this rate is found to get modified. Thus the decay of state due to the interaction in non-commutative space-time is beautifully exhibited in this approach, and this demonstration clearly indicates that the so-called Fermi-Golden rule has to be modified in a nontrivial way.
\section{Discussion}
 A toy-model of  $(1+1)$-dimensional non-relativistic quantum mechanics on quantum space-time have been studied in this chapter, wherein the quantum structure of space-time arising from the canonical type of space-time non-commutative algebra  \eqref{e1} which is relevant to capture the essential features of the new quantum geometry of space-time. In fact, in Appendix A, using Dirac's constraint analysis, we obtain this non-commutative space-time structure from a first-order non-relativistic system Lagrangian.
It is important to note that we did not fix any type of gauge to get the non-commutative space-time algebra; this deformed non-commutative space-time algebra was induced by a Chern-Simons like term in momentum space. Therefore, in this sense, our method is different from some other approaches where suitable gauge fixing was mandatory \cite{rbbc}. Then we show that by introducing coherent state basis \eqref{cohe} it is possible to extract an effective commutative theory starting from an operatorial version of Schr\"odinger equation \eqref{e70} using Hilbert-Schmidt operators. This equation now enjoys a similar form to that of commutative quantum mechanics except that the point-wise product of two functions is necessarily deformed by the Voros star product, ensuring positive definiteness of probability density. Also since the star products involve terms with infinite order derivatives of spatio-temporal variables, this theory becomes highly non-local. We started from a time-re-parametrization invariant form of action, wherein both space and time coordinate are treated as a dynamical variables, in the sense that they are both parts of an enlarged configuration space and eventually are promoted to the level of quantum operators. However, the corresponding Hilbert space is `'bigger' in the sense that we needed to introduce a space of physical states that are equipped with the so-called `induced' inner product to enable us to provide the conventional probabilistic interpretation. And this is clearly less stringent than the one used to define $L^2(R^2)$.\\

Moreover, we have discussed how this formulation can be applied for the case of free Gaussian wave packet and harmonic oscillator. For the free particle case, we show explicitly that, the inherent noncommutativity does not allow to localize the particle to a single point even if one introduces an infinite uncertainty in the momentum space. We then compute harmonic oscillator spectra, where we observed that the spectra remain same as we expect for a usual oscillator. But, the corresponding energy eigenfunctions get deformed. Specifically, the ground state wave functions indicate a parity-violating shift in the origin and a modified width of the wave packet. We also show that for an infinitely large confining potential, it is not possible to squeeze the position of a particle under a threshold scale. This is certainly a non-local feature of the theory arising purely from space-time noncommutativity. We compute the rate of transition probability of a  bound system under time-dependent perturbation, which gives rise to a similar deformation in Fermi's golden rule.\\\\\\

\section{Appendix A } 
\textbf{ Dirac procedure for generating the deformed space-time algebra}\\

Starting with the first-order Lagrangian written below, here we are going to carry out Dirac's constraint Hamiltonian analysis and compute the Dirac's bracket among the configuration space variables where space-time and their corresponding canonical momenta are considered to be configuration space variables of extended phase-space as, 
\begin{equation}
L_{\theta}^{\tau}= p_{\mu}\dot{x}^{\mu}+\frac{\theta}{2}\epsilon^{\mu\nu}p_{\mu}\dot{p}_{\nu}-e(\tau)(p_t+H),\,\,\,\,\,\,\mu,\nu=0,1 \label{e80}
\end{equation}
 where $e(\tau)$  is an arbitrary Lagrange multiplier enforcing the constraint $(p_{t}+H)\approx 0$ and the $\theta$ dependent term can be viewed as a Chern-Simons term in the momentum space. Now we can derive the canonical momenta corresponding to $x^{\mu}, p_{\mu}$ and $e(\tau)$ as follows:
\begin{align}
\pi^x_{\mu}&=\frac{\partial L_{\theta}^{\tau}}{\partial \dot{x}^{\mu}}=p_{\mu};\nonumber\\
\pi_p^{\mu}&=\frac{\partial L_{\theta}^{\tau}}{\partial \dot{p}_{\mu}}=-\frac{\theta}{2}\epsilon^{\mu\nu}p_{\nu}; \nonumber\\
\pi_{e}&=\frac{\partial L_{\theta}^{\tau}}{\partial\dot{e}}=0\label{e81}
\end{align}
We can see that the canonical momenta are not related to the generalized velocities and so that the above equations (\ref{e81}) are interpreted as primary constraints of the theory and can be written as

\begin{equation}
\Phi_{1,{\mu}}=\pi_{\mu}^x-p_{\mu}\approx 0,\,\,\,\,\,\,\,\Phi_2^{\mu}=\pi^{\mu}_p+\frac{\theta}{2}\epsilon^{\mu\nu}p_{\nu}\approx 0,\,\,\,\,\,\,\,\Psi = \pi_{e} \approx 0\label{e82}
\end{equation}
Using the following Poission brackets between the canonical pairs,
\begin{equation}
\{x^{\mu},\pi_{\nu}^x\}=\delta^{\mu}\,_{\nu},\,\,\,\,\,\,\{p_{\mu},\pi^{\nu}_p \} =\delta_{\mu}\,^{\nu}; \{e,\pi_{e}\}=1
\end{equation}
we can derive the Poisson brackets between the constraints (\ref{e82}) as
\begin{align}
&\{\Psi,\Psi\} =\{\Psi,\Phi_{1,\mu}\} = \{\Psi,\Phi_2^{\mu}\} =0\nonumber\\
&\{\Phi_{1,{\mu}},\Phi_{1,{\nu}}\} =0\nonumber\\
&\{\Phi_{1,{\mu}},\Phi_2^{\nu}\} = -\delta_{\mu}\,^{\nu}\nonumber\\
&\{\Phi_2^{\mu},\Phi_2^{\nu}\}= \theta\epsilon^{\mu\nu}\label{e84}
\end{align}
The sign $\approx$ refers to weak equality in the sense of Dirac. By simple Legendre transformation, we can obtain the canonical Hamiltonian as
\begin{equation}
H_c= e(p_t+H)\label{e85}
\end{equation}
Since $\Psi$ has zero brackets with all other constraints, it can be classified as a first-class constraint. The rest of the constraints are classified as second class constraints as the Poisson's bracket between of $\Phi_{1,{\mu}}$ and $\Phi_{2,\mu}$ does not vanish. We can eliminate the second class constraints by calculating the Dirac brackets.\\

Now the total Hamiltonian can be written by adding all  primary constraints to the canonical Hamiltonian as following
\begin{equation}
H_T=e(\tau)(p_t+H)+\lambda_{1,\mu}\Phi_{1,\mu}+\lambda_{2,\mu}\Phi_2^{\mu}+\lambda \Psi \label{e86}
\end{equation}
where $\lambda_{1,\mu},\lambda_{2,\mu}$ and $\lambda$ are the suitable Lagrange's multipliers enforcing the respective constraints.. Now the constraint should be zero at all times. So for a consistency check we are going to derive the time derivative of the constraints using Hamiltonian equation and check whether it gives zero; at least weakly as, 
\begin{equation}
\Phi:=\dot{\Psi}=\{H_T,\pi_{e}\} = p_t+H \approx 0 \label{e87}
\end{equation}
Hence, the consistency condition on the time derivative of $\Psi$ gives a secondary constraint  (\ref{e87}), i.e. the original constraint $\phi$ (\ref{3.18}) appears as a secondary constraint. No further new constraint is generated due to the time conservation of secondary constraint (\ref{e87}) and the iterative process thus terminates. The consistency condition of the other primary second class constraints is used to fix the Lagrange multiplier $\lambda_{1,\mu},\lambda_{2,\mu}$, which are not of our concern in this context.\\

Before deriving the Dirac brackets, we need to write the constraint matrix and its inverse, which are given below.
\begin{equation}
\Lambda_{ab}= \begin{pmatrix}
\{\Phi_{1,{\mu}},\Phi_{1,{\nu}}\}&\{\Phi_{1,{\mu}},\Phi_2^{\nu}\}\\
\{\Phi_2^{\nu},\Phi_{1,{\mu}}\}&\{\Phi_2^{\mu},\Phi_2^{\nu}\}
\end{pmatrix}=\begin{pmatrix}
0&-\delta_{\mu}\,^{\nu}\\
\delta^{\nu}\,_{\mu}&\theta\epsilon^{\mu\nu}
\end{pmatrix}\label{e88}
\end{equation}
\begin{equation}
(\Lambda^{-1})_{ab}= \begin{pmatrix}
\theta \epsilon_{\mu\nu}&\delta^{\mu}\,_{\nu}\\
-\delta_{\nu}\,^{\mu}&0
\end{pmatrix}\label{e89}
\end{equation}
Note that $\Lambda_{ab}$ and $(\Lambda^{-1})_{ab}$ satisfy $\Lambda_{ab}(\Lambda^{-1})_{bc}=\delta_{ac}$.\\

Knowing the inverse of the constraint matrix, we can easily calculate
the Dirac bracket between two phase-space variables as,
\begin{equation}
\{f,g\}_D= \{f,g\}-\{f,\Phi_a\}(\Lambda^{-1})_{ab}\{\Phi_b,g\}
\end{equation}
A straightforward computation then yields
 the relevant Dirac brackets between the congiguration space variables as given below:
\begin{align}
&\{x^{\mu},x^{\nu}\}_D=\theta\epsilon^{\mu\nu}\nonumber\\
&\{p_{\mu},p_{\nu}\}_D=0\nonumber\\
&\{x^{\mu},p_{\nu}\}_D=\delta^{\mu}\,_{\nu}
\end{align}
So we can see that the Dirac bracket between the space-time variable gives a non-zero result. We can now easily quantize the system by elevating the classical variables to the level of operators and elevating the Dirac brackets to the level of the commutator bracket.\\

We will also comment on the status of the primary first-class constraint $\Psi$ and the secondary constraint $\Phi$. The primary first-class constraint $\Psi$ (\ref{e82}) involves the canonical momentum corresponding to the Lagrange's multiplier $e(\tau)$, which is not physical in nature. So we can ignore this constraint. On the other hand, if we compute the Dirac bracket between the secondary constraint $\Phi$ (\ref{e87}), and all other primary constraints we will get vanishing brackets. So $\Phi$ can be treated as a non-trivial first-class constraint that eventually gives rise to the $\tau$ evolution of the system or the generator of the gauge transformation of the theory.\\

\section{Appendix B}
\textbf{ Self adjoint property of quantum space-time operators ($X^{L}_{\theta}$, $T^{L}_{\theta}$) on $\mathcal{A}_{\theta}$}\\

Let us define the 'symbol' mapping

\begin{align}
&\mu: \mathcal{A}_{\theta} \rightarrow \mathbb{C},\nonumber\\
&\mu(\Psi(\hat{t},\hat{x}))=(z,\bar{z}|\Psi)=<z|\Psi(\hat{x},\hat{t})|z>
\end{align}

Hence it can be shown that this is a homomorphism from $\hat{\mathcal{A}}_{\theta}$ to $\mathbb{C}$ with the Voros star product $\star_{V}$ viz.
\begin{equation}
\mu(\Psi_{1} (\hat{x},\hat{t})\Psi_{2}(\hat{x},\hat{t})=(z,\bar{z}|\Psi_{1})\star_{V}(z,\bar{z}|\Psi_{2});~~|\Psi_{1}),|\Psi_{2})\in \hat{\mathcal{A}}_{\theta}
\end{equation}
As $\mu$ is onto and $\mu(I)=<z|I|z>=1$, this implies that $\mu$ is actually an isomorphism.\\

We therefore have 
\begin{equation}
\mu(\Psi_{1}^{\dagger}(\hat{x},\hat{t}) \hat{x}\Psi_{2}(\hat{x},\hat{t}))\rightarrow \Psi^{*}_{1}(x,t)\star_{V} (x\star_{V}\Psi_{2}(x,t))=\Psi^{*}_{1}\star_{V}X^{L}_{\theta}\Psi_{2}(x,t)
\end{equation}

Now the associativity of the Voros star product implies
\begin{equation}
\Psi^{*}_{1}(x,t)\star_{V} (x\star_{V}\Psi_{2}(x,t))=(\Psi^{\star}_{1}(x,t)\star_{V}x)\star_{V}\Psi_{2}(x,t)
\end{equation}
\begin{equation}
\implies \Psi^{*}_{1}\star_{V}X^{L}_{\theta}\Psi_{2}(x,t)= (X^{L}_{\theta}\Psi_{1}(x,t))^{*}\star_{V}\Psi_{2}(x,t)
\end{equation}
 After integrating both sides with respect to $x$ we obtain
\begin{equation}
\int dx~\Psi^{*}_{1}\star_{V}X^{L}_{\theta}\Psi_{2}(x,t)=\int dx~(X^{L}_{\theta}\Psi_{1}(x,t))^{*}\star_{V}\Psi_{2}(x,t)
\end{equation}
Thus the operator $X^{L}_{\theta}$ is  self adjoint with respect to induced inner product (\ref{op1}). A similar calculation for the operator $T^{L}_{\theta}$ indeed gives rise to  the self-adjoint nature for the induced inner product itself.

 

\chapter{Conclusions}
Even after several decades of intellectual investment, quantum gravity still remain, arguably, the most important open problem in fundamental physics. The problem of quantum gravity is being pursued through different approaches, where the primary objective is to combine general relativity with quantum mechanics. Although none of these theories are nowhere near their final stage and many of their results remain tentative in nature, almost all these competing approaches have one result in common and this is some form of granular structure i.e. quantum space-time at Planck length scale,
although there is no consensus regarding the final mathematical structure of such quantum space-time. For example, in the approach of loop quantum gravity \cite{LCqg}, the space-time was shown to be describable through certain spin-network, where each node can be identified as ``quantum of space". Furthermore, it was shown that area/ volume of a surface/ region are quantized in discrete units and which are computable within the scope of the theory. On the other hand, in the approach of non-commutative geometry, the structure of space-time is taken to be non-commutative in nature, where space-time coordinates are promoted to the level of operators, satisfying non-vanishing commutator algebra. This can serve as possible deterrent against the possibility of forming micro black holes, arising in any attempt to localize an event to this scale and thereby effectively removing that part from the observable region.  
This non-commutative structure of space-time was later corroborated by string theory also, as a sort of low energy description.The main objective of this thesis was to study different quantum theories on quantum space-time manifested through its non-commutativity, i.e. of the latter type, taken in its simplest form: the Moyal space/ space-time.\\

In the second chapter, we have shown how spatial noncommutativity emerges in a
planar quantum mechanical problem by taking the example of an interacting electric
dipole subjected to a constant magnetic field in the normal direction. In this context, it is worth-mentioning that the parent dipole model had particular applications in
the context of fractional quantum Hall effect (FQHE) for filling factor $\nu=\frac{1}{2}$ \cite{Vhe}. Here
noncommutativity appeared as an effective way of describing the interaction with external (magnetic) field. Then we have adopted a Hilbert-Schmidt operator formulation
to provide a systematic quantum mechanical analysis of the non-commutative model. A generating functional was constructed through a coherent state path-integral approach. As
we know, quantum mechanical models may be interpreted as (0 + 1) dimensional field
theory and can be regarded as a precursor to genuine higher dimensional (i.e. $(2+1)$ or $(3+1)$- dimensional) field theories. Thus we could have interpreted our planar theory by a 0 + 1 dimensional non-canonical complex scalar field theory. The advantage of this perspective is that it facilitated the use of many of the arsenals of QFT even in our quantum mechanical case. With this, we could analyze the issues related to the time dilatation symmetry on non-commutative spaces
in the chapter. By exploiting the
path-integral analysis, due to Fujikawa, wherein the anomalous terms are identified with the Jacobian factors obtained from the path-integral measure under dilatation transformation, we showed that the Ward-Takahashi identities for such $0 + 1$ dimensional non-commutative fields associated with broken scale invariance contain anomalies in all orders.\\

In next chapter, we had studied emergence of the geometrical phase in a time-dependent
quantum mechanical system in non-commutative phase space. By making use of proper non-canonical transformations we mapped the problem on effective commutative space and determined the
energy spectrum. Thereafter, exploiting the adiabatic circuital theorem, we obtained
an expression of geometrical phase over and above the dynamical phase. Our findings agreed with the group-theoretical analysis of Berry's phase. As far as the physical origin of the spatial non-commutative parameter is concerned, it can be identified with a non-relativistic fractional spin in $2 + 1$ dimension. Given the recent findings of anyonic excitations in some experiments \cite{HKB}, we believe our theoretical results may become more relevant in explaining some of the observations in  anyonic systems in near future.\\

We then take up issues related to space-time noncommutativity in non-relativistic quantum mechanics in chapter 4. Reperametrization  invariant form of $(1 + 1)$-dimensional action for a non-relativistic system is first considered. Here we have shown, how non-commutative space-time can appear naturally through the Dirac bracket of
second class constraints, in presence of a Chern-Simons like term in momentum space. Here Non-commutativity (NC) can be interpreted as
a fundamental constant of space-time. Then introducing appropriate Hilbert space, a time-dependent Schr\"odinger equation in $1 + 1$ dimensional non-commutative space-time was obtained. A salient feature of our approach is that there is no issue of
violating unitarity. The time-independent non-relativistic model of a free particle and a harmonic oscillator living in this non-commutative space-time is then analysed. We found that the spectrum of the system Hamiltonian remains unchanged, making it virtually impossible to see any effect of noncommutativity in the dynamics of the system. Rather, the wave function gets deformed. Additionally, for the case of time-dependent perturbation, the non-commutative correction is shown to appear as a perturbative
expansion in the deformation parameter. By the explicit calculation, we demonstrated
that the leading order correction of non-commutativity  can have quite significant contribution in the rate of transition amplitude.\\

Thus, we studied different aspects of non-commutative space-time in quantum mechanics through the lens of $0 + 1$ dimensional nonlocal quantum field theory. It was quite reassuring to see that in the limit of vanishing deformation parameters the quantum theories reduced smoothly to their usual commutative versions. In this way, we have successfully illustrated that if we revise the notion of space-time at a short length scale by a non-commutative geometry associated with one more deformation parameters, these could then be considered as fundamental ingredients of a
quantum theory of gravity. Moreover, non-commutative quantum theories had many novel properties. In this thesis, we have been strictly confined  our attention to the planar quantum systems. It would thus be an obvious suggestion for future research to generalize
this formalism to higher dimensions. As a future goal, we can extend Hilbert Schmidt-operator formulation of non-commutative quantum mechanics to the coherent state
description of $\mathcal{P}$ representation of mixed density matrix, which may admit an immediate hydrodynamics interpretation \cite{gbd} corresponding to the incompressible quantum
fluid. On the other hand, the use of multi-particle systems as a theoretical model for
searching some possible Planck scale manifestations has become a very fascinating line
of research \cite{1,2,3}. Thus, our present investigation may be the basis to develop
the second quantized description of Schrödinger field theory over non-commutative
space-time through the Hilbert-Schmidt operator approach, which can perhaps provide a natural
background to obtain possible signals caused by the quantum fluctuation of space-time
in low-energy earth-based experiments. Furthermore, quantum space-time structure
arising from the relevant nontrivial commutator of $[\hat{t},\hat{x}]$ also may be extended to
understand the structure of quantum Minkowski, Rindler, and Schwarzschild space-times and study their implications.\\


\end{document}